\documentclass[aps,prb,reprint,groupedaddress,twocolumn,eqsecnum]{revtex4-1}

\usepackage{color} 
\usepackage{bm,bbm}

\usepackage{graphics,graphicx} 

\usepackage{amsthm,amsmath,mathtools,amssymb,amsfonts,wasysym} 
\usepackage[mathscr]{euscript} 

\usepackage[colorlinks=true,citecolor=blue,linkcolor=blue,urlcolor=black]{hyperref} 


\newcommand{\e}{\mathrm{e}}
\newcommand{\diff}{\mathrm{d}}

\newcommand{\rr}{{\bf r}}

\newcommand{\kk}{{\bf k}}
\newcommand{\DD}{ {\bf D}}
\newcommand{\AAA}{ {\bf A}}
\newcommand{\xx}{{\bf x}}
\newcommand{\qq}{{\bf q}}

\newcommand{\lql}{\|q\|}
\newcommand{\half}{\tfrac{1}{2}}

\newcommand{\idk}{\int \frac{\diff^2 k}{(2\pi)^2}\;}
\newcommand{\sk}{\frac{1}{\beta}\sum_{k_0}}

\newcommand{\dtheta}{\int \frac{\diff \theta_\kk}{2\pi}\;}

\newcommand{\kervq}{\frac{v_F\delta q}{v_F \delta q- iq_0}}
\newcommand{\keriq}{\frac{iq_0}{v_F \delta q- iq_0}}
\newcommand{\kertan}{\frac{v_F |\qq|}{\lql + |q_0|}}
\newcommand{\tkertan}{\tfrac{v_F |\qq|}{\lql + |q_0|}}

\newcommand{\feny}[2]
{\begin{array}{l}\includegraphics[#1]{#2}\end{array}}

\newcommand{\snabla}{{\overset{\leftrightarrow}{\nabla}}}
\newcommand{\spartial}{{\overset{\leftrightarrow}{\partial}}}
\newcommand{\sDD}{{\overset{\leftrightarrow}{\DD}}}
\newcommand{\sD}{{\overset{\leftrightarrow}{D}}}

\newcommand{\nl}{ \nonumber \\}

\definecolor{darkred}{RGB}{139,0,0}
\definecolor{crimson}{RGB}{220,20,60}

\def\Hline{\noindent{\color{black}\rule{\textwidth}{0.4pt}}}

\newcommand{\GG}{{\mathcal{G}}}
\newcommand{\OO}{\mathscr{O}}

\begin{document}

\title{Effective field theory of an anomalous Hall metal from interband quantum  fluctuations}
\date{\today}
\author{Victor Chua}
\author{Wathid Assawasunthonnet}
\author{Eduardo Fradkin}
\affiliation{Department of Physics, and Institute for Condensed Matter Theory, University of Illinois at Urbana-Champaign, 1110 West Green Street, Urbana, Illinois 61801-3003, USA}

\begin{abstract}

We construct an effective field theory a two-dimensional two-component metallic system described by a model with two Fermi surfaces (``pockets''). This model describes a translationally invariant metallic system with two types of fermions, each with its own Fermi surface, with forward scattering interactions. This model, in addition to the $O(2)$ rotational invariance, has a $U(1) \times U(1)$ symmetry of  separate charge conservation for each Fermi surface. For sufficiently attractive interactions in the $d$-wave (quadrupolar) channel this model has an interesting phase diagram that includes a spontaneously generated anomalous Hall metal phase. We derive the Landau-Ginzburg effective action of quadrupolar order parameter fields which enjoys an $O(2)\times U(1)$ global symmetry associated to spatial isotropy and the internal $U(1)$ relative phase symmetries respectively. We show that the order parameter theory is dynamically local with a dynamical scaling of $z=2$ and perform a one-loop renormalization group analysis of the Landau-Ginzburg theory. The electronic liquid crystal phases that result from spontaneous symmetry breaking are studied and we show the presence of Landau damped Nambu-Goldstone modes at low momenta that is a signature of non-Fermi liquid behavior. Electromagnetic linear response is also analyzed in both the normal and symmetry broken phases from the point of view of the order parameter theory. The nature of the coupling of electromagnetism to the order parameter fields in the normal phase is non-minimal and  decidedly contains a precursor to the anomalous Hall response in the form of a order-parameter-dependent Chern-Simons term in the effective action. 

\end{abstract}

\maketitle 

\section{Introduction}


Interest in soft electronic liquids\cite{fradkin2010nematic,fradkin2015colloquium} grew out of attempts to understand the rich phase diagram of the cuprates which hosts high temperature superconductivity.\cite{kivelson1998electronic} This approach draws its inspiration from classical soft-condensed matter. Specifically, classical liquid crystals where competing orders and entropic gains from broken translational and/or rotational symmetries can lead to a plethora of energetically competitive phases. Electronic liquid crystal phases, the quantum analogs of classical liquid crystals, have been confirmed in several electronic condensed matter systems such as in quantum Hall fluids,\cite{du1999strongly,lilly1999evidence} stripe and nematic phases in the cuprate superconductors,\cite{kivelson2003detect,fradkin2015colloquium}  and  the iron-based superconductors,\cite{fernandes2014review} and  in S$_3$Ru$_2$O$_7$.\cite{borzi2007formation}  

Among the many possible soft electronic phases, there are those electron fluids that retain their translational symmetry but break orientational isotropy either spatially\cite{oganesyan2001quantum,lawler2006nonperturbative,barci2003strongly} and/or internally i.e. the spin degree of freedom \cite{wu2004dynamic,wu2007fermi,fu2015parity} or a band (`pseudo-spin') degree of freedom.\cite{sun2008time} From the weak-coupling perspective,  this order is driven by a tendency of the Landau Fermi-liquid to develop a Pomeranchuk instability.\cite{pomeranchuk1958stability} Near the transition point into an electronic liquid crystal, the soft modes\cite{soft} of the theory are shape distortions of the Fermi-surface(s) (FS) that may be mixed with internal degrees of freedom.\cite{wu2004dynamic,wu2007fermi} Moreover, the spontaneous symmetry breaking of the orientational isotropy leads to non-Fermi liquid behavior\cite{oganesyan2001quantum} due to Landau damping of the particle-hole soft collective modes near the FS.


In this paper, we revisit a simple model  introduced in Ref.[\onlinecite{sun2008time}] of a soft 2D electron metal that has an interaction-driven unquantized anomalous Hall effect.\cite{nagaosa2010anomalous} In its simplest form the model consists of a 2D metal with nearly degenerate orbital degrees of freedom with two FSs nested inside each other with  forward scattering  attractive inter-band interactions in the $d$-wave channel. This simple model may also be realized in a strongly layered metallic system in a perpendicular (Zeeman) magnetic field with an easy-plane magnetic anisotropy with quadrupolar exchange interactions. In this case the two species of electrons are labelled simply by the spin projection and the forward scattering interactions are the $XY$ quadrupolar exchange interactions. The soft modes can also be thought of as `nematic Stoner excitations' which mediate residual forward $XY$-exchange scattering in the $l=2$ partial wave channel after the establishment of ferromagnetism. Here we analyze this soft electronic liquid using weak coupling perturbation theory in the vicinity of its quantum critical point within the Hertz-Millis approach.\cite{hertz1976quantum,millis1993effect} Due to the $XY$-exchange-type fluctuations which are inter-band and gapped in character, the critical theory is that of a non-relativistic boson with $z=2$  dynamical scaling exponent. Throughout this paper we will use the terminology of a pseudo-spin model but the translation to other cases is straightforward.

The mean-field theory phase diagram of this model has three phases: normal metal and two broken-symmetry states (dubbed $\alpha_1$ and $\beta_1$ in Ref.[\onlinecite{sun2008time}]) representing a nematic-like state ($\alpha_1$) which is invariant under a rotation by $\pi/2$ followed by an orbital exchange, and an anomalous quantum Hall metal ($\beta_1$) with a spontaneously broken time-reversal invariance. All three  phases are gapless. In this work we will consider the effects of quantum fluctuations about these mean field results. To this end we derive an effective low-energy field theory of the soft collective modes in the normal phase near the quantum phase transition to the $\alpha_1$ and $\beta_1$ phases.
At the level of a one-loop renormalization group (RG) treatment we find that the three phases meet  at a quantum tricritical point. We further investigate the electromagnetic response of these phases.
The main results of this paper that follow from this analysis are the electromagnetic linear response kernels that contain the direct coupling between the electromagnetic gauge field and the soft quadrupolar modes (at quadratic order) near the critical point of the theory. We  show that the Ward identities for the electromagnetic $U(1)$ gauge symmetry are necessarily satisfied.  
The non-minimally electromagnetic coupling to the soft quadrupolar modes has a subtle structure but yields in the $\beta_1$ phase the intrinsic anomalous Hall effect in the form of a Chern-Simons term for the order-parameter fields with an unquantized coefficient (as expected for a gapless system). Previously the anomalous Hall conductivity was determined only the DC limit within the self-consistent mean-field approximation.\cite{sun2008time} 


The paper is organized as follows. In Section \ref{sec:MicroModel} we define the model, its various symmetries and coupling to electromagnetism. In Section \ref{sec:Seff} we briefly discuss the effective action for the quadrupolar $XY$-exchange bosons whose derivation is included in Appendix \ref{app:derive_Seff}. Next in Section \ref{sec:SBS} the symmetry broken phases are discussed in detail from the point of the view of the effective action. This is followed up with an RG analysis in Section \ref{sec:RG} that leads to a schematic phase-diagram which takes into account of quantum fluctuations at one-loop order. The electromagnetic linear response is addressed in Section \ref{sec:EMresponse} with special attention placed on the Ward identities. In Section \ref{sec:reduced} we briefly comment of the effects of breaking the $U(1)$ symmetry of the relative phase down to a $\mathbb{Z}_2$ invariance. Possible experimental realizations and material candidates are proposed in Section \ref{sec:experiments}. We summarize and discuss the implications of our findings in Section \ref{sec:conclusions}. 

\section{Fermionic Model}\label{sec:MicroModel}

We begin with the model of Ref. [\onlinecite{sun2008time}] described here as a (2+1) dimensional action of two split Fermi-liquids that contains only a single $XY$-exchange interaction in the quadrupolar ($l=2$) partial wave channel, which we minimally couple to probe electromagnetic (EM) fields $(A_0,\AAA)$, 
\begin{subequations}
\begin{align}
&S[\psi^\dagger,\psi,A] = \int \diff^3 x \mathcal{L}_0 
+\int  \diff^3 x \int \diff^3 x' \mathcal{L}_\text{int}\\
&\mathcal{L}_0 = \psi^\dagger_\alpha \left[
( D_0 + \xi(-i\DD) \delta_{\alpha\beta} + v_F \Delta \sigma^z_{\alpha\beta}
\right] \psi_\beta 
\label{eqn:S_micro2}
\\
&\mathcal{L}_\text{int} = \frac{1}{2} f(\xx-\xx') \delta(x_0-x_0')\sum_{\substack{i=1,2 \\ \mu=x,y} }\Phi_{\mu i }(x) \Phi_{\mu i}(x').
\end{align}
\label{eqn:S_micro}
\end{subequations}\\
Coupling to EM proceeds with the operators $D_0 = \partial_0 -e \phi$ and $\DD = \nabla + i e \AAA(\xx)$ that are covariant derivatives\cite{minimally}
acting on (2+1) Euclidean space-time coordinates $x=(x_0,\xx)$. In this paper, we focus only on the $T=0$ zero temperature limit. {This model  are  may be realized in  itinerant ferromagnets, spintronic half-metal devices or 2D bilayer materials. These possibilities are discussed in full in Sec. \ref{sec:experiments}.}

First we describe terms in the free Lagrangian $\mathcal{L}_0$ followed by the terms in the interaction $\mathcal{L}_\text{int}$. 
The function $\xi(\kk) = \epsilon(\kk)-\epsilon_F$ is the reduced energy with dispersion $\epsilon(\kk)$ and Fermi energy $\epsilon_F$. The fermionic bands are uniformly split by an energy difference of $2v_F \Delta$ in an otherwise isotropic energy-momentum dispersion $\xi(\kk)$. In one interpretation of the model where the different 1,2 bands of $\psi$ are identified with physical spin-1/2 polarizations, then $2v_F\Delta$ is a Zeeman splitting. The bands have equal Fermi momenta ($k_F$) and Fermi velocities ($v_F$) when $\Delta=0$. In particular, $v_F$ and the effective mass $m$ are related to $\xi(\kk)$ by 
\begin{align}
\partial_a \xi(\kk)|_{k_F}=v_F \hat{k}_a, \quad \quad \partial^2_{ab} \xi(\kk)|_{k_F} = \frac{1}{m} \delta_{ab}.
\end{align}


In $\mathcal{L}_\text{int}$, the field operators $\Phi_{\mu i}$ mediate an $XY$-exchange interaction between the energy split fermionic bands and carry an additional $l=2$ quadrupolar orbital angular momentum. Thus, special care needs to be given to the nature of the coupling of the interaction terms to electromagnetism. The gauge invariant composite quadrupolar field operators $\Phi_{\mu i}$ are given in terms of fermionic bilinears as 
\begin{subequations}
\begin{align}
\Phi_{\mu i}(x) &= \psi^\dagger_\alpha(x) \mathscr{O}_{\alpha\beta}^{\mu i}(-i\DD) \psi_\beta(x)\\
\mathscr{O}_{\alpha\beta}^{\mu i}(-i\DD) &= \frac{1}{k_F^2} \sigma^\mu_{\alpha\beta} \tau^i_{ab}    (-i \sD_a)(-i \sD_b) 
\end{align}
\label{eqn:Phi}
\end{subequations}
or in $\kk$-space when $\AAA=0$
\begin{subequations}
\begin{align}
\Phi_{\mu i} (\qq) &= \sum_{\kk} \psi^\dagger_{\alpha, \kk-\qq/2} \,\mathscr{O}^{\mu i}_{\alpha \beta}(\kk) \,\psi^\dagger_{\beta, \kk+\qq/2}  \\
\mathscr{O}_{\alpha\beta}^{\mu i}(\kk) &=  
\frac{1}{k_F^2} \sigma^\mu_{\alpha\beta} \tau^i_{ab}   k_a k_b
\end{align}
\label{eqn:O2}
\end{subequations}
where the matrices $\tau^1\equiv\sigma^z, \tau^2\equiv \sigma^x$ are the invariant tensors of the $l=2$ quadrupole (d-wave) channel viz. $\tau^1_{ab} p_a p_b = p_1^2 - p_2^2 $ and $\tau^2_{ab} p_a p_b = 2p_a p_b$. The division by $k_F^{2}$ where $k_F$ is a Fermi momentum, is meant to make $\mathscr{O}_{\alpha\beta}^{\mu i}$ dimensionless.\cite{comment} The operator 
\begin{align}
\sDD \equiv \snabla + i e \AAA \equiv \tfrac{1}{2}(\overset{\rightarrow}{\nabla}-\overset{\leftarrow}{\nabla}) + i e \AAA
\end{align}
is the symmetrized covariant derivative operator, which is convenient for defining explicitly Hermitian velocity (vertex) operators. A similar coupling to electromagnetism within the quadrupolar density operator is also found in the context of quadratic semi-metals and fractional quantum Hall systems.\cite{you2013field,maciejko2013field}

Note that the repeated summation convention is implied and the earlier Greek indices $\alpha,\beta=1,2$ vary over the different fermionic bands, while the earlier Latin indices $a,b$ sum over spatial coordinate $x_1,x_2$ directions. The other set of Greek and Latin indices are associated to symmetries. The later Greek indices $\mu,\nu=x,y$ are taken to vary over the $x,y$ basis of the Pauli spin-1/2 matrices and can be considered as an internal $XY$-isospin or pseudospin degree of freedom. While the later Latin indices $i,j=1,2$ are associated to the two real $l=2$ quadrupolar harmonic basis functions $e_i(\kk) = \tau^i_{ab}\hat{k}_a \hat{k}_b$ such that $e_1(\kk)= \cos (2 \theta_\kk)$ and $e_2(\kk)= \sin (2 \theta_\kk)$. 

The $\Phi_{\mu i}(x)$ fields mediate the $XY$ pseudospin-exchange interaction between the two bands, and we take the Landau interaction potential $f(\xx)$ for this partial wave to be attractive, $f(\xx) < 0 $, with a Lorentzian profile in momentum space of the form
\begin{align}
f(\qq) = \frac{f(0)}{1+\kappa|f(0)||\qq|^2}.
\label{eqn:fq}
\end{align}
The parameter $\kappa>0$ is the range of this potential. 

We next describe the symmetries of $\Phi_{\mu i}$. The composite quadrupolar field $\Phi_{\mu i}$ which is a tensor in two indices $\mu$ and $i$, transforms under the global action of a $O(2)_\text{iso}\times O(2)_\text{rot}$ symmetry group which acts on the `left' and the `right' of the matrix valued field $\Phi_{\mu i}$. Moreover, the total action Eq.\eqref{eqn:S_micro} is globally $O(2)_\text{iso}\times O(2)_\text{rot}$ invariant when translated appropriately into a group action on $\psi_\alpha$. It will be convenient to regard $\Phi_{\mu i}$ as being composed of two real $XY$-vectors according to $\vec{\Phi}_i = (\Phi_{xi},\Phi_{yi})^T$, $i=1,2$. Furthermore, the angle of the vector $\vec{\Phi}_i$ relative to the $X$-axis is the phase difference between the fermionic fields $\psi_1$ and $\psi_2$ in the $i^\text{th}$ quadrupolar channel. In the case that the fermionic bands 1 and 2 correspond to physical spin polarizations of $S^z$, then $\vec{\Phi}_i$ is the spin-1/2 polarization in the $xy$-plane for the $l=2$ partial wave channel which is part of the spin triplet channel.\cite{wu2007fermi} 
For this paper, we have  chosen to  focus only on the $O(2)_\text{iso}\times O(2)_\text{rot}$ symmetric model although Ref.[\onlinecite{sun2008time}] had also considered broken $O(2)_\text{iso}$ generalizations which break the `relative phase' symmetry, leading to more nuanced electronic liquid crystal phases (see below).
The physical transformations of the symmetry group $O(2)_\text{iso}\times O(2)_\text{rot}$ and time-reversal are described as follows. \\

\emph{O(2)$_\text{iso}$ {symmetry}}:\quad  The action of $O(2)_\text{iso}$ on the $XY$-vectors $\vec{\Phi}_{1},\vec{\Phi}_{2}$ is comprised of a proper rotation
\begin{align}
\Phi_{\mu i} \rightarrow [(\cos\varphi)\delta_{\mu\nu} - (\sin \varphi) \epsilon_{\mu \nu}] \Phi_{\nu i}
\label{eqn:SO2phase}
\end{align}
and a parity transformation 
\begin{align}
(\Phi_{x i},\Phi_{y i}) \rightarrow (\Phi_{x i},-\Phi_{y i}), \quad i =1,2
\label{eqn:O2reflect}
\end{align}
where $\epsilon_{\mu \nu}$ is the totally anti-symmetric Levi-Civita symbol. The first continuous transformation is a $U(1) \cong SO(2)$ phase rotation by an amount $\varphi$ \emph{between} the $\psi_1$ and $\psi_2$ bands, while the second is a discrete complex conjugation of their relative phase. Note that $\epsilon_{\mu\nu}$ is invariant under $SO(2)$ but is odd under the parity transformation. Also this $O(2)$ symmetry is a broken symmetry of $U(2)$ which is present in free part of the theory of Eq. \eqref{eqn:S_micro2} whenever $\Delta = 0$.
Lastly, the combination of the total $U(1)$ phase rotation symmetry of both bands -- which preserves total electric charge -- and the relative $U(1)$ phase symmetry leads to the separate $U(1)$ phase rotation symmetries in each band. Hence the model enjoys separate number conservation in each band.  
\\

\emph{O(2)$_\text{rot}$ {symmetry}}:\quad The other symmetry group factor corresponds to an isometric transformation in Euclidean space $\mathbb{E}^2$ and acts on the quadrupolar or $d$-wave multiplet $(\vec{\Phi}_{1},\vec{\Phi}_{2})$ as a proper rotation
\begin{align}
{\Phi}_{\mu i} \rightarrow [(\cos2\theta)\delta_{ij} - (\sin 2\theta) \epsilon_{ij}] {\Phi}_{\mu j}
\label{eqn:SO2rot}
\end{align} 
and a mirror reflection
\begin{align}
(\Phi_{\mu 1},\Phi_{\mu 2}) \rightarrow (\Phi_{\mu 1},-\Phi_{\mu 2}), \quad \mu =x,y 
\end{align} 
where $\epsilon_{ij}$ is the Levi-Civita symbol now acting on $l=2$ orbital space and $\theta$ is the angle of rotation. Note that being a quadrupolar density, $\Phi_{\mu i}$ is invariant under spatial inversion, but a non-zero expectation value of it may break reflection symmetry.\cite{sun2008time} Like in the previous symmetry, $\epsilon_{ij}$ is rotationally invariant but is odd under mirror reflection. When this symmetry is broken spontaneously, a nematic fluid phase is achieved. Due to the continuous $O(2)_\text{rot}$ group, this is an $XY$-nematic as opposed to an Ising nematic in which $C_4$ is the broken symmetry.
\\

\emph{Time-reversal}:\quad The transformation properties of $\Phi_{\mu i}$ under time-reversal depend crucially on how the fermionic bands are physical interpreted. In the case of spin-less fermions and an internal isospin symmetry, then time-reversal manifest as complex conjugation $K$ on $\psi_\alpha$ which results in a mirror reflection, c.f. Eq.\eqref{eqn:O2reflect}, of $\vec{\Phi}_i$. The inversion of momenta $(\kk \rightarrow -\kk)$ does not affect $\Phi_{\mu i}$ because it is inversion invariant. However, when $\psi_\alpha$ is taken to be spin-1/2 with the 1 and 2 bands having a well defined $S^z$ eigenvalue, then time-reversal on $\psi_\alpha \rightarrow [i \sigma^y]_{\alpha\beta} K \psi_\beta$ produces $\vec{\Phi}_{i}\rightarrow - \vec{\Phi}_{i} $ as an inversion in $XY$-vector space and $\Delta \rightarrow -\Delta$ in the Zeeman splitting. Thus only in the case of an internal isospin symmetry and where $\Phi_{y i}=0$ for $i=1,2$ is $\vec{\Phi}_{1,2}$ time-reversal invariant. The action of Eq. \eqref{eqn:S_micro} is however always time-reversal invariant.\\    

One last symmetry to comment on is the $U(1)$ electromagnetic gauge symmetry. Despite $\Phi_{\mu i}$ being gauge-invariant by definition, it can still depend non-trivially on the vector potential $A$ because the Fermi field $\psi$ is not gauge invariant. As we shall see, this leads to subtle vertex couplings between $A_a$, $\psi_\alpha$ and $\Phi_{\mu i}$ through the quadrupole operator of Eq. \eqref{eqn:Phi}. 
\\

The action of $O(2)_\text{iso}\times O(2)_\text{rot}$ then endows $\Phi_{\mu i}$ with isospin in the internal $XY$-space and an orbital angular of momentum of $2\hbar$ from its $d$-wave character. The term isospin is an appropriate one in this instance because $\Phi_{\mu i}$ facilitates inter-band transitions in analogy to pions in nuclear physics. In the instance that the isospin corresponds to physical spin polarizations of $S^z$, then these inter-band transitions are literally spin-flips.\cite{technically}
We should also remark that a Hartree-Fock treatment\cite{sun2008time} of the model of Eq.\eqref{eqn:S_micro} shows that a non-zero $\langle \Phi_{\mu i} \rangle \neq 0$ will lead to an $XY$-isospin texture on the mean-field corrected Fermi-surfaces of $\psi$. More precisely $\langle \Phi_{\mu i}\rangle $ will determine the strength and orientation of this isospin texture. The specific $O(2)_\text{iso}\times O(2)_\text{rot}$ symmetry broken phases and their properties will be the focus of the subsequent sections. 

Next we perform a Hubbard-Stratonovich (HS) decoupling in the $\Phi \Phi$ channel inside the interaction term. This produces the following HS decoupled action
\begin{subequations}
\begin{align}
&S_\text{tot}[\psi^\dagger,\psi,\Gamma,A] 
= S_0 [\psi^\dagger,\psi,A] + S_1[\Gamma] \nonumber \\& \hspace{2.4cm}+ S_2[\psi^\dagger,\psi,\Gamma,A] \\
&S_0[\psi^\dagger,\psi,A] = \int \diff^3x\; \mathcal{L}_0 \\
&S_1[\Gamma] = \int \diff^3x\int \diff^3x' \, \mathcal{L}_1\\
&\mathcal{L}_1 = -\frac{1}{2} f^{-1}(\xx-\xx') \delta(x_0-x'_0) \Gamma_{\mu i}(x) \Gamma_{\mu i} (x') \\
&S_2[\psi^\dagger,\psi,\Gamma,A] = \int \diff^3 x \; \Phi_{\mu i}(x) \Gamma_{\mu i} (x) 
\end{align}
\label{eqn:Stot}
\end{subequations}
where $\Gamma_{\mu i}(x)$ is the bosonic HS field and 
\begin{align}
f^{-1}(\rr) = \sum_{\qq} \frac{1}{f(\qq)} \e^{i \qq \cdot \rr}, \quad
f(\rr) = \sum_{\qq} f(\qq) \e^{i\qq \cdot \rr}.
\end{align}
Note that $\Gamma_{\mu i}$ has the same (Cartesian) indices as $\Phi_{\mu i}$ and hence transforms identically under $O(2)_\text{iso}\times O(2)_\text{rot}$. Thus, $\Gamma_{\mu i}$ carries the same types of conserved quantities as $\Phi_{\mu i}$ ie. $l=2$ orbital angular momentum, and $XY$-isospin. 
{Other decoupling channels may also be considered but lie beyond the intended scope of this work. Here we expressly focus on the physics of the $l=2$ quadrupolar $XY$-exchange fluctuations and the phases associated with these. It is however worth mentioning that an account of superconducting order parameters intertwined with electronic liquid phases leads to a rather complex and rich phase-diagram with a plethora of possible phases, see, e.g. Ref. [\onlinecite{soto2014pair}].}

For the purposes of studying the linear EM response, we need only expand the gauge fields $(A_0,\AAA)$ to second order; which is the same order as in the interaction vertex $\Phi \Gamma$. This yields the following truncated expansion for the electromagnetically coupled portions of the action,
\begin{align}
&S_0[\psi^\dagger,\psi,A] = S_0[\psi^\dagger,\psi,0] 
+\int\diff^3 x\,\tfrac{e^2}{2m} [\psi_\alpha^\dagger \psi_\alpha] A_a A_a \nonumber \\
&+\int\diff^3 x \,e v_F \left[\psi^\dagger_\alpha \left( \tfrac{\spartial_a}{|\nabla|}\right) \psi_\alpha\right] A_a 
- \int \diff^3 x \, e[\psi^\dagger_\alpha\psi_\alpha] A_0   
\label{eqn:Stot2}\\
&S_2[\psi^\dagger,\psi,\Gamma,A] = S_2[\psi^\dagger,\psi,\Gamma,0] \nonumber \\
&+ \int\diff^3 x\,\Gamma_{\mu i} \left(\tfrac{-2ie \sigma^\mu_{\alpha\beta} \tau^i_{ab}}{k_F^2}\right)[\psi^\dagger_\alpha\spartial_b\psi_\beta] A_a \nonumber \\
&+\int\diff^3x\,\Gamma_{\mu i } \left(\tfrac{e^2\sigma^\mu_{\alpha\beta} \tau^i_{ab}}{k_F^2}\right)[\psi^\dagger_\alpha \psi_\beta]A_a A_b.
\label{eqn:Stot3}
\end{align}
Now $S_0[\psi^\dagger,\psi,A]$ contains the conventional interaction vertices between $\psi$ and the gauge field $A$ which includes the $(1/2m)|\AAA|^2$ diamagnetic contribution. However, additional interaction vertices arise from the $S_2[\psi^\dagger,\psi,\Gamma,A]$ functional that linearly couples to $\Gamma$, while linearly as well as quadratically to $\AAA$. These may interpreted as perturbative anisotropic corrections to the group velocity and band curvature due to $\Gamma$. More importantly, the presence of the $\sigma^\mu$ tensor implies that non-trivial transitions between fermionic bands are involved in these $\Gamma$ dependent vertex operators. The reason for this can be traced back to the definition of Eq.\eqref{eqn:Phi} of the $\Phi$ field. Physically, this allows the gauge field $A$ to couple to the distortions of the Fermi-surface due to to $\Gamma$. This fact will prove to be crucial in understanding the EM response properties of this model. 

Finally we have the partition function formally expressed as
\begin{align}
\mathcal{Z}_\text{tot} = \int \mathscr{D}\Gamma\mathscr{D}\psi\mathscr{D}\psi^\dagger\, 
\mathrm{e}^{
-S_0[\psi^\dagger,\psi,A] - S_1[\Gamma] - S_2[\psi^\dagger,\psi,\Gamma,A]
}.
\end{align} 
Notationally we can compactly express the action by suppressing the integrations and contraction of indices, where the interpretation is obvious. This gives
\begin{subequations}
\begin{align}
&S_0[\psi^\dagger,\psi,A] = - \psi^\dagger \hat{\mathcal{G}}^{-1}_0 \psi  
- eA_0 \psi^\dagger \psi 
+e v_F \AAA \psi^\dagger \hat{\kk} \psi \nonumber \\
&\hspace{2cm}+ \tfrac{e^2}{2m} \AAA^2 \psi^\dagger \psi \\
&S_1[\Gamma] = -\tfrac{1}{2f}\Gamma\Gamma \\
&S_2[\psi^\dagger,\psi,\Gamma,A] = \psi^\dagger \hat{\OO}_2 \psi \Gamma + 2e\AAA \psi^\dagger \hat{\OO}_1 \psi \Gamma \nonumber \\
&\hspace{2.4cm}-e^2 \AAA^2 \hat{\OO}_0 \psi^\dagger \psi \Gamma 
\end{align}
\label{eqn:Seff_compact}  
\end{subequations}
where $\hat{\mathcal{G}}_0^{-1}$ in the inverse free propagator and $\hat{\OO}_n$ is the vertex operator with $n$ internal momenta operators $-i\spartial$ contracted with the quadrupolar coupling constant $\sigma \tau /(k_F)^2$. Then, upon formally integrating-out the field
\cite{caveat}
$\psi$ first, we get the partition function as
\begin{align}
&\mathcal{Z}_\text{tot} = \mathcal{Z}_{\psi_0} \int \mathscr{D}\Gamma\;  \e^{-S_1[\Gamma]} 
\mathrm{e}^{\text{Tr}\ln (\hat{1}+ \hat{\mathcal{M}}[\Gamma,A])} \\
&\hat{\mathcal{M}}[\Gamma,A]=-\hat{\GG}_0 \hat{\OO}_2 \Gamma + e\hat{\GG}_0 A_0 -e v_F \AAA\hat{\GG}_0 \hat{\kk} -\tfrac{e^2}{2m} \AAA^2 \hat{\GG}_0 \nonumber \\
&\hspace{1.6cm} -2e\AAA \hat{\GG}_0\hat{\OO}_1 \Gamma + e^2 \AAA^2 \hat{\GG}_0 \hat{\OO}_0 \Gamma
\end{align}
where $\mathcal{Z}_{\psi_0}$ is the free fermion partition function, ie. $\det[-\hat{\GG}_0^{-1}]$. This then leads to the effective action  
\begin{align}
S_\text{eff}[\Gamma,A] = -\tfrac{1}{2f}\Gamma \Gamma -\text{Tr}\ln (\hat{1}+ \hat{\mathcal{M}}[\Gamma,A])
\label{eqn:Seffall} 
\end{align}
which will be the focus of our discussion in the following sections. 

\section{Effective Action}\label{sec:Seff}

Initially, we shall limit our discussion to $A=0$ to focus on the order parameter $\Gamma_{\mu i}$, the Landau-Ginzburg effective action and the symmetry broken phases. We proceed to expand Eq.\eqref{eqn:Seffall} in order of perturbation and formally derive an expression for the effective Lagrangian $\mathcal{L}_\text{eff}(\Gamma,\partial \Gamma)$ in the same vein as Hertz-Millis\cite{hertz1976quantum,millis1993effect} theory. The details of this calculation which involve a single fermionic loop are presented in Appendix \ref{app:derive_Seff}. The form of the Landau-Ginzburg effective Lagrangian up to $O(\Gamma^4)$ is 
\begin{align}
&\mathcal{L}_\text{eff} = 
-\frac{i \beta }{2} \epsilon_{\mu\nu}\Gamma_{\mu i}\partial_\tau\Gamma_{\nu i} 
+\frac{\kappa}{2}\, \nabla {\Gamma}_{\mu i} \cdot \nabla{\Gamma}_{\mu i}
\nonumber \\
&+ \frac{\rho}{2}(\Gamma_{\mu i} \Gamma_{\mu i}) 
+\frac{\alpha}{4}(\Gamma_{\mu i}\Gamma_{\mu i})^2 - \frac{\lambda}{4} (\epsilon_{\mu\nu}\epsilon_{ij}\Gamma_{\mu i}\Gamma_{\nu j})^2. 
\label{eqn:L_eff}
\end{align}
The form of this Lagrangian could have been determined purely from the requirements of $O(2)_\text{iso}\times O(2)_\text{rot}$ symmetry. Note that the oddness of $\epsilon_{\mu\nu}$ under time-reversal compensates for the change in sign arising from $\partial_\tau$ under time-reversal. However the specific coefficients at one loop order in terms of the microscopic theory of Eq.\eqref{eqn:S_micro} have to be computed from a one loop calculation. These couplings are given as
\begin{subequations}
\begin{align}
& r = \frac{1}{2v_F \Delta}\int_{-v_F \Delta}^{v_F \Delta} N(\xi) \diff \xi \approx N(0)+ \tfrac{(v_F\Delta)^2}{6}N''(0) \\
&\beta = \frac{r}{2v_F\Delta},\quad  \rho = \frac{1}{|f(0)|}-r\\
&\alpha = \frac{N'(0)^2}{2N(0)} - \frac{N''(0)}{8},\quad  \lambda = -\frac{N''(0)}{24}. 
\end{align}
\label{eqn:couplings}
\end{subequations}
where 
\begin{align}
N(\xi) = \int \frac{\diff^2 k}{(2\pi)^2}\; \delta(\xi -\xi_\kk )
\end{align}
is the density of states (DOS) induced from the dispersion $\xi(\kk)$ and the quantity $r$ is defined as the average DOS in the region between the split Fermi-surfaces. For stability purposes, it is also required that $\alpha> |\lambda| > 0$. 

The free Gaussian portions of $\mathcal{L}_\text{eff}$ are entirely local with a dynamical scaling of $z=2$ that makes the $\alpha$ and $\lambda$ couplings marginal in (2+1) dimensions. The coupling $\rho$ is relevant and controls the phase of the model such that when $\rho<0$, spontaneous symmetry breaking occurs. This is the Pomeranchuk instability\cite{pomeranchuk1958stability} from the point of view of the fermionic theory that occurs at sufficiently negative $f(0)$.
Noticeably absent is the dynamical term in the form of Landau-damping that is typically associated with Hertz-Millis theories of itinerant magnetism.
This can be attributed to the \emph{gapped} inter-band fluctuations of the microscopic interactions. 
In the next section we will introduce and discuss the different symmetry broken phases that follow from Landau-Ginzburg theory. 

\section{Broken Symmetry Phases}\label{sec:SBS}

On a classical level, the theory described by Eq.\eqref{eqn:L_eff} experiences spontaneous symmetry breaking whenever $\rho<0$. The specific type of symmetry broken phase will depend on the coupling $\lambda$, which we take here to be either positive or negative. Notice that the symmetry group $SO(2)_\text{iso}\times SO(2)_\text{rot}$ is isomorphic to $U(1) \times U(1)$, and that the form of the dynamical term is highly suggestive that the $O(2)_\text{iso}$ plays a more important role than $O(2)_\text{rot}$. In fact, for the purposes of deriving the effective action after spontaneous symmetry breaking, we have found it useful to elevate the $U(1)\times U(1)$ isomorphism by \emph{complexifying} the 4-real components of $\Gamma_{\mu i}$ into a $\mathbb{C}^2$ valued field according to
\begin{align}
\phi_j = \Gamma_{x j } - i\Gamma_{y j}, \quad \quad j = 1,2.
\end{align}
This choice is made to simplify the form of the dynamical term. Thereafter by re-scaling time and space, and dropping a boundary term, we have the much simplified Lagrangian density
\begin{align}
\mathcal{L}_\text{eff} & =  \phi^\dagger \partial_\tau \phi + \nabla \phi^\dagger \cdot \nabla \phi + \rho\, \phi^\dagger \phi \nonumber \\
&+ \frac{\alpha}{2} (\phi^\dagger \phi)^2 -\frac{\lambda}{2}(\phi^\dagger \sigma^y \phi)^2  
\label{eqn:Leffphi}
\end{align}
where the couplings $\rho,\alpha,\lambda$ are related to the ones in Eq.\eqref{eqn:couplings} by proportionality constants. Note that the $\sigma^y$ Pauli-matrix here acts on the two component `spinor' $\phi$ and hence mixes the $\vec{\Gamma}_1$ and $\vec{\Gamma}_2$ vectors. In particular, in the internal isospin interpretation of $O(2)_\text{iso}$, the operations of time-reversal and mirror reflection are now given by $\phi \rightarrow \phi^*$ and $\phi \rightarrow \sigma^z \phi$ respectively. In the case of the magnetic spin-1/2 interpretation, time-reversal is instead implemented as $\phi \rightarrow -\phi$.

Firstly, it is easily recognized that the effective action now describes a two component non-relativistic boson whose total number is conserved. Hence, the route to spontaneous symmetry breaking is very much like the physics of (bosonic) superfluidity where the $O(2)_\text{iso}$ symmetry has been promoted to a global $U(1)$ charge conserving symmetry. Secondly, note that the free Gaussian part has a $U(2)$ symmetry which is more symmetric than the $U(1)\times U(1)$ symmetry of the full action with the interaction terms. But the $U(2)$ symmetry is only restored in the fine-tuned limit where $\lambda=0$. Furthermore, the $\lambda$ dependent term is proportional to the square of the cross product of $\vec{\Gamma}_{1,2}$ vectors according to
\begin{align}
\epsilon_{\mu\nu} \epsilon_{ij} \Gamma_{\mu i }\Gamma_{\nu j } =
2(\vec{\Gamma}_1 \wedge \vec{\Gamma}_2)
=-(\phi^\dagger \sigma^y \phi). 
\end{align}
Thus, the distinguishing feature between the different possible phases whenever $\lambda \neq 0$, is whether or not $\langle \phi^\dagger \sigma^y \phi\rangle$ is zero or extremal. In the following subsections, we will discuss each of the broken symmetry phases. 

Given a non-zero order parameter $\phi_0 = \langle \phi \rangle$, it will prove convenient in our subsequent discussion to define the following unit isospin vector  
\begin{align}
\vec{s} = \frac{\phi_0^\dagger \vec{\sigma} \phi_0}{n_0} 
\label{eqn:vec_s}
\end{align}
where $n_0 = \phi_0^\dagger \phi_0 >0 $ quantifies of the symmetry breaking; essentially the density of condensed $\phi$ bosons. The mapping $\phi_0 \mapsto \vec{s}$ is simply the Hopf map of $S^3 \rightarrow S^2$. Moreover, the Hopf fiber which is topologically $S^1$ and transforms under $O(2)_\text{iso}$. Hence, fluctuations in the Hopf fiber are really fluctuations of the internal $XY$-isospin directions. While fluctuations in $\vec{s}$ are a mixture of (spatial) orientational and orbital fluctuations.

\subsection{\texorpdfstring{$\alpha_1$}{}-phase}

\begin{figure}
\includegraphics[width=0.4\textwidth]{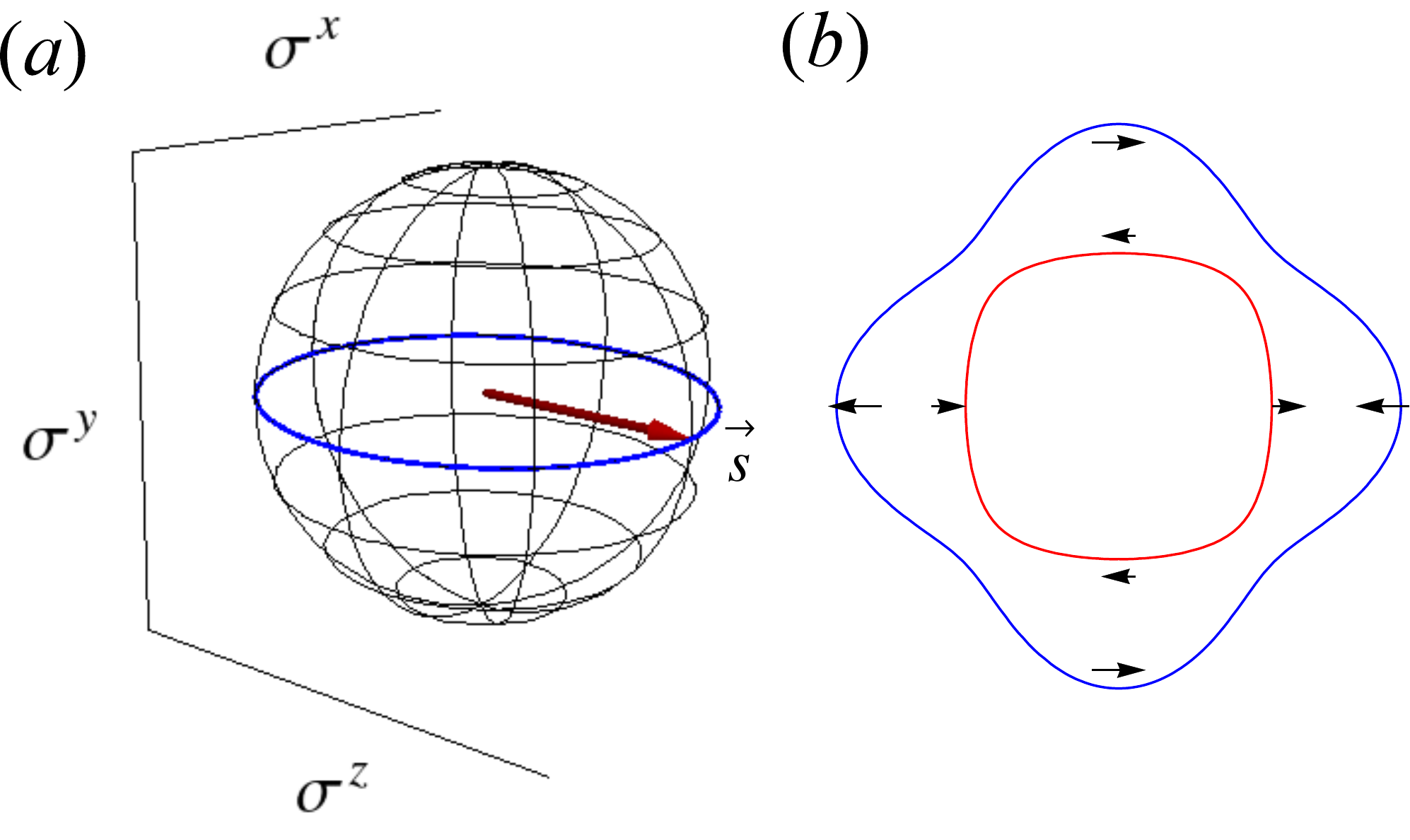}
\caption{Visualization of the $\alpha_1$-phase. ($a$) The vacuum manifold (thick blue line) of $\vec{s}$ defined in Eq.\eqref{eqn:vec_s} which lies on the $xz$ great circle of the Bloch sphere such that $s^y =0$. The position of $\vec{s}$ depends on $\arg(u/ v)$ defined from Eq.\eqref{eqn:alpha_vacuum}. Not shown is the remaining Hopf fiber component $\arg(u v)$ that takes values in $S^1$. ($b$) The nematic Fermi surfaces in the $\alpha_1$-phase with the isospin texture plotted. The $\langle\psi^\dagger_\kk {\sigma}^{x,y} \psi_\kk \rangle$ texture does not wind around the Fermi surface and is shown here as being horizontal and vanishes along the high symmetry directions $\pm\pi/4,\pm3\pi/4$.}
\label{fig:alpha1}
\end{figure}

Conditions that characterize this phase, whenever $\lambda <0$, are
\begin{align}
\phi_0^\dagger \phi_0 = n_0, \quad n_0 = \frac{|\rho|}{\alpha},\quad \phi_0^\dagger \sigma^y \phi_0 = 0
\label{eqn:alpha}
\end{align}
which also means $s^y =0$. This translates to collinear $XY$-vectors $\vec{\Gamma}_1 \parallel \vec{\Gamma}_2$ with a squared sum $(|\vec{\Gamma}_1|^2+|\vec{\Gamma}_2|^2)$ that is invariant under the $O(2)_\text{rot}$ rotation. In Ref.[\onlinecite{sun2008time}] this phase was called the $\alpha_1$-phase  by analogy with the He$^3$ system, and in keeping with the naming convention of Ref.[\onlinecite{wu2007fermi}] in regard to spin-nematics. Additional $\alpha_2$ and $\alpha_3$ phases were also defined in Ref.[\onlinecite{sun2008time}] and they arise from explicitly broken $O(2)_\text{iso}$ symmetry. Superficially, the $\alpha_1$-phase breaks time-reversal symmetry $\phi_0 \rightarrow \phi_0^*$, whenever $\phi_0$ is not real. However this is a `broken' time-reversal symmetry in the same vein that an ordinary BCS s-wave superconductor breaks time-reversal whenever the order parameter is complex valued. The key point is that, one can always transform using the $O(2)_\text{iso}\times O(2)_\text{rot}$ group to choose a   `gauge' where $\phi_0$ is real. Likewise, in a BCS superconductor unless there is a topological obstruction, one can always choose a gauge where the order parameter is real. Said another way, in the space of all symmetry broken vacua, there exist a representative vacuum state, degenerate in energy with all others, which is time-reversal invariant. In a similar fashion chiral or reflection symmetry is also preserved\cite{sun2008time} due to the existence of a mirror symmetric plane. However, in the magnetic spin-1/2 interpretation, any non-zero $\phi_0$ will always break time-reversal symmetry.  Lastly, from the point of view of the fermionic mean-field theory, the $\alpha_1$-phase will exhibit quadrupolar Fermi-surface distortions but no isospin textures that wind around the Fermi-surfaces.  

Next we describe the space of broken symmetry vacua associated to the $\alpha_1$-phase. Let us generally express $\phi_0$ in the basis where $\sigma^y$ is diagonal as
\begin{align}
\phi_0 = 
\frac{u}{\sqrt{2}}\begin{pmatrix}1 \\ i \end{pmatrix}
+\frac{v}{\sqrt{2}}\begin{pmatrix}1 \\ -i \end{pmatrix}
\label{eqn:alpha_vacuum}
\end{align}
such that $|u|^2=|v|^2=n_0/2$. The $U(1)\times U(1)$ symmetry acts on $u$ and $v$ by the following phase rotations
\begin{align}
u \mapsto \e^{i\theta_1} u, \quad v \mapsto \e^{i\theta_2}v,\quad  \theta_1,\theta_2 \in \mathbb{R}.
\end{align}        
By choosing $u=v$, we can make $\phi_0$ manifestly time-reversal invariant. More importantly this means that the space of vacua is topologically a 2-torus, $S^1\times S^1$. One then predicts that there should be 2 Nambu-Goldstone modes and topological defects in the form of vortices with $\mathbb{Z}\oplus\mathbb{Z}$ charges. This is indeed the case but the non-relativistic form of Eq.\eqref{eqn:Leffphi} and the $U(2)$ symmetry of the free theory complicates the broken symmetry analysis a bit. Finally, shown in Fig.\ref{fig:alpha1} is a visualization of the manifold of possible $\vec{s}$ and an example of a nematic Fermi-surface in the $\alpha_1$-phase.     

The derivation of the Nambu-Goldstone action is best understood by considering the $U(2)$-symmetric, $\lambda=0$, point first. Now, taking the group quotient of $U(2)$ by the stabilizer of any general $\phi_0 \in \mathbb{C}^2$, reduces $U(2)$ to $SU(2)$.\cite{convenience}
Since $SU(2)$ is topologically a 3-sphere, one naively expects that there are 3 Nambu-Goldstone modes to start out with in the $\lambda=0$ limit. This is however not the case because of the non-relativistic symmetry and the non-commutativity of $SU(2)$.\cite{nielsen1976count} There are in fact 2 Nambu-Goldstone modes which satisfies the Brauner-Watanabe-Murayama\cite{brauner2010spontaneous,watanabe2012unified} relation
\begin{align}
n_\text{BG}-n_\text{NGB} &=\tfrac{1}{2}\text{rank}\; \rho \label{eqn:Haruki}\\
\rho^{i j} &= -i \langle [\sigma^i,\sigma^j] \rangle 
\end{align}
where $n_\text{BG}=3$ is the number of symmetry broken generators, $n_\text{NBG}=2$ is the number of massless Nambu-Goldstone modes and $\text{rank} \;\rho =2$ is the rank of the commutator matrix of the $SU(2)$ Lie algebra. 

To confirm this, we begin by parameterizing the order parameter fluctuations by 
\begin{align}
\phi &= \sqrt{\tfrac{n_0 + \delta n}{n_0}} \e^{i \vec{\pi} \cdot \vec{\sigma}} \phi_0 \\
\Rightarrow \delta \phi &= i (\vec{\pi}\cdot \vec{\sigma}) \phi_0 + \left(\tfrac{\delta n}{ 2 n_0}\right)\phi_0.
\end{align}
Then we determined the symmetry-broken action\cite{derivation}
in the $\lambda=0$ limit to be 
\begin{align}
\mathcal{L}_\text{NG}^{(0)} &= 
\tfrac{1}{2\alpha} \left(\vec{s}\cdot \partial_\tau \vec{\pi}\right)^2 
+ i n_0\left[\vec{s} \cdot (\vec{\pi}\times \partial_\tau \vec{\pi})\right] \nonumber \\
&+ n_0 \nabla \vec{\pi} \cdot \nabla \vec{\pi}
\label{eqn:LNG0}
\end{align}
which follows from an integration over the $\delta n$ amplitude fluctuations. This in turn leads to the following equations of motion for the Nambu-Goldstone fields $\vec{\pi}$
\begin{subequations} 
\begin{align}
(\partial_\tau^2 + 2|\rho| \nabla^2 ) {\pi_\parallel} &= 0 \\
i (\vec{s}\times \partial_\tau \vec{\pi}) + \nabla^2 \pi_\perp &= 0 
\end{align}
\end{subequations}
where $\pi_\parallel = \vec{s}\cdot \vec{\pi}$ and $\pi_\perp = (\mathbbm{1}-\vec{s}\otimes \vec{s})\vec{\pi}$, are the parallel and perpendicular isospin directions of $\vec{\pi}$ with respect to $\vec{s}$. Hence the $\pi_\parallel$ field is gapless and relativistic, while the second equation describes a \emph{single} gapless non-relativistic mode with $z=2$ dynamical scaling. To appreciate this, we can take as a concrete example $\vec{s}= \hat{\bf z}$ which then yields a relativistic $\pi_z$ mode but produces the following complimentary equations of motion
\begin{subequations}
\begin{align}
i \partial_\tau \pi_x + \nabla^2 \pi_y = 0, \quad 
-i \partial_\tau \pi_y + \nabla^2 \pi_x = 0.
\end{align}
\end{subequations}
This coupling between $\pi_x$ and $\pi_y$ modes that leads to a single gapless non-relativistic mode stems from their mutual identification with conjugate momenta fields from the `Berry phase' term in Eq.\eqref{eqn:LNG0}. 

Returning now to the $\alpha_1$-phase, we include the $\lambda<0$ term in the effective Lagrangian which gives the following Nambu-Goldstone action
\begin{align}
\mathcal{L}_\text{NG}^{(\alpha)}  
&= 
\tfrac{1}{2\alpha} \left(\vec{s}\cdot \partial_\tau \vec{\pi}\right)^2 
+ i n_0\left[\vec{s} \cdot (\vec{\pi}\times \partial_\tau \vec{\pi})\right] \nonumber \\
&+ n_0 \nabla \vec{\pi} \cdot \nabla \vec{\pi} + 2|\lambda| n_0^2 (\vec{s}\times\pi)_y^2.
\label{eqn:L_alpha}
\end{align} 
Remarkably, the additional $\lambda$ dependent term does not remove any of the previously two discussed gapless modes. This is so because the equations of motion that follow from $\mathcal{L}_\text{NG}^{(\alpha)}$ are
\begin{subequations}
\begin{align}
(\partial_\tau^2 + 2|\rho|\nabla^2) \pi_z &= 0 \\
i \partial_\tau \pi_x + \nabla^2 \pi_y &=0  \\
-i\partial_\tau \pi_y + \nabla^2 \pi_x &= 4 |\lambda|n_0^2 \pi_x. 
\end{align}
\label{eqn:ELalpha}
\end{subequations}\\
where we have again taken $\vec{s} = \hat{\bf z}$ for concreteness. Taking another time-derivative of the non-relativistic modes gives
\begin{subequations}
\begin{align}
\partial_\tau^2 \pi_x - \nabla^4 \pi_x + 4 |\lambda|n_0^2 \nabla^2 \pi_x &=0 \\
\partial_\tau^2 \pi_y - \nabla^4 \pi_y + 4 |\lambda|n_0^2 \nabla^2 \pi_y &=0  
\end{align}
\end{subequations}
which results in a higher order dispersion relation in the gapless modes $\pi_x$,$\pi_y$. Thus the addition of the $\lambda<0$ dependent term preserves the number of gapless modes and is still in agreement with the counting relation of Eq.\eqref{eqn:Haruki}. This is so because although there are now 2 broken symmetry generators from the $U(1)\times U(1)$ symmetry group, the rank of $(\rho^{ij})=0$ since the group is now Abelian. Finally in the static limit, $\pi_x$ and $\pi_y$ decouple in Eq.\eqref{eqn:ELalpha}, but $\pi_x$ is short-ranged because of the $\lambda$ dependent term. Thus in the vicinity of a vortex, the order parameter winding occurs through slow gradients of solutions to the harmonic equations $\nabla^2 \pi_y=0$ and $\nabla^2 \pi_\parallel=0$. The latter fluctuation is recognized as the winding of the vector $\vec{s}$, while the former with the $O(2)_\text{iso}$ symmetry.

\subsection{\texorpdfstring{$\beta_1$}{}-phase}

\begin{figure}
\includegraphics[width=0.4\textwidth]{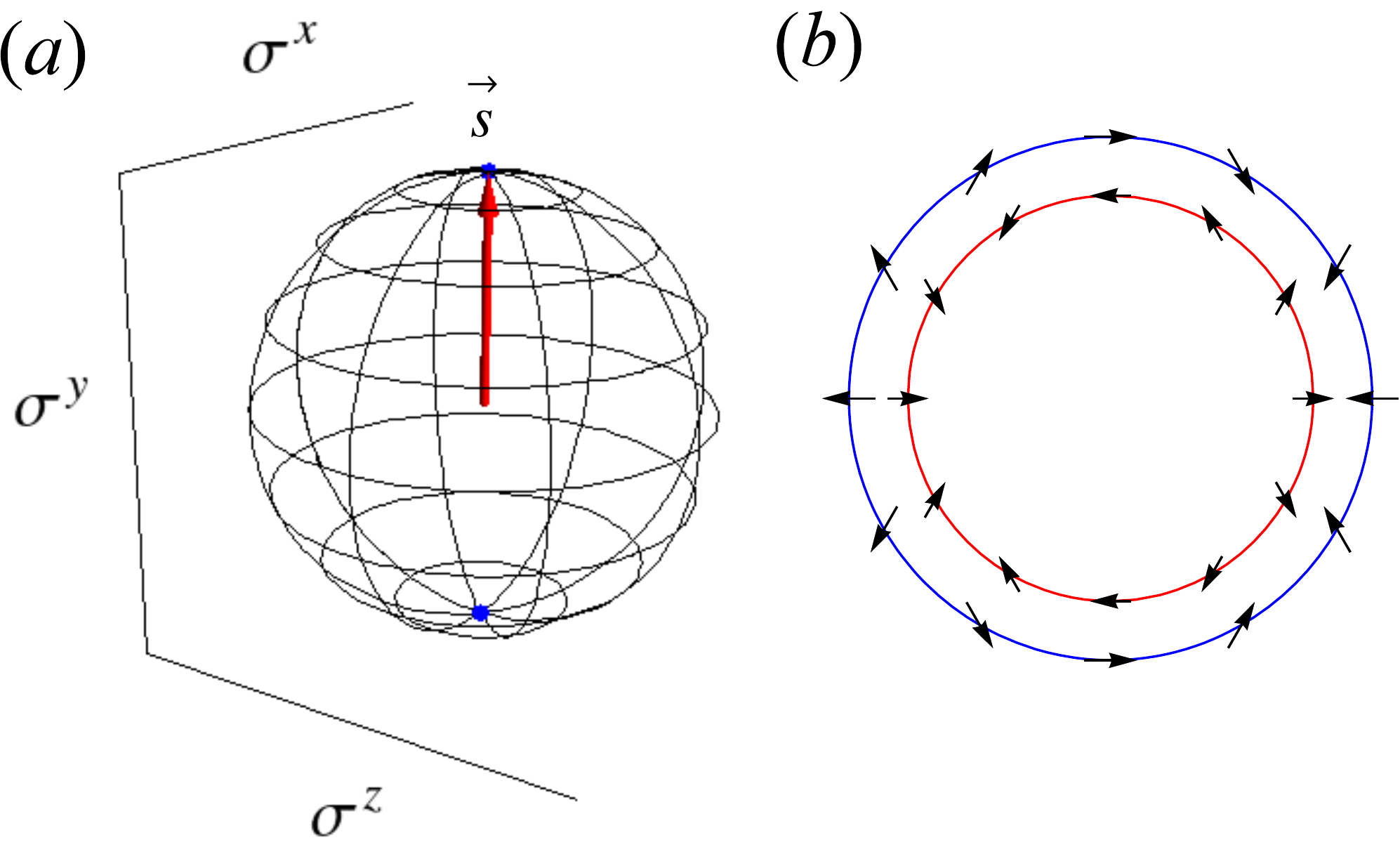}
\caption{Visualization of the $\beta_1$-phase. ($a$) The vacuum manifold (dark blue dots) of $\vec{s}$ defined in Eq.\eqref{eqn:vec_s} which lies on the $\pm y$ poles of the Bloch sphere. Not shown is the remaining Hopf fiber component $\arg(u)$ from Eq.\eqref{eqn:beta_vacuum} that takes values in $S^1$. ($b$) The Fermi surfaces in the $\beta_1$-phase with the isospin texture plotted. The $\langle\psi^\dagger_\kk {\sigma}^{x,y} \psi_\kk \rangle$ texture winds around the Fermi surface twice in this case.}
\label{fig:beta1}
\end{figure}

In the opposite limit, $\lambda>0$, the order parameter is required to satisfy
\begin{align}
\phi_0^\dagger \phi_0 = n_0,\quad \phi_0^\dagger \sigma^y \phi_0 = s^yn_0,\quad
n_0 = \frac{|\rho|}{\alpha-\lambda}
\label{eqn:beta}\end{align}
such that
$\vec{s}=(0,s^y,0)^T, s^y=\pm 1 $. In this instance the expectation value $\langle \phi^\dagger \sigma^y \phi \rangle $ is extremal and there is a $\mathbb{Z}_2$ degree of freedom that determines the direction of $s^{y}$. In the $XY$-vector language, this corresponds to vectors that are perpendicular ($\vec{\Gamma}_1 \cdot \vec{\Gamma}_2 = 0$) and equal in magnitude ($|\vec{\Gamma}_1| = |\vec{\Gamma}_2|$). The sign of $s^y$ determines whether or not $\vec{\Gamma}_1$ and $\vec{\Gamma}_2$ describes a left-handed or right-handed system of axes. The phase is called the $\beta_1$-phase in Ref.[\onlinecite{sun2008time}], again in analogy with the pattern of symmetry breaking in He$^3$. Similarly to the $\alpha$ phases, $\beta_2$ and $\beta_3$ phases are also defined in Ref.[\onlinecite{sun2008time}] and also arise from explicitly broken $O(2)$$_\text{iso}$ symmetry. In the $\beta_1$-phase, time-reversal symmetry and parity symmetry are always broken. This is because either transformation causes the exchange $s^y \rightarrow -s^y$. From the point of view of the fermionic mean-field description, the $\beta_1$-phase will not experience a Fermi-surface distortion, but an isospin texture that winds non-trivially around the Fermi surfaces.\cite{sun2008time} This produces a non quantized Berry phase Wilson loop on the mean-field Fermi surfaces which is the geometrical obstruction to time-reversal and parity symmetries. A fact that gives the $\beta_1$-phase a non-zero anomalous Hall response.   

The symmetry broken manifold for the $\beta_1$-phase is more constrained than the $\alpha_1$ and is given by    
\begin{align}
\phi_0 =\frac{u}{\sqrt{2}} \begin{pmatrix} 1 \\ s^y\,  i \end{pmatrix}
\label{eqn:beta_vacuum}
\end{align}
with $|u|^2 = n_0$ and $s^y = \pm 1$. The phase ambiguity in the coordinate $u$ and the sign choice of $s^y$ means that the total vacuum manifold is topologically $\mathbb{Z}_2\times S^1$. Thus, one should expect a single Nambu-Goldstone mode and the possibility of vortices with $\mathbb{Z}_2\oplus \mathbb{Z}$ charge, and this is indeed the case. Shown in Fig.\ref{fig:beta1} is a visualization of the manifold of possible $\vec{s}$ and an example of a (iso-)spin textured Fermi-surface in the $\beta_1$-phase. 

The derivation of the Nambu-Goldstone action follows closely with the $\alpha_1$-phase case, but with the difference that the equilibrium density $n_0$ is now $\lambda$ dependent. The final form of the Nambu-Goldstone Lagrangian is 
\begin{align}
\mathcal{L}_\text{NG}^{(\beta)} &= 
\tfrac{1}{2(\alpha-\lambda)}(\partial_\tau \pi_y )^2 
+ n_0\nabla \vec{\pi} \cdot \nabla\vec{\pi} 
+ in_0 \, s^y(\vec{\pi}\times \partial_\tau \vec{\pi})_y \nonumber \\
&\quad + 2\lambda n_0^2 (\pi^2_x+\pi^2_z)
\label{eqn:L_beta}
\end{align}
which does indeed contain only a single relativistic Nambu-Goldstone mode in $\pi_y$. The other $\pi_x,\pi_z$ modes have acquired a mass which is now controlled by $\lambda>0$. Again this counting conforms with the relation Eq.\eqref{eqn:Haruki} since symmetry is reduced to the Abelian $U(1)$ group in this phase. \\

As a final point, we remark that the order parameter manifold of the $\alpha_1$-phase Eq.\eqref{eqn:alpha_vacuum} does not contain in it the smaller order parameter manifold of the $\beta_1$-phase Eq.\eqref{eqn:beta_vacuum} as a sub-manifold. Hence, on the basis of Landau symmetry breaking, there is a first order transition that separates the $\alpha_1$ and $\beta_1$ phases. 

\subsection{Over-damped Nambu-Goldstone Modes}

The preceding analyses superficially indicate that the Nambu-Goldstone modes in the $\alpha_1$ and $\beta_1$ phases are gapless, coherent and long-lived. This is incorrect due to a general argument by Watanabe and Vishwananth\cite{watanabe2014criterion} which determines a necessary criterion for the stability of such modes. This criterion being that the broken symmetry generators, $\vec{s}\cdot\vec{\pi}$ and $\pi_y$ in $\alpha_1$ and $\beta_1$ respectively, do not commute with the total momentum operator, and in this instance leads to the non-vanishing $l=2$ quadrupolar matrix elements at $\qq=0$. In this subsection, we discuss the damping experienced by these modes in the broken symmetry phases at the level of the mean-field approximation. We show that the damping is parametrically small in its pre-factor such that Landau damping is only experienced at very small momenta.  

Generally, inter-band transitions lead to coherent undamped excitations and intra-band transitions lead to Landau damping. However, in the normal phase only inter-band transitions are involved in the computation of the effective action of Eq.\eqref{eqn:L_eff} which leads to the effective actions of Eq.\eqref{eqn:L_alpha} and Eq.\eqref{eqn:L_beta} after symmetry breaking. Hence they predict no damping for their respective Nambu-Goldstone modes because only inter-band transitions within the fermionic loop are involved. Nevertheless, the presence of a non-zero order parameter $\langle \Gamma_{\mu i} \rangle \neq 0$ opens up an inter-band channel that ultimately produces damping for the Nambu-Goldstone modes. 

The calculation of this Nambu-Goldstone damping is given in Appendix \ref{app:NBG_damping} and we just quote here the final result. Denoting the fluctuation correlator in the broken symmetry phase as
\begin{equation} 
J^{(2)}(q)_{\mu i \nu j} =  \langle \delta\Gamma_{\nu j}(-iq_0,-\qq)\; \delta\Gamma_{\mu i}(iq_0,\qq)\rangle_\text{MF}
\end{equation}
the intra-band contribution is given to leading order by
\begin{align}
&(J_\text{intra}^{(2)})(q)_{\mu i \nu j} - (J_\text{intra}^{(2)})(0)_{\mu i \nu j} \nonumber \\
&\approx-\frac{2 \bar{r}\;C_{ijlp}\langle\Gamma_{\mu l}\rangle \langle \Gamma_{\nu p}\rangle}{(v_F \Delta)^2}\left(\frac{|q_0|}{\sqrt{v_F^2|\qq|^2+q_0^2}}\right) 
\end{align}
where we have defined the tensor
\begin{align}
C_{ijlp} = \frac{1}{8} \left( \delta_{ij}\delta_{lp} + \delta_{il}\delta_{jp} + \delta_{ip}\delta_{jl}\right). 
\end{align}
Analytically continuing to real frequencies $iq_0 \rightarrow \omega + i0^+$ and in the branch where $s= \frac{|\omega|}{v_F |\qq|}<1$, yields to $O(s)$ 
\begin{align}
&(J_\text{intra}^{(2)})(\omega,\qq)_{\mu i \nu j} - (J_\text{intra}^{(2)})(0)_{\mu i \nu j} \nonumber \\
&\approx \left(\frac{-i\omega}{v_F |\qq|}\right) \left[ \frac{2\bar{r}}{(v_F \Delta)^2}\right]C_{ijlp}\langle\Gamma_{\mu l}\rangle\langle \Gamma_{\nu p}\rangle
\label{eqn:damping}
\end{align}
which is an imaginary self-energy correction to the correlator $\langle \delta\Gamma_{\nu j}(-\omega,-\qq)\; \delta\Gamma_{\mu i}(\omega,\qq)\rangle$. 

Next we express the Goldstone fluctuations as $\delta \Gamma = i  \pi_A T^A \langle\Gamma\rangle$ where $\{T^A\}$ are an orthonormal basis of the broken symmetry generators and $\pi_A$ is the Nambu-Goldstone field. This implies that inverse lifetime deduced from Eq.\eqref{eqn:damping} is of order $O(\langle\Gamma\rangle^4) = O(n_0^2)$ in terms of the Goldstone modes $\pi$, where $n_0$ is the density of condensed $\Gamma$ bosons. Comparing terms other terms of order $O(n_0)$ and $O(n_0^2)$ within the effective actions Eq.\eqref{eqn:L_alpha} and Eq.\eqref{eqn:L_beta}, we conclude that 
\begin{itemize}
\item[ ($\alpha_1$) ] {The relativistic mode $\vec{s}\cdot \vec{\pi}$ mode is lightly damped when $|\qq|^2 \gg \frac{\bar{r}}{(v_F\Delta)^2}\sqrt{\frac{|\rho|}{\kappa\beta}}$ but eventually becomes over-damped when $|\qq|^2 \ll \frac{\bar{r}}{(v_F\Delta)^2}\sqrt{\frac{|\rho|}{\kappa\beta}}$. By contrast, the mode $\vec{s}\times\vec{\pi}$ is always over-damped since its equations of motion are of $O(n_0^2)$ as well.}
\item[ ($\beta_1$) ] {Likewise the relativistic $\pi_y$ mode is lightly damped when $|\qq|^2 \gg \frac{\bar{r}}{(v_F\Delta)^2}\sqrt{\frac{|\rho|}{\kappa\beta}}$ and is otherwise over-damped at lower momentum.}
\end{itemize}
Note that the fact that the damping rate is proportional to $n_0^2 \propto |\rho|^2$ means that it is \emph{parametrically small} compared to the leading order $n_0$ terms that arise from inter-band dynamics. Thus near the critical point, where $|\rho|$ and $n_0$ are small, inter-band damping is suppressed leading to the stated lightly damped behavior. Nevertheless, strictly speaking the broken symmetry phases are non-Fermi liquid in accordance to Ref.[\onlinecite{watanabe2014criterion}]. 

\section{Renormalization Group Analysis}\label{sec:RG}

In this section, we present a renormalization group analysis at one-loop order for the effective action of Eq.\eqref{eqn:Leffphi} to determine approach to symmetry breaking from the normal phase prior to the Pomeranchuk instability. This is relatively straightforward given the similarity of Eq.\eqref{eqn:Leffphi} to a superfluid theory.\cite{sachdev2007quantum} What makes this possible and interesting is the $z=2$ dynamical scaling such that $d=2$ becomes the upper critical dimension. We note that the renormalization group method can also be applied within the broken symmetry phases\cite{castellani1997infrared} but we shall not pursue that matter here. At the classical level, the normal phase would correspond to $\rho\geq 0$ or a negative chemical potential for the $\phi$ bosons. 

\begin{figure}
	\boxed{\includegraphics[width=0.4\textwidth]{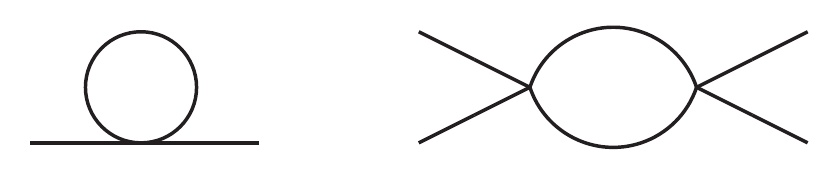}}
	\caption{The first order loop corrections to the quadratic and quartic interactions.}
	\label{fig:RG_loops}
\end{figure}

In accordance to the $z=2$ dynamical scaling, we begin by parameterizing the space-time scaling 
\begin{align}
{\bf x} \rightarrow {\bf x}\;\mathrm{e}^{-u}, \quad 
\tau \rightarrow \tau \; \mathrm{e}^{-2u}, \quad \phi \rightarrow \phi \;\mathrm{e}^{-u}  
\end{align}
where $u \in \mathbb{R}_{>0}$. We then proceed to carry out a Wilsonian renormalization group analysis by using a hard momentum cutoff $\Lambda$ to define the regularized theory. Formally, we then divide the theory into fast and slow modes, where modes with momentum $|\kk|< \Lambda e^{-u}$ are deemed slow and those with $\Lambda e^{-u}\leq |\kk|<\Lambda$ are fast.\cite{work}
Then integrating over the fast modes produces quantum corrections to the effective action functional $\Gamma[\varphi^\dagger,\varphi]$ for the slow modes $\varphi$. This is conveniently done using the background field method\cite{zinn2002quantum} where the slow field is the background field. Using 
\begin{align}
\e^{-\Gamma[\varphi^\dagger, \varphi]} &= 
\int \mathscr{D}\phi^\dagger \mathscr{D}\phi \exp
\left(
-S[\phi^\dagger+\varphi^\dagger, \phi+\varphi]
\right) 
\end{align} 
the one-loop order quantum contributions to $\Gamma$ is\cite{divided}
\begin{align}
\delta \Gamma[\varphi^\dagger,\varphi] &=
\text{tr} \left[
\ln \delta^2_{\phi^\dagger \phi} S[\varphi^\dagger,\varphi] 	
-\ln \delta^2_{\phi^\dagger \phi} S[0,0] 
\right] \nonumber \\
&=\text{tr} \left\{(\partial_\tau -\nabla^2 + \rho)^{-1} \hat{M}[\varphi^\dagger,\varphi](\alpha,\lambda) \right\} \nonumber \\
&-\tfrac{1}{2} \text{tr}\left\{\left[
(\partial_\tau -\nabla^2 +\rho)^{-1} \hat{M}[\varphi^\dagger,\varphi](\alpha,\lambda)
\right]^2
\right\}
\end{align}
with the matrix operator 
\begin{align}
\left(\hat{M}[\varphi^\dagger,\varphi](\alpha,\lambda)\right)_{ij} &=
\alpha [(\varphi^\dagger \varphi)\delta_{ij}+\varphi^\dagger_{j}\varphi_j] \nonumber \\
&-\lambda[ (\varphi^\dagger \sigma^y \varphi)\sigma^y_{ij} +\sigma^y_{ik}\sigma^y_{lj}\varphi^\dagger_l \varphi_k
].
\end{align}
These one-loop contributions correspond to the usual tadpole or Hartree bubble, and the four point bubble diagrams that are shown in Fig.\ref{fig:RG_loops}. Because the propagators are $U(2)$ symmetric, the form of UV and IR divergences are identical to usual Bose gas case. The only differences arise from the complicated form of the coupling which is $U(1) \times U(1)$ symmetric that is encoded in $\hat{M}$. 

It turns out that the Hartree bubble is zero and does not introduce any quantum corrections to the relevant coupling $\rho$. This is due to the simple reason that in the normal phase, the boson number density is zero at $T=0$. Since the scaling dimension $[\rho]=2$ we have the renormalized $\rho$ coupling as $\rho_R = \rho_0\e^{2u}$ and the RG equation
\begin{align}
\frac{\partial \tilde{\rho}_R}{\partial u } = 2 \tilde{\rho}_R \label{eqn:beta_rho}
\end{align}
where $\tilde{\rho}_R = \Lambda^{-2} \rho_R$ is the dimensionless form of the coupling. Additional higher order terms if present on the RHS of Eq.\eqref{eqn:beta_rho} will act to shift the critical value $\rho_c$ from zero.

Moving on to the marginal couplings, we instead have the following one-loop corrections from the four-point bubble
\begin{align}
\delta \alpha  &= - \frac{2u}{8\pi}(4\alpha^2 + (\alpha -\lambda)^2 ) \\
\delta \lambda &= + \frac{4u}{8\pi} \lambda (\alpha+\lambda)
\end{align}
which is independent of $\rho$. This then yields the following RG equations
\begin{align}
\frac{\partial \alpha_R}{\partial u } &= 
- \frac{4\alpha_R^2 +(\alpha_R-\lambda_R)^2}{4\pi} \label{eqn:beta1} \\
\frac{\partial \lambda_R}{\partial u } &= 
+ \frac{2\lambda_R(\alpha_R+\lambda_R)}{4\pi} \label{eqn:beta2}
\end{align}
that remain independent of $\rho_R$. The fate of the symmetry broken phase is then dictated by Eq.\eqref{eqn:beta1} and Eq.\eqref{eqn:beta2}. 

\begin{figure}
\includegraphics[width=0.3\textwidth]{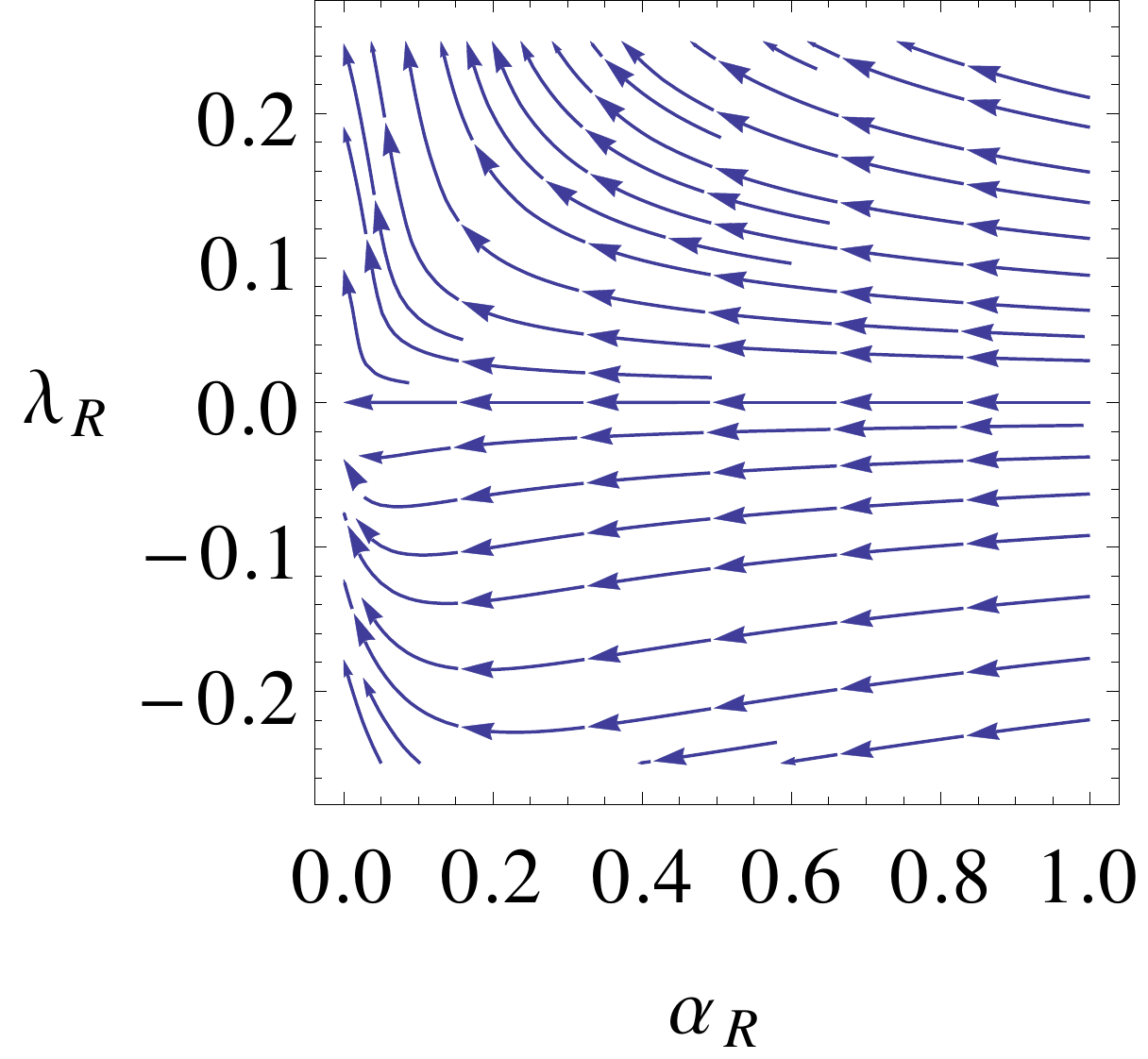}
\caption{Integrated RG flow of the marginal couplings.}
\label{fig:RGflow}
\end{figure}

The phase portrait for the flow equations of Eq.\eqref{eqn:beta1} and Eq.\eqref{eqn:beta2} is shown in Fig.\ref{fig:RGflow}. In the regime where $ \alpha_R , -\lambda_R > 0$, we have the $\alpha_1$ phase whenever $\rho < \rho_c$. In Fig.\ref{fig:RGflow}, this phase corresponds to the flows on the lower half plane $\lambda_R<0$ which are attracted to the Gaussian fixed line at $(\alpha_R,\lambda_R)=0$. Tuning $\rho$ across $\rho_c$ in this case corresponds to a conventional second order phase transition. Conversely, the $\beta_1$ phase occurs whenever $\alpha_R < \lambda_R < 0$ and $\rho < \rho_c$. So this is positive $\lambda_R$ coupling and corresponds to the flows in the upper plane in Fig.\ref{fig:RGflow}. However the phase-portrait predicts a flow to large positive $\lambda \gg \alpha$ where the naive free energy implied by $\mathcal{L}_\text{eff}$ is no longer stable. Then we require $\phi^6$ terms to stabilize the free energy and this generally leads to a fluctuation-induced first order transition. Another first order transition line lies between the $\alpha_1$ and $\beta_1$ phases since their pattern of symmetry breaking is unrelated. However the transition from the normal to the $\alpha_1$ phase is second order since this is controlled by the Gaussian fixed point. 

\begin{figure}
\includegraphics[width=0.3\textwidth]{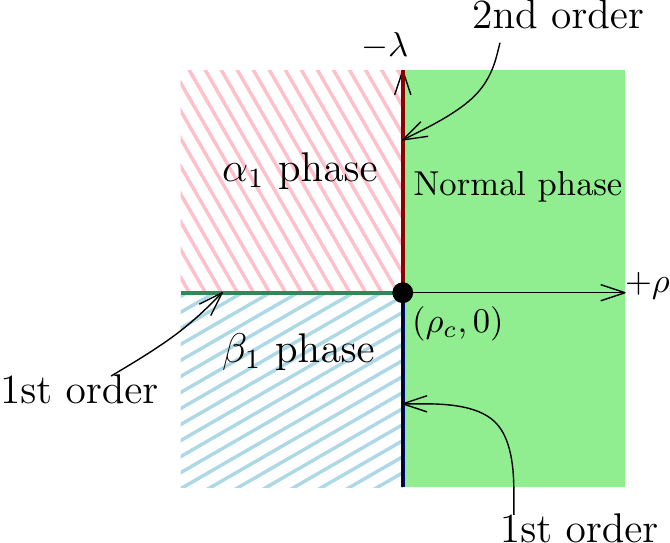}
\caption{Schematic phase diagram implied by the RG equations at one-loop order. Not shown is the $\alpha$ axis where it is required that $\alpha>|\lambda|>0$ for stability.}
\label{fig:phase_diag}
\end{figure}

We then summarize the phase diagram implied by this one-loop RG analysis in Fig.\ref{fig:phase_diag}. This differs from the phase diagram of the mean-field analysis of Ref.[\onlinecite{sun2008time}], in that a tricritical point is predicted to exist between the normal and $\beta_1$-phase. Again, this physical feature is unattainable within our simple $\phi^4$ theory, and would require additional higher order terms to capture this tricritical behavior.  

\section{Electromagnetic Response}\label{sec:EMresponse}

We now turn our discussion to the electromagnetic response properties of the model. It is expected that the different symmetry broken phases will respond differently to EM probe fields. For this discussion, we prefer to revert back to the $\Gamma_{\mu i}$ representation. In principle the entire EM response is contained in the functional expansion in $A$ of Eq.\eqref{eqn:Seffall}. In practice, it is much easier to organize this expansion in terms of the linear response current which can be divided into separately conserved contributions. We shall only concern ourselves with the linear response currents; equivalently the expansion of $S_\text{eff}[\Gamma,A]$ to order $O(A^2)$. Also, we will regard the $\Gamma$ fields to move much slower than $A$ such that we neglect their gradient terms $\partial \Gamma_{\mu i}$. 

From the expansion in Eq.\eqref{eqn:Seff_compact}, the total gauge invariant 3-current density $j^\mu(x) $ to order $O(A)$ can be determined from the variation of $S[\psi^\dagger,\psi,\Gamma,A]$ by $\delta A$ as
\begin{subequations}
\begin{align}
j^0 &= -e \psi^\dagger \psi, \quad  j^a[A] = j_1^a[A] + j_2^a[\Gamma,A] \\
\nonumber \\
j_1^a[A] &= e \psi^\dagger (v_F \hat{k}_a) \psi + e^2 \left(\tfrac{\delta_{ab}}{m}\right) \psi^\dagger \psi \, A_b  \nonumber \\
&= \feny{width=1.6cm}{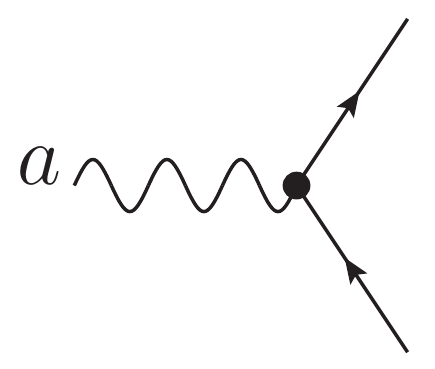} - \left( \feny{width=1.2cm}{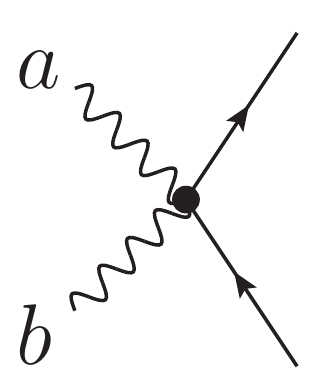} \right)A_b \\
\nonumber \\
j_2^a[\Gamma,A] &= \tfrac{2e}{k_F^2}\Gamma_{\mu i} \tau^i_{ab}(\psi^\dagger\sigma^\mu k_b \psi) 
+ \tfrac{2e^2}{k_F^2} \Gamma_{\mu i} \tau^i_{ab} (\psi^\dagger\sigma^\mu \psi) A_b \nonumber \\
&= \feny{width=1.6cm}{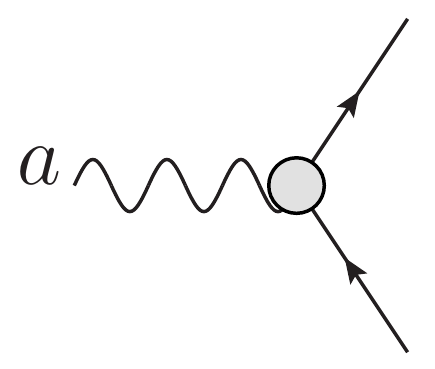} -\left( \feny{width=1.2cm}{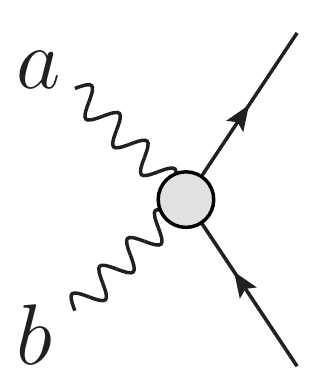}\right)A_b 
\end{align}
\end{subequations}
where we have used the diagrammatic convention of Fig.\ref{fig:FeynmanRules} and have suppressed the $\Gamma$-lines for clarity. Also, the wiggly photon lines are understood to be amputated. The total current density $j^a$ has been separated into contributions that are $\Gamma$-independent ($j^a_1$) and $\Gamma$-dependent ($j^a_2$). Note that both these current density vertices $j^a_{1}$ and $j^a_{2}$, are separately gauge invariant\cite{technically2}
because they include the linear in $A$ diamagnetic terms. As we shall show, this leads to separate sets of Ward identities that result from the total gauge invariance of the action.  

Now the total \emph{linear} response 3-current is given as
\begin{align}
&\langle j^\mu [A] \rangle_{A,\Gamma} \nonumber \\
&= \mathcal{Z}^{-1} \int \mathscr{D} \psi^\dagger \mathscr{D} \psi \, j^\mu [A] \; \e^{-S[\psi^\dagger,\psi,\Gamma,A]} \nonumber \\
&= \mathcal{Z}^{-1} \int \mathscr{D} \psi^\dagger \mathscr{D}\psi  \,(j^\mu[A]) (1- A_\nu j^\nu[A])\; \e^{-S[\psi^\dagger,\psi,\Gamma,0]} \nonumber \\
&= \mathcal{Z}^{-1} \int \mathscr{D} \psi^\dagger \mathscr{D}\psi \left(
-j^\mu[0]j^\nu[0] + \tfrac{\delta j^\mu[0]}{\delta A_\nu}
\right) A_\nu \, \e^{-S[\psi^\dagger,\psi,\Gamma,0]} \nonumber \\
&= A_\nu 
\left( 
\feny{width=2.5cm}{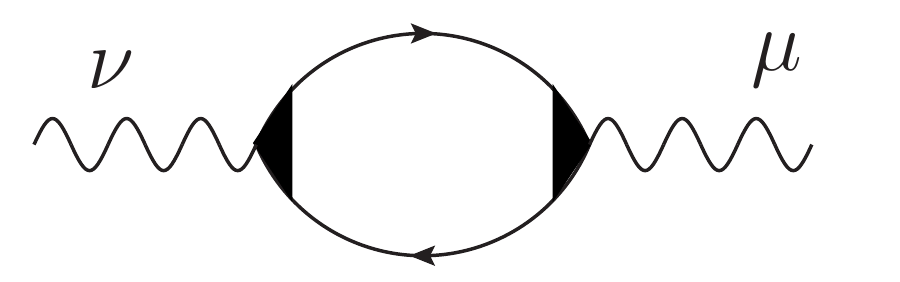}
-\feny{width=1.6cm}{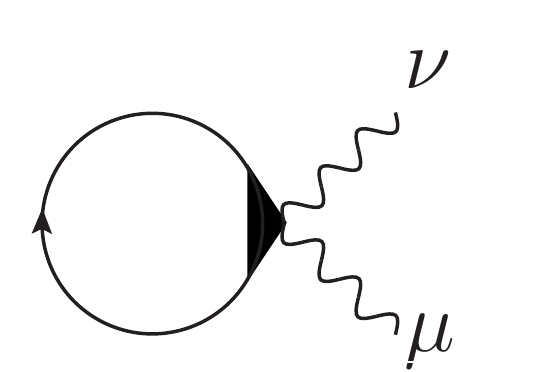}
\right) 
\label{eqn:Jtot}
\end{align} 
where the $\Gamma$ corrected current vertices are
\begin{align}
&\feny{width=1.4cm}{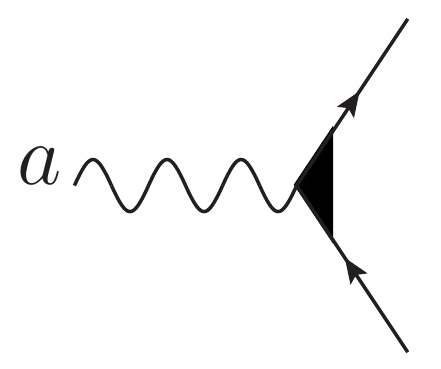} =
\feny{width=1.4cm}{diagrams/feynman/j.pdf} +
\feny{width=1.4cm}{diagrams/feynman/jg.pdf}\\
&\feny{width=1.4cm}{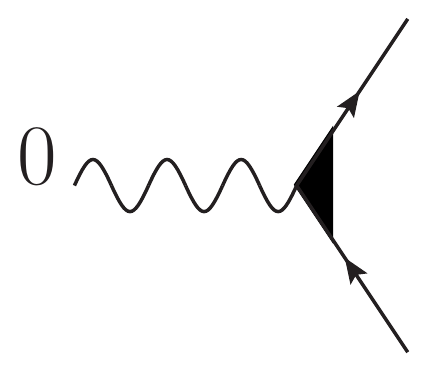} =
\feny{width=1.4cm}{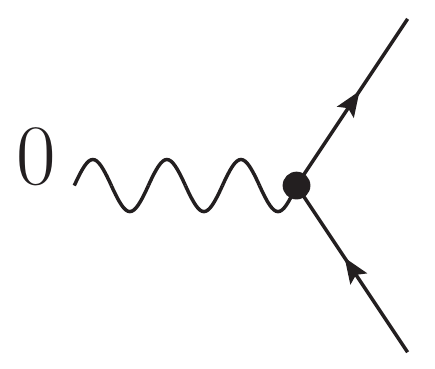}\\
&\quad\feny{width=1.2cm}{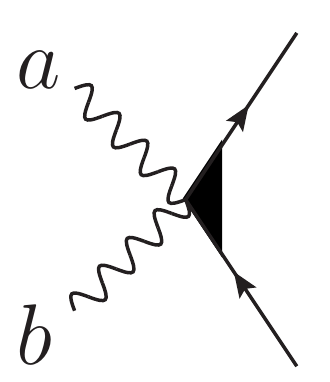}= \feny{width=1.2cm}{diagrams/feynman/dA_j.pdf} + \feny{width=1.2cm}{diagrams/feynman/dA_jg.pdf}
&
\end{align}
Note that in the third line of Eq.\eqref{eqn:Jtot}, we have only retained terms up to linear order in $A$, and the second `tadpole' term in the final line is only relevant for the spatial 2-current since there are no direct charge-current vertices. We have also  omitted the spontaneous current density $\langle j^\mu [0] \rangle_{(A=0),\Gamma}$, since it is zero and we are only interested in linear response current. 

Hence, diagrammatically the different contributions to the gauge invariant linear response currents are
\begin{align}
\langle j_1^b[A]\rangle &= A_a \left(
\feny{width=2.5cm}{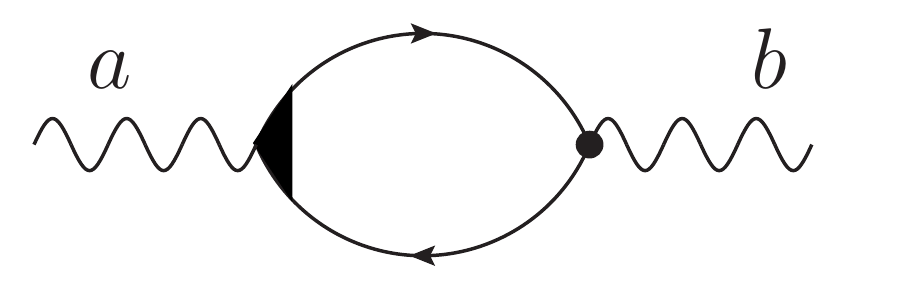} -
\feny{width=1.4cm}{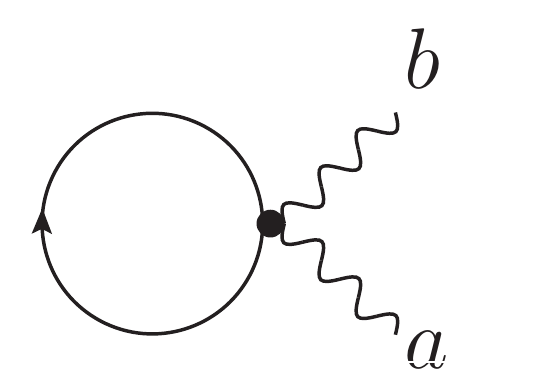}
\right) \nl 
& + A_0 \left(
\feny{width=2.5cm}{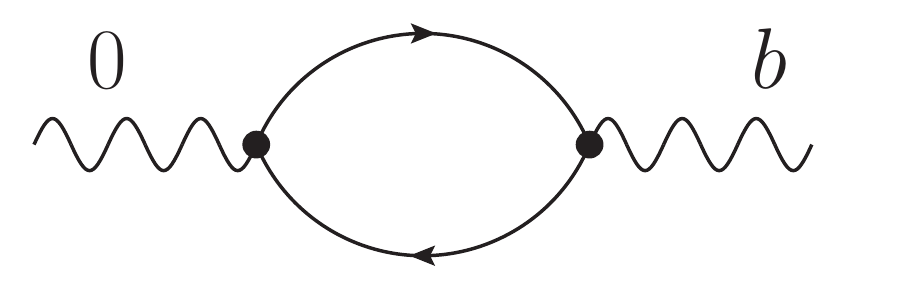}
\right)
\end{align}
\begin{align}
\langle j_2^b[A]\rangle &= A_a \left(
\feny{width=2.5cm}{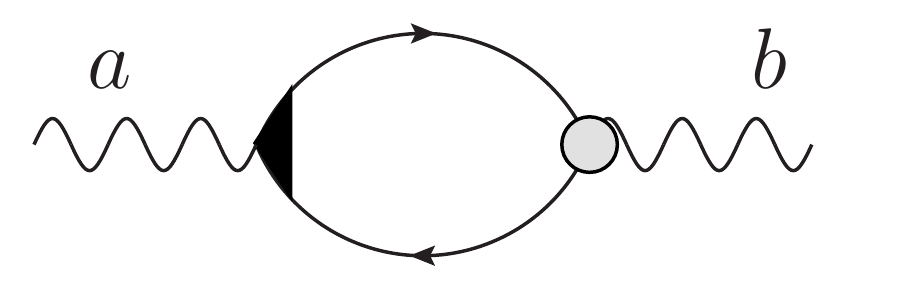} -
\feny{width=1.4cm}{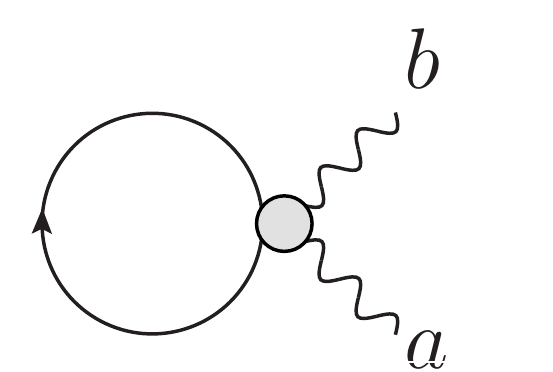}
\right) \nl & + A_0 \left(
\feny{width=2.5cm}{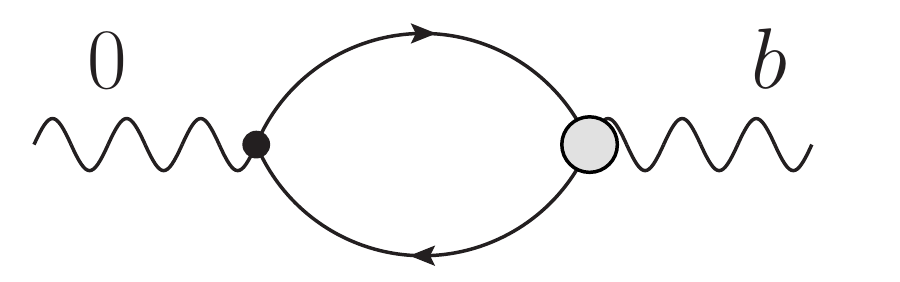}
\right) 
\end{align}
\begin{align}
\langle j^0 \rangle &= A_a \left(
\feny{width=2.5cm}{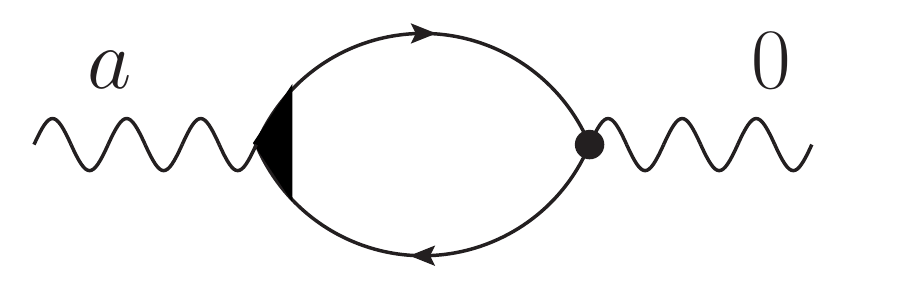}
\right) + A_0 \left(
\feny{width=2.5cm}{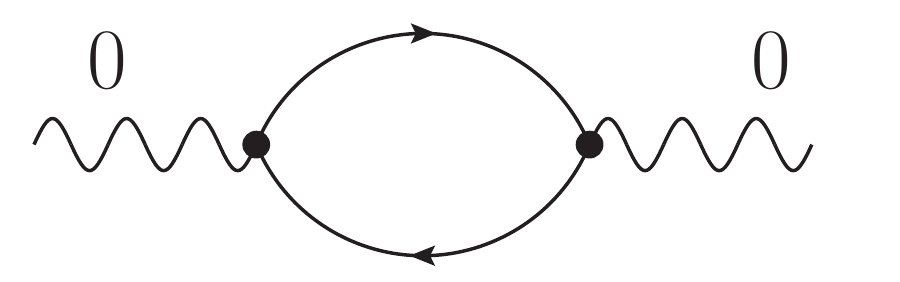}
\right)
\end{align}
It is also understood that the fermionic propagator lines here include perturbative $\Gamma$ corrections in the form of a self-energy.


To appreciate the gauge invariance of $\langle j_1^b[A]\rangle$, $\langle j_2^b[A]\rangle$ and $\langle j^0 \rangle$, we first consider $j_2[A]$
\begin{align}
&\langle j_2^b[A]\rangle \nl
&= \mathcal{Z}^{-1} \int \mathscr{D}\psi^\dagger\mathscr{D}\psi  \left(\tfrac{2e}{k_F^2}\right)
\tau^i_{ab}\; [\psi^\dagger \sigma^\mu (k+eA)_a \psi]
\e^{-S[\psi^\dagger,\psi,\Gamma,A]}.
\end{align}
Then applying the gauge transformation $A \rightarrow A + d\varphi$, $\psi \rightarrow \psi \e^{-i\varphi}$ we see that the current remains unchanged because the action and the current density are gauge invariant. Thus
\begin{align}
&\langle j_2^b[A]\rangle = \langle j_2^b[A+d\varphi] \rangle = \langle j_2^b[A]\rangle + \int \; \partial_\mu \varphi \frac{\delta}{\delta A_\mu} \langle j_2^b[A]\rangle 
\end{align}
to linear order in $d \varphi$. Then by an integration by parts, this produces the following Ward identity 
\begin{align}
\partial_\mu \left(\frac{\delta}{\delta A_\mu} \langle j_2^b[A]\rangle\right) 
&= \partial_a \left(
\feny{width=2.5cm}{diagrams/feynman/Jjg.pdf}-
\feny{width=1.4cm}{diagrams/feynman/tadpole_g.pdf}
\right) \nl & + i\partial_0 \left( 
\feny{width=2.5cm}{diagrams/feynman/qjg.pdf}
\right)\nl 
&=0.
\end{align}
By the same arguments, the following Ward identities can be proven as well.
\begin{align}
&\partial_a \left(
\feny{width=2.5cm}{diagrams/feynman/Jj.pdf}-
\feny{width=1.4cm}{diagrams/feynman/tadpole_m.pdf}
\right) \nl &+ i \partial_0 \left(
\feny{width=2.5cm}{diagrams/feynman/qj.pdf}
\right) =0 \\ \nl
&\partial_a \left(
\feny{width=2.5cm}{diagrams/feynman/Jq.pdf} 
\right) + i\partial_0 \left(
\feny{width=2.5cm}{diagrams/feynman/qq.pdf}
\right) =0
\end{align} 


This then suggests the following definitions of separate response kernels ($q=(q_0,\qq)$ is the incoming momentum)
\begin{subequations}
\begin{align}
K_0^{\mu b}(q) &= \left.\frac{\delta \langle j_1^b[A]\rangle_{\Gamma=0,A}}{\delta A_\mu(q)}\right|_{A=0} \\
K_0^{\mu 0}(q) &= \left.\frac{\delta \langle j^0\rangle_{\Gamma=0,A}}{\delta A_\mu(q)}\right|_{A=0} \\
K_1^{\mu b}(q) &= \left.\frac{\delta \langle j_1^b[A]\rangle_{\Gamma,A}}{\delta A_\mu(q)}\right|_{A=0}  - K_0^{\mu b}(q)\\
K_1^{\mu 0}(q) &= \left.\frac{\delta \langle j^0\rangle_{\Gamma,A}}{\delta A_\mu(q)}\right|_{A=0} - K_0^{\mu 0}{q}\\
K_2^{\mu b}(q) &= \left.\frac{\delta \langle j_2^b[\Gamma,A]\rangle_{\Gamma,A}}{\delta A_\mu(q)}\right|_{A=0} 
\end{align}
\end{subequations}
which are all individually transverse $\partial_\mu K_m^{\mu\nu} (q) = 0$. The total linear response current is then given as 
\begin{align}
j^{\nu}(-q) = \left[K_0^{\mu\nu}(q)+K_1^{\mu\nu}(q)+K_2^{\mu\nu}(q)\right]A_{\mu}(q).
\end{align}
The first and second kernels $K_0^{\mu b},K_0^{0 b}$ are the conventional linear responses when $\Gamma=0$ and are transverse by themselves. This makes the next pair $K_1^{\mu b},K_1^{0 b}$ mutually transverse. In the first set $K_0^{\mu\nu}$ we have 
\begin{subequations}
\begin{align}
& K_0^{ab}(q) =
\feny{width=2.5cm}{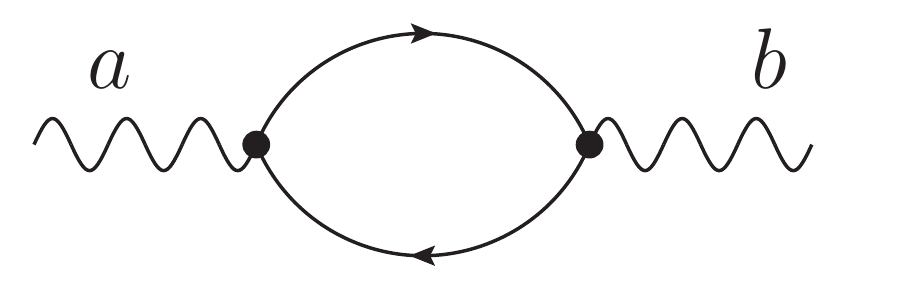} -
\feny{width=1.4cm}{diagrams/feynman/tadpole_m.pdf} \\ 
& K_0^{0b}(q) = 
\feny{width=2.5cm}{diagrams/feynman/qj.pdf}\\
& K_0^{00}(q) =
\feny{width=2.5cm}{diagrams/feynman/qq.pdf} 
\end{align}
\end{subequations}
which can be recognized as the conventional polarization tensors. In the next set of response kernels $K_{1,2}^{\mu\nu}$, we expand to order $O(\Gamma^2)$ to give
\begin{subequations}
\begin{align}
K_1^{ab}(q) &= 
\feny{width=2.5cm}{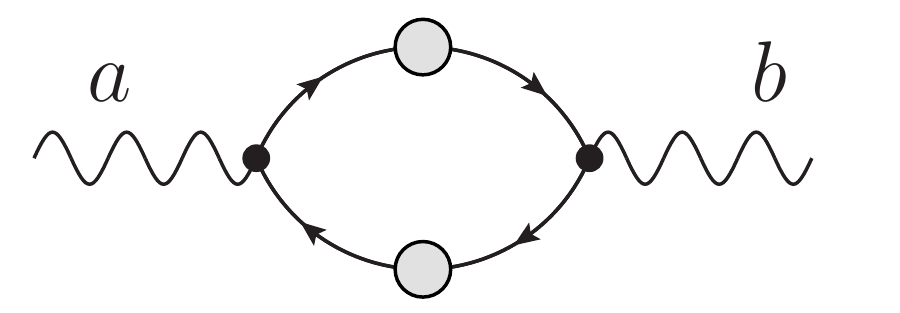}-
\feny{width=1.6cm}{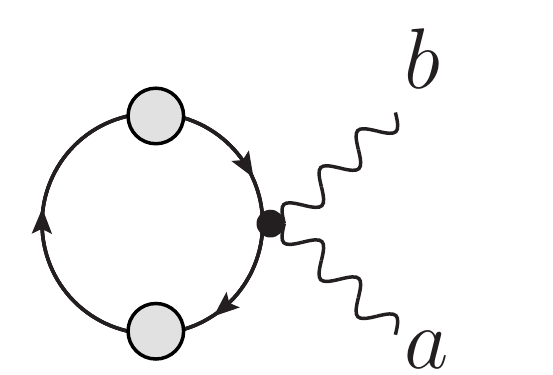} \nl &+ 
\feny{width=2.5cm}{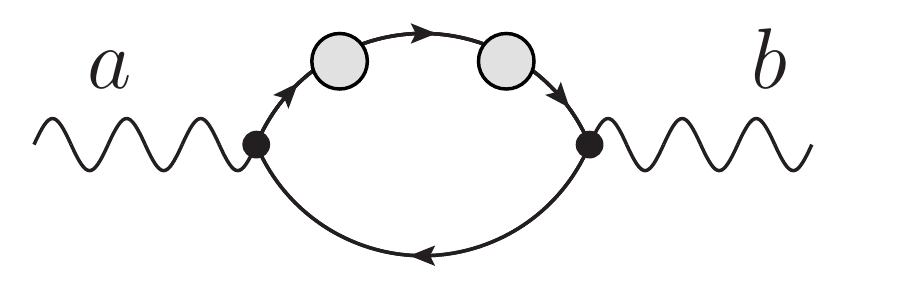} +
\feny{width=2.5cm}{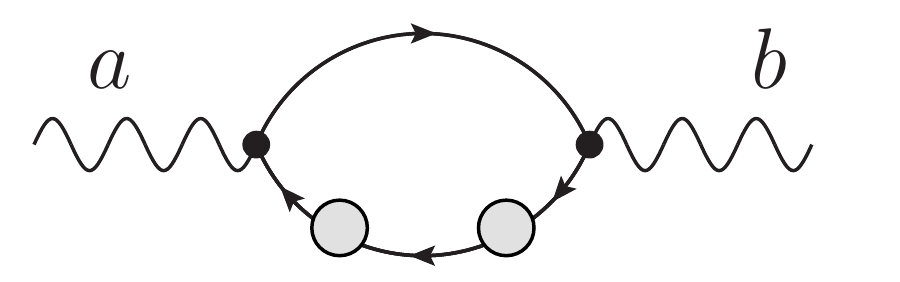} \nl &+
\feny{width=2.5cm}{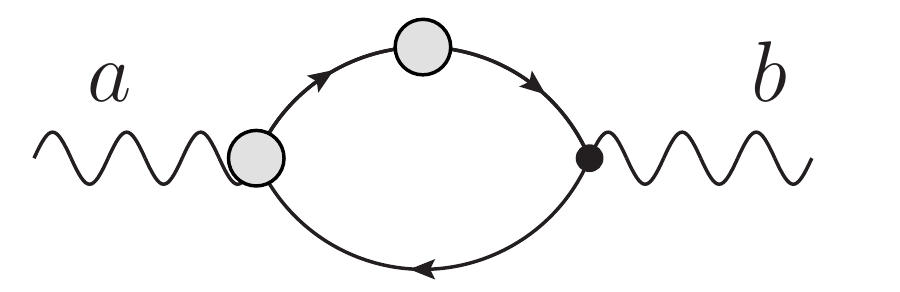} + 
\feny{width=2.5cm}{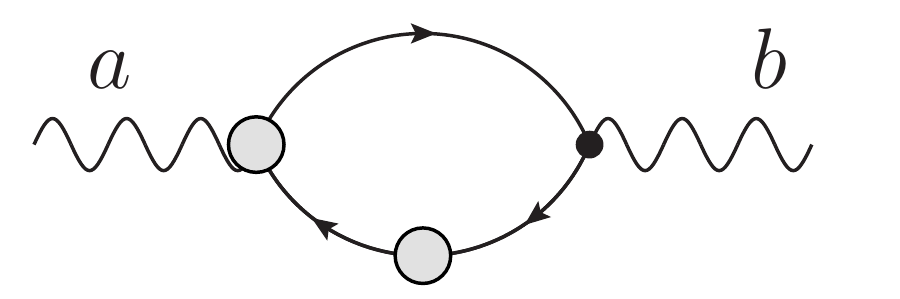}\\
K_1^{0b}(q) &=
\feny{width=2.5cm}{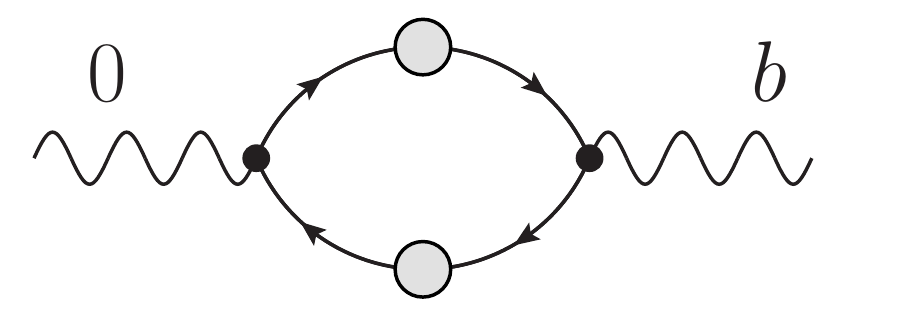}+
\feny{width=2.5cm}{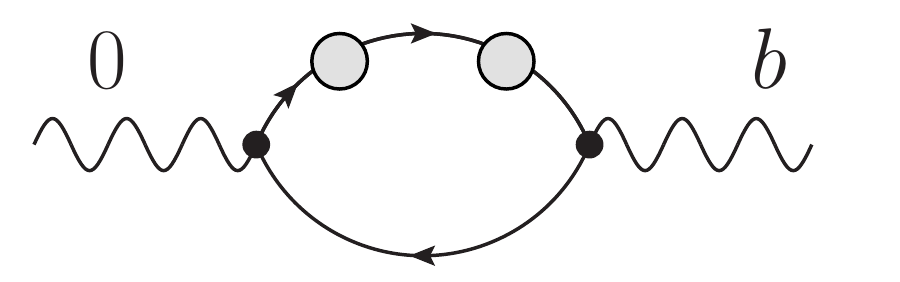} \nl &+ 
\feny{width=2.5cm}{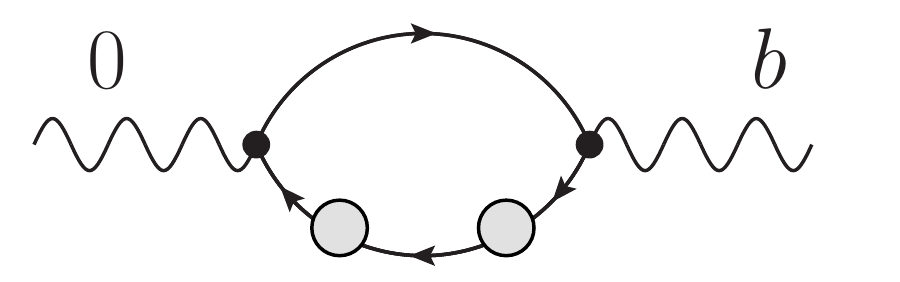} \\
K_1^{00}(q) &=  
\feny{width=2.5cm}{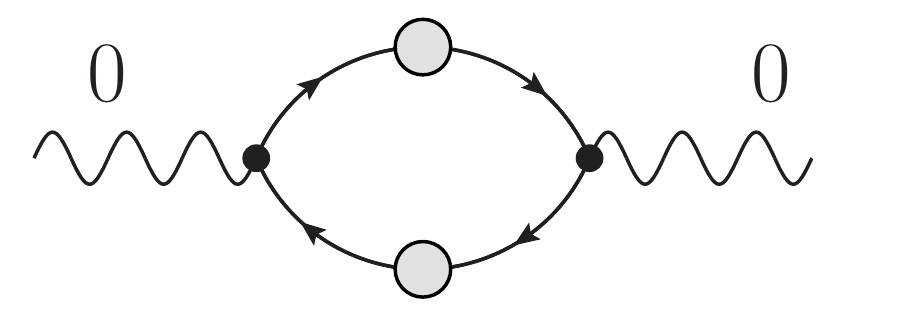}+
\feny{width=2.5cm}{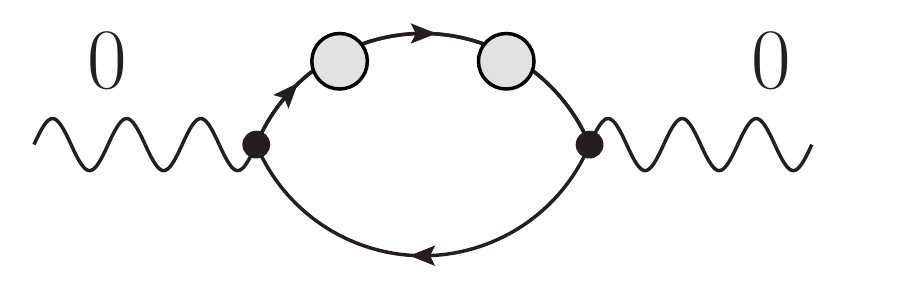} \nl &+ 
\feny{width=2.5cm}{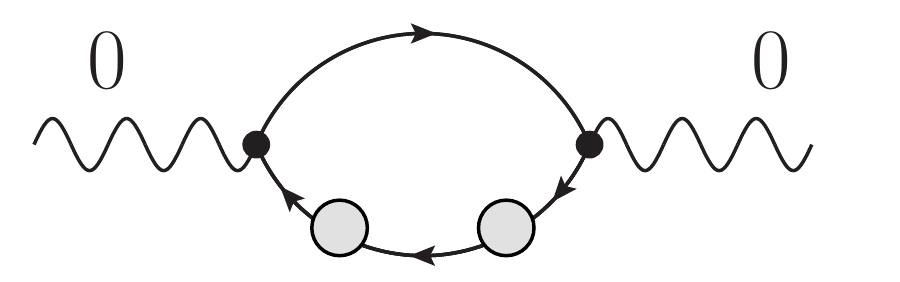}
\end{align}
\label{eqn:K1}
\end{subequations}
and 
\begin{subequations}
\begin{align}
K_2^{ab}(q) &= 
\feny{width=2.5cm}{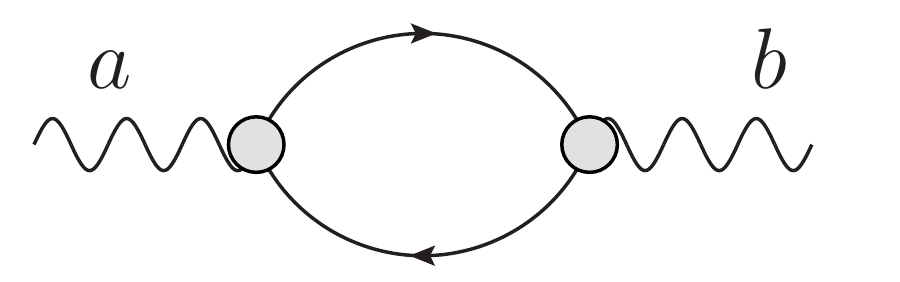}-
\feny{width=1.6cm}{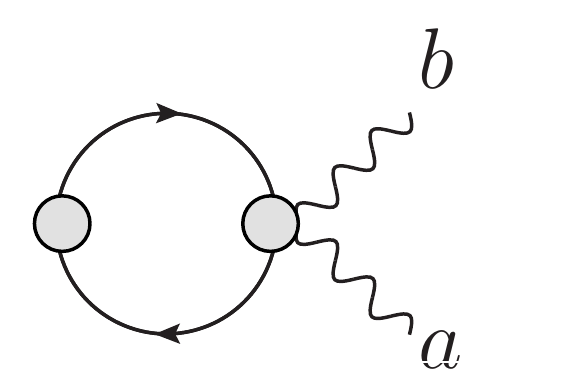} \nl &+ 
\feny{width=2.5cm}{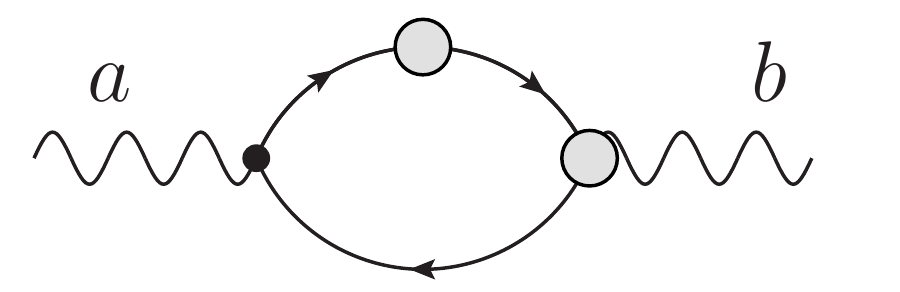} +
\feny{width=2.5cm}{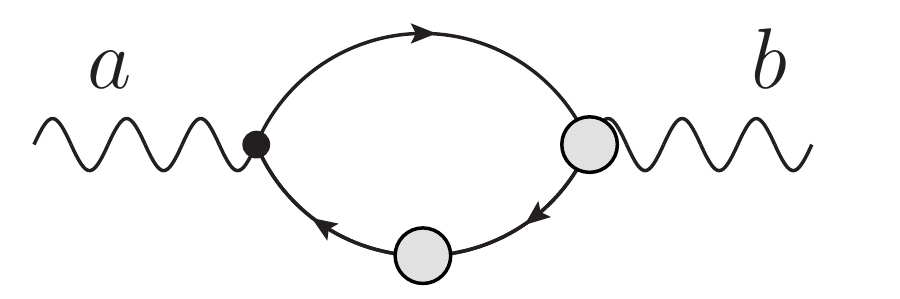}
\\
K_2^{0b}(q) &= \feny{width=2.5cm}{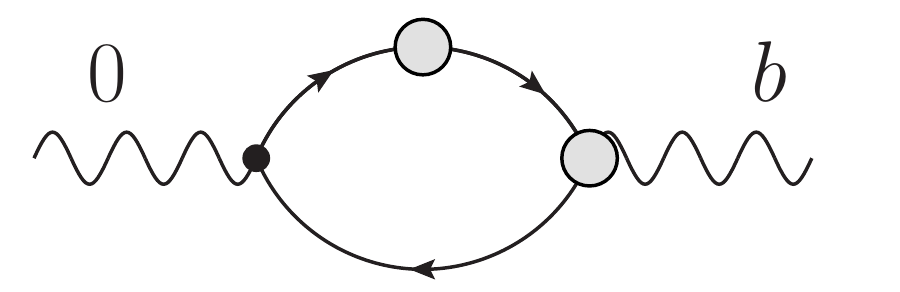} +
\feny{width=2.5cm}{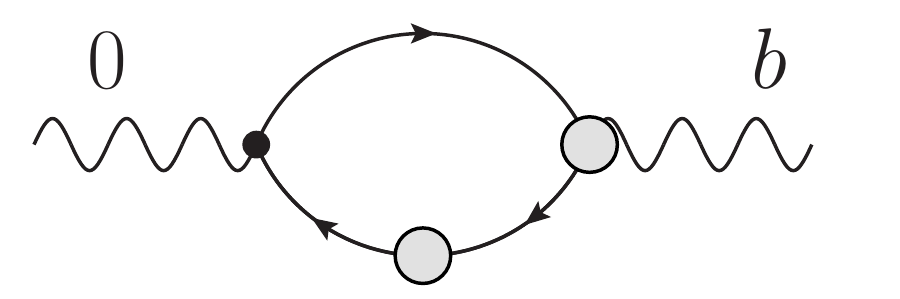}
\end{align} 
\label{eqn:K2}
\end{subequations}
The kernels $K_{1}^{b0}$ are similarly defined to $K_{1}^{0b}$, and $K_2$ does not have a density-density response counterpart $K_2^{00}$ because the quadrupole operator $\mathscr{O}$ does not couple to $A_0$. These kernels being all transverse means that they satisfy the following Ward identities
\begin{subequations}
\begin{align}
& q_a K^{a b}_m (q) + iq_0 K^{0 b}_m(q) = 0 \\
& q_a K^{a 0}_m (q) + iq_0 K^{0 0}_m(q) = 0
\end{align}
\label{eqn:Ward}
\end{subequations}
with $m=0,1,2$ where defined. Recall that by construction $\Gamma$ is gauge invariant, however the response kernels $K_{1,2}^{\mu\nu}$ depend very much on $\Gamma$. This can be traced back the minimally coupling to $A$ in the quadrupolar definition of $\Gamma$, which leads to these very non-trivial forms of $K_{1,2}^{\mu\nu}$. 

We remark that a further integration over the $\Gamma$-fields can be carried out by connecting the $(\feny{width=0.17cm}{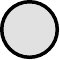})$ vertices by $\Gamma$-propagator lines $f(\qq)$ in Eq.\eqref{eqn:K1} and Eq.\ref{eqn:K2}. This yields linear response kernels that include Gaussian \emph{fluctuations} in $\Gamma$. The diagrams in Eq.\eqref{eqn:K1} and Eq.\eqref{eqn:K2} are then recognized as EM susceptibilities computed with vertex and self-energy corrections that satisfy Ward identities.\cite{metzner1998fermi,fradkin2013field} This is another way to interpret the equations Eq.\eqref{eqn:Ward}. However, in this paper we will not include the effects of $\Gamma$ fluctuations in the normal phase which are known to introduce non-analyticities\cite{belitz1997nonanalytic,chubukov2003nonanalytic} at higher order in perturbation theory for the $l=0$ channel. The case of the $l=2$ channel at first order in Gaussian fluctuations has so far not yielded any non-analyticities\cite{zacharias2009multiscale} that destabilize the zeroth order $\Gamma-\Gamma$ susceptibility, but more study is required to clarify the issue.  

To proceed, we need to evaluate the Feynman diagrams in above expressions Eq.\eqref{eqn:K1} and Eq.\eqref{eqn:K2} for $K^{\mu\nu}_m(q)$; which is very formidable for finite $q$. Since we are only interested in the response to slow $A$ fields, we can look for a gradient expansion in small $q$. This is however more subtle than it seems because the $\psi$ fluctuations are now generally gapless, unlike in the derivation of the effective action of Eq.\eqref{eqn:L_eff}. Hence we will expect non-local forms for $K^{\mu\nu}_m(q)$ which contains singularities at $q_0=|\qq|=0$. This precludes a Taylor expansion in $q$ and also a Laurent expansion for reasons that will become clear. Our resolution of this technical obstacle involves extracting the `leading order' singularities as $q\rightarrow 0$. Due to complex contours methods that are involved in the internal Matsubara frequency summations, these expansions resemble very much the Operator Product Expansions (OPEs)\cite{wilson1969non-lagrangian,kadanoff-1969,Polyakov-1970} in  particle physics and statistical physics whenever two operators are brought asymptotically close to one and another. The details of this gradient expansion method is described in Appendix \ref{app:gradient}.     

The explicit forms of the response kernels $K^{\mu\nu}_m(q)$ are collected in Appendix \ref{app:all_responses}, after a gradient expansion in $q$ has been applied. As was claimed, $K^{\mu\nu}_0(q)$ reduces to the usual forms of the polarization tensor and density-density response in (2+1) dimensions for an isotropic Fermi-liquid. They include both longitudinal and transverse components. Chiefly, the Landau damping remains present in these response kernels. For example, the density-density response $K_0^{00}(q)$ is given as
\begin{align}
K^{00}_0(q) = 2e^2 \bar{r} \left(1-\frac{|q_0|}{\sqrt{q_0^2 + v_F^2 |\qq|^2}}\right) 
\end{align}
where $\bar{r}=(N(v_F\Delta)+N(-v_F\Delta))/2$ is a DOS parameter. Note that the RHS is very much sensitive to the order of limits of $q_0,|\qq|\rightarrow 0$ due to the branch-cut singularity of the square root. This is the reason for why a Laurent expansion is not suited for these class of response kernel functions. Moreover, the other kernels $K^{\mu\nu}_{1,2}(q)$ also contain singularities of this sort. The key point is that, although the effective action Eq.\eqref{eqn:L_eff} for just the $\Gamma$ dynamics is local, the full EM response of the theory remains that of a metal and the coupling of $\Gamma$ to $A$ highly non-trivial and non-local. 

The order $O(\Gamma^2)$ corrections to the linear response is contained in $K^{\mu\nu}_{1,2}(q)$, where the anisotropy of the Fermi-surface due to $\Gamma \neq 0$ play out in the response. These are of lower order than $K_0$ due to their $\Gamma$ dependence and are only accurate near the critical point of theory. More importantly, they determine the form of the non-minimal EM coupling to the soft modes $\Gamma$ near the critical point. 

Now, the more important response kernel of the two is $K_2^{\mu\nu}(q)$ which contains the following two contributions
\begin{align}
&K_2^{ab} = -\left(\tfrac{4e^2}{k_F^2}\right) \left(\tfrac{r}{2v_F\Delta}\right) (\epsilon_{ij}\epsilon_{\mu \nu}\Gamma_{\mu i}\Gamma_{\nu j}) \epsilon_{ab} \, q_0 + \ldots \\
&K_2^{0b} = -\left(\tfrac{4e^2}{k_F^2}\right) \left(\tfrac{r}{2v_F\Delta}\right) (\epsilon_{ij}\epsilon_{\mu \nu}\Gamma_{\mu i}\Gamma_{\nu j}) \epsilon_{ab} \, iq_a  + \ldots
\end{align}    
which are linear in $q$ and hence local. In fact, they contribute to the effective action as a Chern-Simons Lagrangian
\begin{align}
\mathcal{L}_\text{CS} = \left(\tfrac{8e^2}{k_F^2}\right) \left(\tfrac{r}{2v_F\Delta}\right)
(\vec{\Gamma}_1\wedge\vec{\Gamma}_2) \epsilon_{ab} \left[
A_a \partial_t A_b +   A_0 \partial_a A_b
\right]
\end{align}  
where $\partial_t$ is the real-time derivative. The constant prefactor in this case is \emph{not} quantized, because of the gaplessness of the system. This does lead to a $\Gamma$ dependent intrinsic anomalous Hall response whenever $(\vec{\Gamma}_1\wedge\vec{\Gamma}_2) \neq 0$. In the normal and $\alpha_1$ phases  $\langle \Gamma_{\mu i}\rangle=0$. However, in the $\beta_1$ phase, we have $\langle \vec{\Gamma}_1\wedge\vec{\Gamma}_2\rangle  \neq 0$ and hence a non-zero anomalous Hall response co-exists with the other metallic EM response contributions that is strongly anisotropic in the $\alpha_1$ phase. This is in total agreement with the previous mean-field study\cite{sun2008time} of this model; albeit by a more complicated method. Furthermore, this Chern-Simons term distinguishes the $\beta_1$ phase from the $\alpha_1$ and normal phases. In fact, the sign of the non-universal Chern-Simon coefficient can used to distinguish between different chiral $\beta_1$-phase domains that differ in their $\mathbb{Z}_2$ order component (cf. Fig.\ref{fig:beta1}).    

\subsection{Broken Symmetry  Phases}

We next consider the EM response deep inside either symmetry broken phase. Here, we may determine the EM response using the mean-field approximation for the theory and where the dependence on $\langle \Gamma_{\mu i}\rangle \neq 0$ is no longer perturbatively quadratic. The mean-field action stems from the following manipulations of the partition function in the limit $A=0$.
\begin{align}
\mathcal{Z}_\text{tot} &= \mathcal{Z}_{\psi_0} \int \mathscr{D}\Gamma\; \e^{-S_1[\Gamma] + 
\text{Tr}\ln\left[\hat{1}- \hat{\mathcal{G}}^{-1}_0 \hat{\mathscr{O}}_2 \Gamma \right]  }\nl
&= \mathcal{Z}_{\psi_0} \int \mathscr{D}\Gamma\; \e^{-S_\text{eff}[\Gamma]} \nl
&\approx \mathcal{Z}_{\psi_0}\int \mathscr{D}(\delta\Gamma)\; \e^{
-S_\text{eff}[\Gamma_\text{class}] -S_\text{eff}^{''}[\Gamma_\text{class}]\delta \Gamma^2} \nl
&= \mathcal{Z}_{\psi_0} \e^{ -S_\text{eff}[\Gamma_\text{class}]}
\int \mathscr{D}(\delta\Gamma) \e^{-S_\text{eff}^{''}[\Gamma_\text{class}]\delta\Gamma^2} \nl
&= \mathcal{Z}_{\psi_0} \e^{
-S_1[\Gamma_\text{class}] + \text{Tr}\ln \left[\hat{1}- \hat{\mathcal{G}}^{-1}_0 \hat{\mathscr{O}}_2 \Gamma_\text{class} \right]} 
\mathcal{Z}_{\delta \Gamma,\Gamma_\text{class}} \nl
&= \det \left[ -\hat{\mathcal{G}}^{-1}_0 + \mathscr{\hat{O}}_2 \Gamma_\text{class} \right] 
\e^{-S_1[\Gamma_\text{class}]} \mathcal{Z}_{\delta \Gamma,\Gamma_\text{class}} 
\end{align}
where $\mathcal{Z}_{\psi_0} = \det[-\mathcal{\hat{G}}_0^{-1}]$
is the partition function for the free fermion model and $\Gamma_\text{class}$ the uniform classical mean-field that is a saddle point solution of $S_\text{eff}[\Gamma]$. The other partition function $\mathcal{Z}_{\delta \Gamma,\Gamma_\text{class}}$ is the partition function that results from integrating out the fluctuations $\delta \Gamma$ about the saddle point solution $\Gamma_\text{class}$. The final fermionic determinant factor contains in it the mean-field theory of the fermion through
\begin{align}
\det \left[ -\hat{\mathcal{G}}^{-1}_0 + \mathscr{\hat{O}}_2 \Gamma_\text{class} \right] = 
\int \mathscr{D}\psi\mathscr{D}\psi^\dagger\; \e^{
-\psi^\dagger [-\hat{\mathcal{G}}^{-1}_0 + \hat{\mathscr{O}}_2 \Gamma_\text{class} ]\psi
}
\end{align}
which then yields the following mean-field Lagrangian for the fermionic theory
\begin{align}
\mathcal{L}_\text{MF} &= \psi^\dagger_\alpha( \partial_0 + \xi(-i\nabla)\psi_\alpha \nonumber \\
\nonumber \\
&+\psi^\dagger_{\alpha}\left[v_F \Delta \sigma^z_{\alpha\beta}
+(\Gamma_\text{class})_{\mu i}\mathscr{O}^{\mu i}_{\alpha\beta}
\right] \psi_\beta.
\end{align}
We then define the mean-field Hamiltonian in $\kk-$space as 
\begin{align}
H_\kk = \xi_\kk + v_F \Delta \sigma^z + (\Gamma_\text{class})_{\mu i}\, \mathscr{O}^{\mu i}(\kk)
\end{align}
using the definitions of Eq.\eqref{eqn:O2}. Then minimally coupling to EM yields the following Lagrangian
\begin{align}
\mathcal{L}_\text{MF} &= -\psi^\dagger G^{-1} \psi - e A_0 \psi^\dagger \psi + e A_a\psi^\dagger \partial_a H \psi \nonumber \\
&+ \tfrac{e^2}{2}(\psi^\dagger \partial^2_{ab}H \psi) \, A_a A_b  
\end{align}
where $G = -[\partial_0 + H]^{-1}$ is the mean-field corrected Green's function and the derivatives are $\partial_a \equiv \partial_{k_a}$. Again for the purposes of linear-response, we only coupled to $O(A^2)$. The Green's function $G$ can be further expressed in spectral form as
\begin{align}
G(ik_0,\kk) &= \frac{1}{ik_0 - E_{1\kk}} P_1(\kk) + \frac{1}{ik_0 - E_{2\kk}} P_2(\kk) \nonumber \\ \nonumber \\
&= G_1(ik_0,\kk) P_1(\kk) + G_2(ik_0,\kk)  P_2(\kk)
\end{align} 
where $E_{1 \kk},E_{2 \kk}$ are the energy bands of $H_\kk$ such that $H_\kk= E_{1\kk}P_{1}(\kk)+E_{2\kk}P_{2}(\kk)$. The linear response kernels are then given by the same type of Feynman diagrams as in equation Eq.\eqref{eqn:Jtot}. Explicitly they are 
\begin{subequations}
\begin{align}
K^{ab}(q) &= -e^2 \idk \sk \sum_{m,n=1}^2 \nonumber \\
&\times\left\{ 
G_m(k^+)G_n(k^-)\text{tr}\left[
\partial_a H_\kk P_m(\kk^+)\partial_b H_\kk P_n(\kk^-)
\right] \right.\nonumber \\
&\left.\;\;\;-G_m(k)G_n(k)\text{tr}\left[
\partial_a H_\kk P_m(\kk)\partial_b H_\kk P_n(\kk)
\right]
\right\} \\
K^{0b}(q) &= +e^2 \idk \sk \sum_{m,n=1}^2 \nonumber \\
&\times 
G_m(k^+)G_n(k^-)\text{tr}\left[
P_m(\kk^+)\partial_b H_\kk P_n(\kk^-)
\right] \\
K^{b0}(q) &= K^{0b}(-q) \\
K^{00}(q) &= -e^2 \idk \sk \sum_{m,n=1}^2 \nonumber \\
&\times
G_m(k^+)G_n(k^-)\text{tr}\left[P_m(\kk^+) P_n(\kk^-)
\right].
\end{align}
\end{subequations}
The second line on the RHS of $K^{ab}(q)$ comes from the diamagnetic tadpole term of order $O(\AAA^2)$ which is necessary for gauge invariance. The different contributions to the linear-response can be organized into intra-band ($m=n$) and inter-band $(m\neq n)$ contributions as 
\begin{align} 
K^{\mu\nu}(q)= K^{\mu\nu}_\text{inter}(q) + K^{\mu\nu}_\text{intra}(q).
\end{align}
In the case of the inter-band terms, we can Taylor expand the $G_{1,2}(k^{\pm})$ as a series in $q/|E_{1\kk}-E_{2\kk}|$ due to the gap between the mean-field bands $E_{1,2 \kk}$. This then yields, after much simplification, the `topological' response \cite{nagaosa2010anomalous} at the lowest order in $q$
\begin{subequations}
\begin{align}
K^{ab}_\text{inter}(q) &= -q_0 e^2 \idk \sum_{m=1}^2 n_F(E_{m\kk}) \mathcal{F}_m^{ab}(\kk) \\
K^{0b}_\text{inter}(q) &= -iq_a e^2 \idk \sum_{m=1}^2 n_F(E_{m\kk}) \mathcal{F}_m^{ab}(\kk) \\
K^{b0}_\text{inter}(q) &= K^{0b}_\text{intra}(-q) \\
K^{00}_\text{inter}(q) &= 0
\end{align}  
\end{subequations}
where $\mathcal{F}_m (\kk) = \nabla_{\kk}\wedge \mathcal{A}_m(\kk)$ is the Berry curvature for energy band $m$ and $\mathcal{A}_m(\kk)$ is the associated Berry connection. Finally, using the expansion of Eq.\eqref{eqn:GpGm}, the intra-band contributions are determined to be 
\begin{subequations}
\begin{align}
K^{ab}_\text{intra}(q) &= e^2 \idk \sum_{m=1}^2 \left(
\tfrac{iq_0 }{q_c \, v^c_{m\kk} - iq_0}
\right) v^a_{m\kk}v^b_{m\kk}\;\delta(E_{m\kk}) \\
K^{0b}_\text{intra}(q) &= -e^2 \idk \sum_{m=1}^2 \left(
\tfrac{q_c \, v^c_{m\kk}}{q_c \, v^c_{m\kk} - iq_0}
\right) v^b_{m\kk}\;\delta(E_{m\kk}) \\
K^{b0}_\text{intra}(q) &= K^{0b}_\text{intra}(-q) \\
K^{00}_\text{intra}(q) &= e^2 \idk \sum_{m=1}^2 \left(
\tfrac{q_c \, v^c_{m\kk}}{q_c \, v^c_{m\kk} - iq_0}
\right) \;\delta(E_{m\kk}) 
\end{align}
\end{subequations}
where $v^a_{m\kk} = \partial_{k_a}E_{m\kk}$ is the band velocity. Note that the Ward identities are satisfied separately for the inter-band and intra-band contributions. Lastly, the inter-band response kernel $K^{\mu\nu}_\text{inter}(q)$ contains the Hall conductivity which is zero in the $\alpha_1$-phase but is non-zero in the $\beta_1$-phase as is expected. 

\section{Reduced Symmetry}\label{sec:reduced_sym}
\label{sec:reduced}

Next, we discuss the consequences of explicit symmetry breaking of the $O(2)_\text{rot}\times O(2)_\text{iso}$ group. The findings of Ref.[\onlinecite{sun2008time}] remain largely unchanged compared to our analysis of the effective field theory. The breaking of rotational $O(2)_\text{rot}$ down to the discrete rotational group $C_n$ may be caused by the lattice and will have the effect of removing an orientational Goldstone mode from either $\alpha_1$ or $\beta_1$ phases. More importantly, because it is an orientational Goldstone mode, the remaining Goldstone mode in the $\alpha_1$ phase will no longer be over-damped at low energy. This is because the spontaneously broken continuous symmetry $O(2)_\text{iso}$ is an internal symmetry, and in accordance to the criterion of Ref.[\onlinecite{watanabe2014criterion}] will not suffer Landau damping. Meanwhile, the $O(2)_\text{iso}$ symmetry which ensures separate fermionic number conservation in each band may be broken by the inclusion of additional interaction terms of the form $\Phi_{xi}\Phi_{yi}$ and/or $[\Phi_{xi}\Phi_{xi}- \Phi_{yi}\Phi_{yi}]$ to the Lagrangian Eq.\eqref{eqn:S_micro}. The new lower symmetry of the microscopic model is $\mathbb{Z}_2$ which conserves the relative number parity of each band $(-1)^{N_1-N_2}$ and is affected by the transformation $\Phi_{\mu i} \rightarrow (-\Phi_{\mu i})$. This yields additional terms in the effective action in Eq.\eqref{eqn:Leffphi} of the forms $\phi_i\phi_i$, $\phi_i^*\phi_i^*$ which explicitly breaks number conservation in the $\phi$ boson and gaps out the $\vec{s}\cdot\vec{\pi}$  mode in the symmetry broken phases. These $\text{O}(2)_\text{iso}$ symmetry breaking terms may have a microscopic origin in the details of their atomic orbitals that leads to the reduced symmetry of the Landau interaction parameters. The final condensed $\phi_i$-phases in the presence of reduced symmetry models are noticeably more complicated and were termed the $\alpha_{2}$ and $\beta_{2}$ phases in Ref.[\onlinecite{sun2008time}]. In particular, the $\beta_2$ phase  -- like the $\beta_1$ -- breaks the requisite chiral (mirror) and time-reversal symmetries such that an anomalous Hall effect is present. Finally, the $O(2)_\text{rot}\times O(2)_\text{iso}$ symmetric model and its three possible reduced symmetry relatives $C_n\times O(2)_\text{iso}$, $O(2)_\text{rot}\times \mathbb{Z}_2$ and $C_n\times \mathbb{Z}_2$ will have qualitatively different phenomenological finite temperature phase diagrams that are discussed at length in Ref.[\onlinecite{sun2008time}].

\section{Experimental Realizations}\label{sec:experiments}

In this section, we describe possible experimental realizations of the model of Eq.\eqref{eqn:S_micro}. But before that, we point out key challenges that are faced when identifying suitable candidate materials. We focus only on electronic liquids in the solid state and avoid cold atomic systems altogether.\\

\noindent\emph{\color{black}2D/Quasi-2D Fermi Fluids-}{ A two dimensional or quasi two dimensional metallic electron liquid is needed. This requirement may be met either with a legitimate 2D electron gas (2DEG) or stacked two dimensional metals {\color{black}in 3D} with weak interlayer coupling that leads to quasi-two dimensional metals.}\\

\noindent\emph{Broken $SU(2)$ Symmetry-}{ The electron liquid needs to possess two internal flavors with an energy splitting that lifts their degeneracy. Mathematically, this amounts to an internal $SU(2)$ symmetry that is either broken spontaneously or explicitly. The electron's spin-1/2 degree of freedom is the most natural way to realize this and hence points to itinerant ferromagnets as likely candidates. In fact, there are many materials with spin-polarized metallic bands due to significant Hund's coupling at the atomic level.  Most of these materials contain partially filled transition metal ions. {\color{black} Alternatively, atomic orbitals from different layers in a 2D bilayer system can realize a broken $SU(2)$ symmetry. In this scenario, rotational spin-1/2 degeneracy remains intact but interlayer coupling leads to bonding, anti-bonding orbitals which yields an energy splitting. Effectively, a spin-less electron picture can be adopted with the effects of spin symmetry appearing only as the occasional spin-degeneracy factor of two in some physical quantities like electrical conductivity.}}\\

\noindent\emph{Ferromagnetic Exchange Interactions-}{ After the previous two conditions are met, then a sizable `\emph{exchange}' interaction between the two energy split electron liquids is required. Such interactions are generally present and can be modeled semi phenomenologically with the Landau parameters. More rigorously they can be argued to exist from renormalization group\cite{shankar1994renormalization} arguments starting from the bare Coulomb interaction. However determining whether or not the interaction is ferromagnetic or anti-ferromagnetic is more difficult. But ferromagnetic tendencies are more common in itinerant systems. Enhancing these interactions requires a large density of states or equivalently narrow energy bands. Generally speaking, heavy fermion materials {\color{black}could} meet this criterion.}\\

\noindent\emph{Approximate Isotropy-}{ Perfectly isotropic Fermi surfaces are only possible in liquid He$^3$. But there exists well studied materials like the doped cuprate system {\color{black}or GaAs quantum wells at low doping which can meet} this criterion quite well. Moreover, crystals with the $D_4$ point subgroup would be more desirable because its $d_{xy}$ and $d_{x^2-y^2}$ irreducible representations are quadrupolar like. Note that the quadrupolar channel needs to dominate over the other partial wave channels, and in this way the point group of the lattice may play a crucial role. Technically, the effects of the lattice can be accounted for by additional terms in the effective action of Eq.\eqref{eqn:L_eff} that degrade the continuous $O(2)_\text{rot}$ symmetry to favor alignment to the crystallographic directions. 

From these consideration, we have three broad proposals and each has their own challenges in terms of finding realistic material candidates. 

\subsection{Itinerant Ferromagnets}

In this scenario, an exchange splitting between otherwise spin-1/2 degenerate Fermi liquids originates from a Stoner instability. This is just the particular case of the Pomeranchuk instability into the antisymmetric $l=0$ partial wave channel that fits into our overall narrative of strong ferromagnetic interactions within the Fermi-liquid formalism. The exchange interaction in the quadrupolar channel then represents part of the residual interactions of the unbroken $O(2)$ symmetry with the $\Gamma_{\mu i}$ bosons interpreted as spin flip Stoner excitations in the $l=2$ partial wave channel. This is all supposed to occur after the first symmetry breaking which defines the magnetization direction and is accompanied by conventional magnetic $l=0$ Goldstone modes. 

However, it is now well established that the Hertz-Millis approach to Stoner ferromagnetism is incorrect\cite{lohneysen2007fermi} due to the appearance of non-analytic momentum-frequency terms in the effective action invalidating the initial assumptions of the approach and destabilizing the quantum critical point. Hence, we have to imagine instead a scenario where the itinerant ferromagnetism has been established by means other than the Stoner mechanism and consider those itinerant ferromagnets whose magnetism is not attributed to magnetic constituent ions because these lead to oppositely spin-polarized bands which are too dissimilar.  

Most ferromagnetic metals like elemental Fe and Ni are actually made of magnetic constituent ions which are spin-split due to Hund's coupling. Nevertheless, there are known `Stoner ferromagnetic' materials. But as was mentioned these are not meant to be taken literally as ferromagnets whose ordering is attributed to the Stoner mechanism. Rather, they represent known itinerant ferromagnets whose isolated constituent ions are non-magnetic to begin with. The two known itinerant ferromagnets without magnetic ion constituents are Sc$_3$In\cite{matthias1961ferromagnetism} and ZrZn$_2$\cite{matthias1958ferromagnetism}. In fact, non-Fermi liquid behavior has recently been reported in chemically doped Sc$_3$In\cite{svanidze2015non} and it possesses a crystal structure of stacked hexagonal layers. Hence an $l=6$ hexapole (hexatic) generalization of our model may be a more appropriate treatment of this system. By contrast, ZrZn$_2$ is a less likely candidate with its diamond like lattice structure and it is known to have magnetic order that is strongly three dimensional\cite{brown1984magnetisation,uhlarz2004quantum} and superconductivity\cite{pfleiderer2001coexistence} at temperatures below the Curie point. Nevertheless, we do not yet know of any thin-film experiments with Sc$_3$In or ZrZn$_2$, but hope that this work will encourage future work with these and related materials.   

\subsection{Spin Polarized Junctions}

\begin{figure}
\includegraphics[width=8cm]{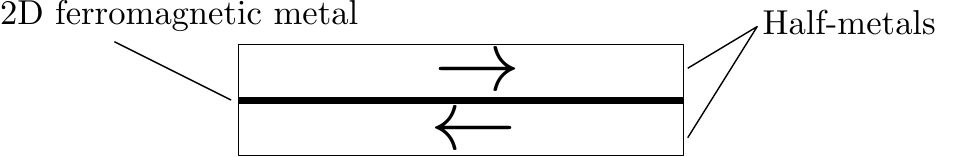}
\caption{Schematic of a layered structure comprised of two thin films of completely spin-polarized metals (``half-metals'') that are oppositely polarized. At the interface between the layers lies a two dimensional electron liquid that remains ferromagnetically ordered due to the imbalance of spin polarized 2D bands that may be controlled by gating. Here the bulk half-metals are assumed to be magnetically hard or pinned by hard magnets (not shown) above and below.}\label{fig:half_metal}
\end{figure}

``Half-metals''\cite{pickett2001half} are special itinerant ferromagnets with completely spin-polarized metallic bands. The name refers to the fact that the partially filled bands at the Fermi-level are of a single spin-polarization because bands with the opposite polarization are insulating or semiconducting. We imagine that a heterojunction of two oppositely polarized half-metals as shown in Fig.\ref{fig:half_metal} could serve as an experimental realization for the model. The oppositely spin-polarized two dimensional electronic liquids at the interface originate from different completely spin-polarized bulk metals. Generically an energy splitting between the two dimensional spin polarized Fermi liquids can be expected and may even be controlled by gating one side of the two sides of the junction. This leads to a metallic ferromagnetic two dimensional liquid at the interface in complete analogy with the previously proposed itinerant Stoner ferromagnet scenario. 

Of all the known half-metals, CrO$_2$\cite{schwarz1986cro2} is the most well studied\cite{korotin1998cro,coey2002half} and continues to be investigated actively.\cite{heffernan2016role,bisti2016site} Its rutile crystal structure and previously measured bulk electronic structure\cite{bisti2016site} are somewhat encouraging for our purposes. Additionally, synthesis of thin film CrO$_2$ has also been reported.\cite{desisto2000highly} Other known half-metals include La$_{0.67}$Sr$_{0.33}$MnO$_3$\cite{park1998direct} and NiMnSb,\cite{ristoiu2000surface}. We refer the reader to Ref.[\onlinecite{katsnelson2008half}] for a more exhaustive list.  

\subsection{2D materials}

In this last scenario, we maintain $SU(2)_\text{spin}$ symmetry and appeal to the isospin interpretation of the model. The effects of spin-1/2 rotational symmetry of the electron if present then manifests as a spin-degeneracy factor. We imagine the $SU(2)_\text{iso}$ degree of freedom originates from a layer index in a 2D bilayer material. Inter-layer coupling then provides the necessary energy splitting between the bonding and anti-bonding bands. 

Ref.[\onlinecite{sun2008time}] had suggested this as a possible interpretation of the model as well as several proposed lattice models. Of note are honeycomb lattice models which are physically realized in graphene. Moreover in Ref.\onlinecite{grushin2013charge}, an extended Hubbard model of spinless electrons on the honeycomb lattice was shown to have metallic anomalous Hall phases and ($C_6$ breaking) Pomenranchuk phases within the mean-field approximation.            

Lately there have been many advances in the synthesis and isolation of atomically thin monolayer materials besides graphene. 
The most promising of which for our purposes are the transition metal dichalcogenide (TMDC) bilayers. Some currently well studied examples include MoS$_2$,\cite{mak2010atomically} WS$_2$, MoSe$_4$ and MoTe$_2$\cite{coleman2011two} to name a few. These 2D materials are hexoganal crystal sheets with a semiconducting gap. Doping is required to produce metallic bands which in the bilayer system (AA stacking) would lead to a version of our proposed model with some differences. For one, the hexagonal symmetry of these 2D materials would likely mean that the $l=6$ hexapolar (hexatic) generalization of our model would be more suited. Secondly, there is significant spin-orbit interaction in the conduction and valence bands that breaks the rotational spin-1/2 degeneracy and an additional 2-fold valley degeneracy of the honeycomb lattice. The latter is an exact symmetry which exhibit an incipient anomalous valley Hall effect. However unless a net valley polarization is present, contributions from each valley cancels exactly.

\section{Summary and Discussion}\label{sec:conclusions}

In this paper, we have re-visited a deceptively simple model of a two component soft electron liquid, first proposed in Ref.[\onlinecite{sun2008time}] as an interaction driven quantum anomalous Hall liquid. In the absence of interactions, the 2D electron liquid is split in its bands and can be interpreted as an itinerant ferromagnet. Interactions are incorporated in the form of $XY$-exchange forward scattering terms in the quadrupolar partial wave channel. Due to the band splitting, fluctuations induced by interactions are gapped and quadrupolar in character. This then leads to a relatively simple effective Landau-Ginzburg action functional for the bosonic quadrupolar order parameter fields. Chiefly, the quantum critical theory resembles that of a two component theory of non-relativistic superfluidity with a non $SU(2)$-symmetric self-interaction which allows for a straightforward RG analysis. 

The broken symmetry phases of model are the $\alpha_1$ and $\beta_1$ electronic liquid crystal phases whose Nambu-Goldstone modes we have analyzed. Despite the apparent simplicity of the Landau-Ginzburg theory, the eventual fate of the broken symmetry phase is really that of a non Fermi-liquid due to Landau damping of the Nambu-Goldstone modes from low energy particle-hole pairs at small momentum. 

We have analyzed the electromagnetic linear response in the normal phase near the quantum transition and in the broken symmetry phases. We find that the non-minimal EM coupling to the quadrupolar boson in the normal phase is surprisingly complicated and strongly anisotropic. Also, despite the deceptively local form of the Landau-Ginzburg action, the electronic system remains very much a metal and Landau damping is present in the EM response. More interestingly, precursors to the quantum anomalous Hall phase are visible even in normal phase and manifest as a Chern-Simons action with \emph{non-universal} coefficient that is proportional to the amount local chiral and time-reversal symmetry breaking due to the quadrupolar boson. Finally, we propose several experimental candidates that might realize the model or a related hexapolar variant. 

The richness and complexity of the order-parameter theory may be traced to the non-trivial form of the quadrupolar interaction and the metallicity of the fermionic sector. Although we have mostly limited ourselves to an order parameter only formulation, much in the spirit of the Hertz model, it is likely that a complete formulation low energy-long wavelength will always require both fermionic and bosonic sectors. In fact, the damping of Nambu-Goldstone was a situation where mean-field corrected fermionic fields had to be re-introduced that eventually led to non Fermi-liquid behavior at very low momenta. This parallels much of the recent revisions of Hertz-Millis theories.\cite{lohneysen2007fermi} 
{In addition, there is the open question of whether or not non-analyticities in the effective action are encountered at higher order in expansion of the effective action in Eq.(\ref{eqn:Seffall}), as it does in Stoner ferromagnetism.\cite{belitz1997nonanalytic,chubukov2003nonanalytic} If present,  they may invalidate  aspects of the Hertz-Millis like approach, as in the case of Stoner theory. On the other hand, the finite $\Delta$ splitting of the Fermi surfaces may serve as an energy/momentum scale that could suppress some of the potential non-analyticities at higher order, as was in the case of our effective action at lowest order. At any rate, the main focus of this paper is on the structure and properties of the phases and not on the validity of the Hertz-Millis approach to quantum criticality which by now is well known to be qualitative. Nevertheless, we should note that a full solution of the problem for quantum criticality in metals, a subject of intense research in recent years, is still a largely unsolved problem (for a recent insightful discussion, with references, see Ref.[\onlinecite{sslee-2017}]).} 

It would be desirable to know if non-perturbative methods such as multi-dimensional bosonization\cite{houghton1993bosonization,neto1994bosonization,lawler2006nonperturbative,haldane2005luttinger} may be brought to bear to this and related Fermi-liquids where inter-band and topological effects are non-trivial.  For example, topological Fermi-liquids\cite{volovik1991new,horava2005stability} are a class systems that might profit from a bosonized formulation. We note that Ref.[\onlinecite{kwon1995theory}] had developed a fairly complete EM response theory couched in the multi-dimensional bosonization framework which is RPA exact. However situations with broken $SU(2)$ Fermi-liquids and interaction driven Berry phase responses remains largely unexplored. Moreover, we believe it likely that the concerns raised in the previous paragraph are intimately related to feasibility of such a bosonized formulation. 
{For example, the issue of non-analyticities in the effective action, if encountered at higher order may are not taken into account by the existent higher-dimensional bosonization theories which focus only on intra-patch forward scattering processes. Since  the non-analytic terms arise from a subtle interplay of large and small momentum transfers, the bosonization approach will have to be modified to include the inter-patch processes. This is an open and interesting problem.}
Lastly, extensions of this and related models to 3D is certainly possible with the rotational $O(2)$ group is replaced by $O(3)$. This could allow for richer broken symmetry theories with monopole defects like those already considered in Ref.[\onlinecite{wu2007fermi}].     

\begin{acknowledgments}

This work was supported in part by the Gordon and Betty Moore Foundation EPiQS Initiative through Grant GBMF4305 (VC), and by the National Science Foundation through the grant  DMR-1408713 at the University of Illinois (WA,EF). 
{We thank one of our referees for comments on the role of nonanalytic contributions to the effective action.}
%

\end{acknowledgments}

\appendix

\begin{figure*}
	\boxed{\includegraphics[width=0.9\textwidth]{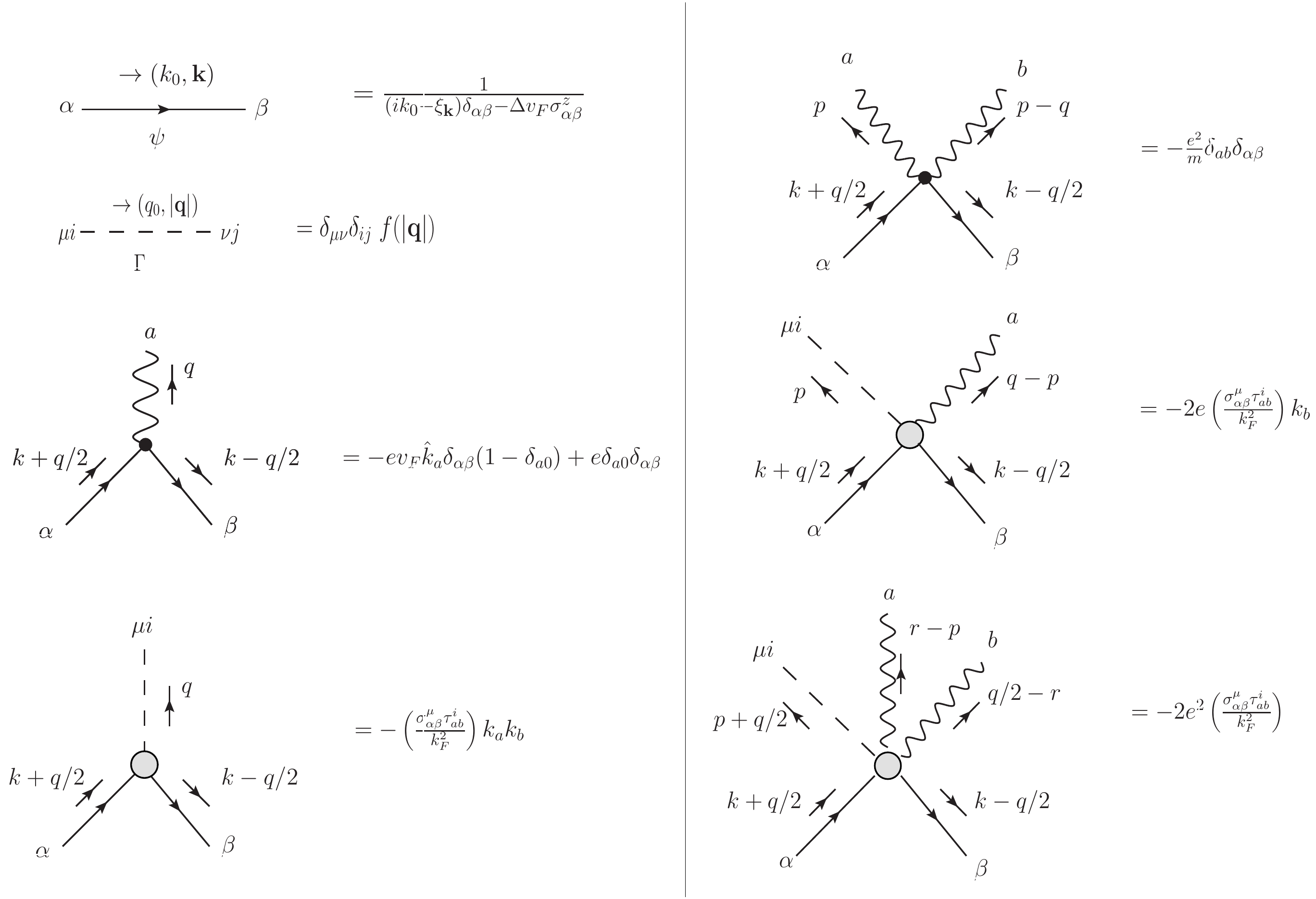}}
	\caption{Diagrammatic lines and fermions for the Feynman rules of the Hubbard-Stratanovich decoupled action and with coupling to electromagnetism. The form of the EM coupling in Eq.\eqref{eqn:Stot2} and Eq.\eqref{eqn:Stot3} yields two additional vertices that couple to $\AAA$ that are due to distortions and isospin-textures on the Fermi-surfaces due to $\Gamma$.}
	\label{fig:FeynmanRules}
\end{figure*}

\section{Derivation of the effective Lagrangian}\label{app:derive_Seff}

In this appendix, we describe the derivation of the effective Lagrangian of Eq.\eqref{eqn:L_eff}. From the total gauge invariant action of equation Eq.\eqref{eqn:Stot} and Eq.\eqref{eqn:Stot2} and Eq.\eqref{eqn:Stot3}, we can read off the various propagators lines and interaction vertices in $(k_0,\kk)-$space. These are listed in Fig.\ref{fig:FeynmanRules}. Expanding Eq.\eqref{eqn:Seffall} in order of $\hat{\mathcal{M}}$ yields  
\begin{align}
S_\text{eff}[\Gamma,A] =&  -\frac{1}{2f}\Gamma\Gamma - 
\text{Tr}\;\hat{\mathcal{M}}[\Gamma,A]+\tfrac{1}{2} \text{Tr}\;\hat{\mathcal{M}}[\Gamma,A]^2 \nonumber \\
&-\tfrac{1}{3}\text{Tr}\; \hat{\mathcal{M}}[\Gamma,A]^3 +\tfrac{1}{4}\text{Tr}\; \hat{\mathcal{M}}[\Gamma,A]^4 +\ldots
\label{eqn:loop_expansion}
\end{align} 
Although this expansion in terms of Feynman diagrams (See Fig.\ref{fig:FeynmanRules}) involves an enormous amount of terms, there is one important simplification. Namely that the $\hat{\OO}_n$ vertices involve transitions between $\psi$ bands but $\mathcal{G}_0$ remains diagonal in the band basis. Hence only diagrams with $\hat{\OO}_n$ appearing an even number of times in a fermion loop are non-zero. Also, we shall only pursue the expansion up to the fourth order, since this -- as we will show --  produces all the marginal couplings that are needed to complete an RG analysis of the theory. Next we set $A=0$, which yields $\hat{M} = - \hat{\mathcal{G}_0} \hat{\mathscr{O}}_2 \Gamma$. Thus only the quadratic $O(\Gamma^2)$ and quartic $O(\Gamma^4)$ kinetic and self-interaction terms generated by the single fermion loop contribute to the effective action. One other factor that constraints the form of the effective action is the adherence to $O(2)_\text{iso}\times O(2)_\text{rot}$ symmetry which requires that the invariant symbols $\epsilon_{\mu\nu},\epsilon_{ij}$ appear in appropriate combinations. 

Before that, we point out  a key difference of this derivation of $S_\text{eff}$ compared to the derivation of the free-energy in Refs.[\onlinecite{wu2007fermi,sun2008time}] within the self-consistent mean-field approximation. Ordinarily the effective action and the free-energy should agree in the case of static mean-fields. However the self-consistency method of Refs.[\onlinecite{wu2007fermi,sun2008time}] uses mean-field ansatz solutions from the saddle point equations of Eq.\eqref{eqn:Seffall} that includes the \emph{full} trace-log functional. Technically this means that effects of the field $\Gamma_{\mu i}$ on $\psi_\alpha$ have been Dyson re-summed. Hence the resulting free-energy functional should be more accurate deep inside the symmetry broken phase. However, the free-energy couplings that are determined this way will be ansatz dependent and may be misleading near the critical point of the theory. Furthermore, the form of the couplings to be derived below can always be associated to specific loop diagrams and hence to specific virtual processes undertaken by fluctuations in $\psi_\alpha$.

\subsection{Zeroth Order}

To the lowest order we have in momentum space
\begin{align}
&-\frac{1}{2f(\qq)}\Gamma(-q)_{\mu i}\Gamma(q)_{\mu i}  \nonumber \\
&= \frac{1}{2 |f(0)|}\Gamma_{\mu i}(-q) \Gamma_{\mu i}(q) 
+\frac{\kappa |\qq|^2}{2} \Gamma_{\mu i}(-q)\Gamma_{\mu i}(q)
\end{align}
where we use the form Eq.\eqref{eqn:fq}. We thus identify a bare `mass' of and a gradient term that derives from the form of $f(|\qq|)$. In real-space this yields the following zeroth order contribution the effective action
\begin{align}
\mathcal{L}^{(0)}_\text{eff} = \frac{1}{2|f(0)|} \Gamma_{\mu i} \Gamma_{\mu i} + \frac{\kappa}{2} \nabla \Gamma_{\mu i} \nabla \Gamma_{\mu i}.
\label{eqn:Szeroth}
\end{align}

\subsection{Quadratic Interaction}

\begin{figure}
	\boxed{\includegraphics[width=0.4\textwidth]{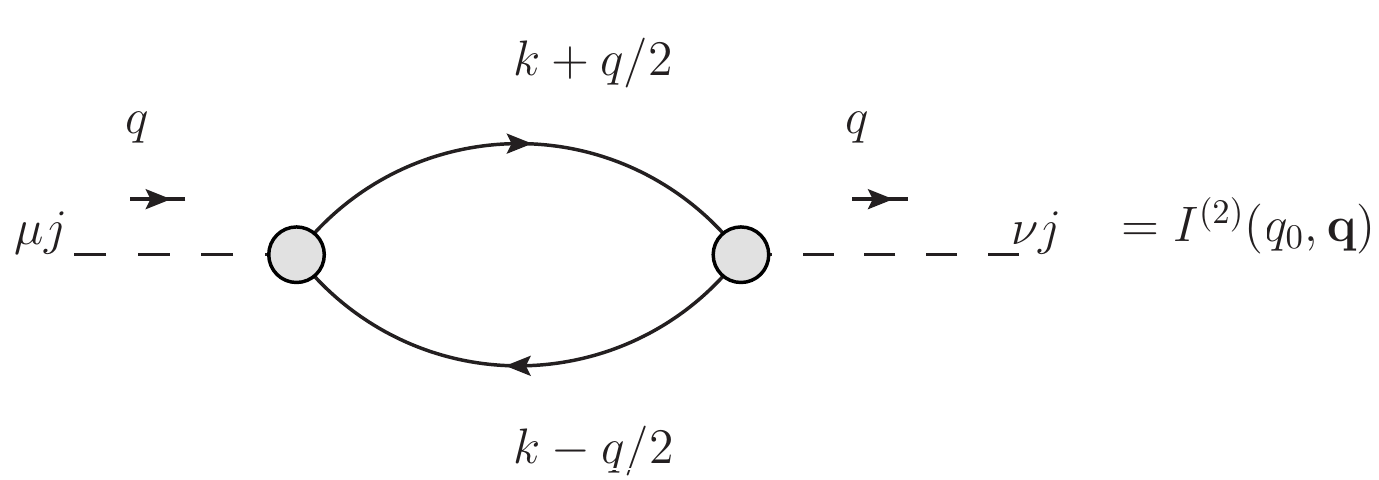}}
	\caption{The susceptibility bubble for the $\Gamma_{\mu i}$ field which also determines its temporal dependence under a gradient expansion of $(q_0,\qq)$.}
	\label{fig:Gamma2}
\end{figure}

The simplest diagram to consider is the $O(\Gamma^2)$ bubble diagram shown in Fig.\ref{fig:Gamma2}. Applying the standard rules gives
\begin{align}
I^{(2)}_{\mu i \nu j}(q) &=\frac{(-1)}{k_F^4} \idk \sk (\kk^T \tau^i \kk)(\kk^T \tau^j \kk) \nonumber \\
&\times \sum_{m,n = 1,2} \sigma_{mn}^\mu \sigma_{nm}^\nu G_m(ik_0^+,\kk^+)G_n(ik_0^-,\kk^-) \nonumber \\
\label{eqn:I2} 
\end{align}
where $k_0^\pm = k_0 \pm q_0/2$, $\kk^\pm = \kk \pm \qq/2$, the Green's function
\begin{align}
G_n(ik_0,\kk) = \frac{1}{ik_0 - \xi_n(\kk)}
\end{align}
and the indices $m,n$ sum over the fermionic bands. 
Then performing the Matsubara frequency summation and linearizing $\kk \approx k_F\hat{\kk}$ yields,
\begin{align}
I^{(2)}_{\mu i \nu j}(q) & = -\idk (\hat{\kk}^T\tau^i \hat{\kk}) (\hat{\kk}^T\tau^j \hat{\kk}) \nonumber \\
&\times \sum_{m,n=1,2} \sigma^\mu_{mn}\sigma^\nu_{nm}
\left[
\frac{n_F(\xi_{n\kk^-})-n_F(\xi_{m\kk^+})}{iq_0 + \xi_{n\kk^-}-\xi_{m\kk^+}}
\right].
\end{align}
Then after Taylor expanding $\xi_{m \kk^\pm}$ and $n_F(\xi_{m\kk^\pm})$ to linear order in $\qq$, using the identities 
\begin{align}
\sigma_{12}^\mu \sigma_{21}^\nu = \delta_{\mu \nu} + i \epsilon_{\mu \nu}, \quad 
\sigma_{21}^\mu \sigma_{12}^\nu = \delta_{\mu \nu} - i \epsilon_{\mu \nu},
\label{eqn:sigma1221}
\end{align}
with a change of variables $\idk \rightarrow \int \frac{\diff \theta_\kk}{2\pi} \int \diff \xi \; N(\xi)$, and some further simplification, finally yields
\begin{align}
I^{(2)}_{\mu i \nu j}(q) &= -2r \dtheta \left[\frac{(\hat{\kk}^T\tau^i \hat{\kk}) (\hat{\kk}^T\tau^j \hat{\kk})}{(iq_0 - v_F \delta q)^2 - (2v_F\Delta)^2}\right] \nonumber \\
&\times\left\{
\left[(2v_F\Delta)^2 + v_F\delta q (iq_0 -v_F\delta q) \left(\frac{\bar{r}}{r}\right)\right]\delta_{\mu\nu} \right. \nonumber \\
&\left.\hspace{0.7cm}+ 2v_F\Delta \left[ (iq_0 -v_F\delta q )+v_F\delta q \left(\frac{\bar{r}}{r}\right)\right] i\epsilon_{\mu \nu}
\right\}.
\end{align} 
where we have defined the following averaged DOS quantities
\begin{align}
r = \frac{1}{2v_F\Delta} \int_{-v_F\Delta}^{v_F\Delta} N(\xi) \diff \xi,\quad 
\bar{r}  = \frac{N(v_F\Delta)+N(-v_F\Delta)}{2}
\end{align}
and defined $\delta q = \hat{\kk}\cdot \qq$. 
\begin{widetext}
In the case where the DOS is constant, for a quadratic dispersion, one has $r=\bar{r}$. Implicit in the above derivation is the zero temperature limit of $n_F(\xi) = \Theta(-\xi)$ which is taken after the Taylor expansion in $\qq$. We separate out the static $(q_0=0)$ and dynamic contributions $(q_0\neq 0)$ by re-arranging
\begin{align}
I^{(2)}_{\mu i \nu j}(q) &= r \delta_{\mu \nu}\delta_{ij} - 2r \dtheta \left[
\frac{(\hat{\kk}^T \tau^i \hat{\kk})(\hat{\kk}^T \tau^j \hat{\kk}) 
	\left\{ (iq_0 - v_F\delta q)\delta_{\mu\nu} + (2v_F\Delta)i\epsilon_{\mu\nu} \right\} \left[iq_0 + \left(\frac{\bar{r}}{r}-1\right) v_F\delta q\right] 
	} {(iq_0 -v_F\delta q)^2 -(2v_F\Delta)^2}
\right].
\end{align}
\end{widetext}
This form of the susceptibility bubble is tremendously useful for studying the two limits where $0 < |\qq |,q_0/v_F< \Delta$ and $0 < \Delta < |\qq|,q_0/v_F $. In the former, we are justified in making an expansion of the denominator in the integrand in $\frac{|\qq|}{\Delta}$ and $\frac{q_0}{v_F\Delta}$, which yields after performing the angular integral
\begin{align}
&I_{\mu i \nu j}^{(2)}(q)  = r \delta_{\mu \nu}\delta_{ij}+ r (i\epsilon_{\mu \nu} \delta_{ij}) \left(\frac{iq_0}{2v_F\Delta }\right) \nonumber \\
&\quad +r \delta_{\mu \nu}\delta_{ij} \left(\frac{iq_0}{2v_F\Delta}\right)^2 + \delta_{ij}\delta_{\mu\nu} \left(\frac{\bar{r}-r}{4}\right)\left(\frac{|\qq|}{2\Delta}\right)^2 + \ldots.
\end{align}
In real-space these lead to the following contributions to the effective action which are entirely \emph{local}
\begin{align}
&\delta \mathcal{L}^{(2)}_\text{eff} =-\tfrac{r}{2}\Gamma_{\mu i}\Gamma_{\mu i}
+\tfrac{1}{2} \left(\tfrac{r}{2v_F\Delta}\right) i\epsilon_{\mu\nu} \Gamma_{\mu i}\partial_\tau \Gamma_{\nu i} \nonumber \\
&+ \tfrac{1}{2}\left[
\tfrac{r}{(2v_F\Delta)^2}
\right](\partial_\tau \Gamma_{\mu i})(\partial_\tau \Gamma_{\mu i}) + \tfrac{1}{2}\left[\tfrac{r-\bar{r}}{16 \Delta^2}\right] \nabla \Gamma_{\mu i} \nabla \Gamma_{\mu i}.
\end{align}
The second term on the RHS is the ``Berry phase" term, which is crucial in giving the theory its $z=2$ character. The third term is higher order temporal gradient term and the last term is a renormalization of the parameter $\kappa$ in Eq.\eqref{eqn:Szeroth}. For simplicity, we have absorbed this latter renormalization into a re-definition of $\kappa$ and we neglect the $q_0^2$ term to leading order. This contributes to Eq.\eqref{eqn:L_eff} with the couplings Eq.\eqref{eqn:couplings}. Note that the $2v_F\Delta >0$ gap is absolutely crucial in producing these local terms. This is so because at small $q$, the fermionic fluctuations within the fermionic loop are gapped intra-band in character.  

Now in the other extreme limit, we can set $\Delta=0$ and for simplicity take $\bar{r}=r$. We will have instead the following non-local expression\cite{oganesyan2001quantum}
\begin{align}
&I_{\mu i \nu j}^{(2)}(q)- I_{\mu i \nu j}^{(2)}(0) \nonumber \\
&= 2r \dtheta (\hat{\kk}^T \tau^i \hat{\kk})(\hat{\kk}^T \tau^j \hat{\kk})\left(\keriq\right)\delta_{\mu\nu} \nonumber \\
&= -r \tfrac{|q_0|}{\lql} \left\{
\delta_{ij} + \left(\tkertan\right)^4 
\begin{pmatrix}
\cos 4\theta_\qq & \sin 4\theta_\qq \\ \sin 4\theta_\qq & -\cos 4\theta_\qq
\end{pmatrix}_{ij} 
\right\}\delta_{\mu\nu} 
\end{align}
where $I^{(2)}_{\mu i \nu j}(0) = r \delta_{ij}\delta_{\mu\nu} $ and $\lql^2 = q_0^2 + v_F^2|\qq|^2$. The angular integrals are carried out using the harmonic expansion identity Eq.\eqref{eqn:harmonic_exp}). Diagonalization of this susceptibility kernel then leads to the usual transverse and longitudinal polarization modes.\cite{oganesyan2001quantum} Also note that 2D Lindhard function\cite{kwon1995theory} is contained in this function as as a factor since 
\begin{align}
\frac{|q_0|}{\lql} = \frac{|q_0|}{\sqrt{v_F^2 |\qq|^2 + q_0^2} } = \frac{|x|}{\sqrt{x^2 +1}} 
\end{align}
where $x = \frac{q_0}{v_F|\qq|}$.

\subsection{Quartic Interaction}

\begin{figure}
	\boxed{ \includegraphics[width=5.5cm]{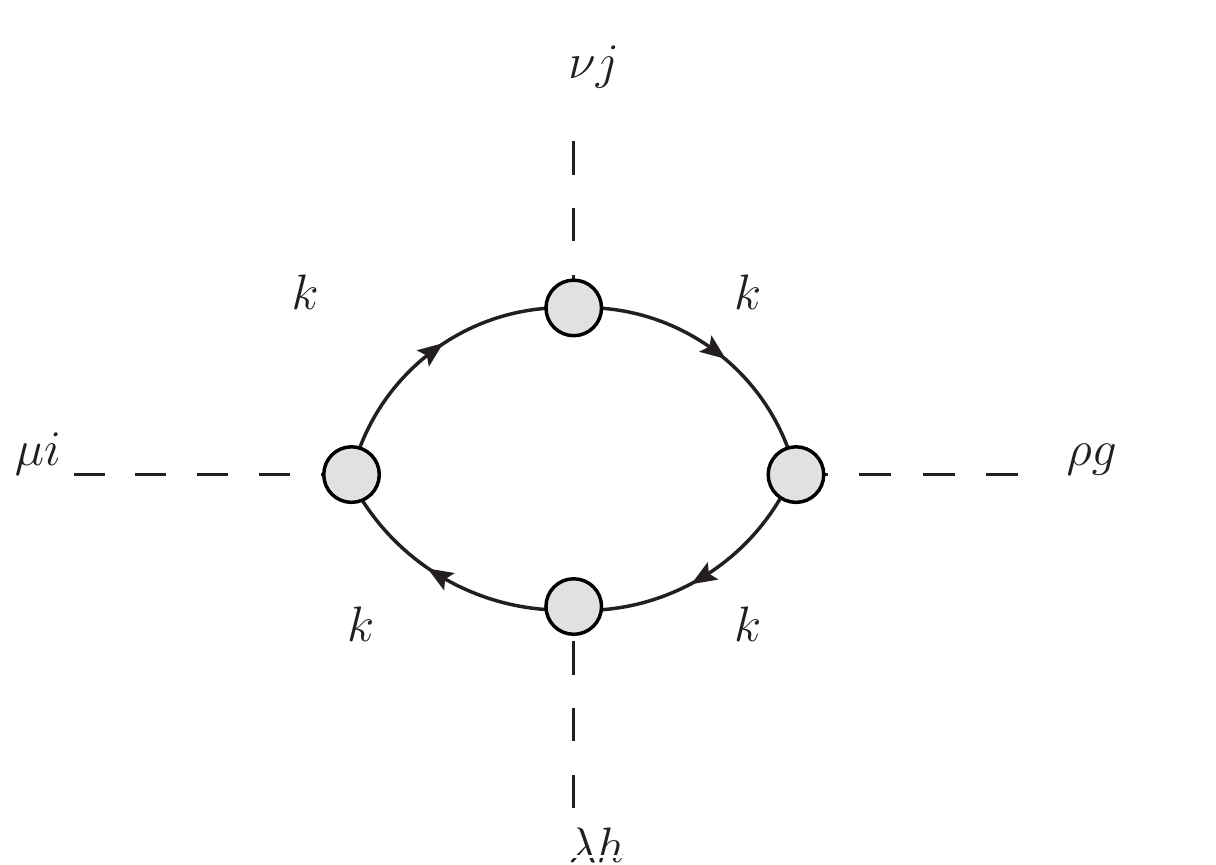}}
	\caption{The quartic interaction bubble $I^{(4)}_{\mu i, \nu j , \rho g , \lambda h}$ for the static-homogeneous mode $\Gamma(q=0)_{\mu i}$.}
	\label{fig:quartic_gamma}
\end{figure}

The next term to consider is the quartic interaction of $\Gamma$ shown in Fig.\ref{fig:quartic_gamma}, specialized to the $q=0$ limit. By the standard application of the Feynman rules and linearizing $\kk \approx k_F \hat{k\kk}$ gives
\begin{align}
I^{(4)}_{\mu i, \nu j , \sigma g , \lambda h} =&  
-\idk \sk\sum_{m,n,l,o} \sigma^{\mu}_{mn}\sigma^\lambda_{nl}\sigma^\rho_{lo}\sigma^\nu_{om} \nonumber \\
&\times e_i(\theta_\kk) e_j(\theta_\kk) e_g(\theta_\kk) e_h(\theta_\kk) \nonumber \\ 
&\times G_m(ik_0,\kk)G_n(ik_0,\kk)G_l(ik_0,\kk)G_o(ik_0,\kk).
\end{align}
where $e_i(\theta_\kk) := \hat{\kk}^T \tau^i \hat{\kk}$. To proceed, we simplify the product of Green's functions by partial fractions and carry out the Matsubara sum over $k_0$ by a contour integral. This last step involves double poles which produces factors of the derivative of the Fermi function $n'_F(\xi_\kk)$. This yields at the end, after a $\int \diff \xi N(\xi)$ integration
\begin{align}
&I^{(4)}_{\mu i, \nu j , \rho g , \lambda h} \nonumber \\
&=4\left(
\delta_{\mu \lambda} \delta_{\rho \nu} -\epsilon_{\mu \lambda} \epsilon_{\rho \nu}
\right) 
\left[
\tfrac{\bar{r}-r}{(2v_F\Delta)^2} 
\right] 
\int_{-\pi}^\pi \frac{\diff \theta_\kk}{2\pi} e_i e_j e_g e_h \nonumber \\
&= \half \left(
\delta_{\mu \lambda} \delta_{\rho \nu} -\epsilon_{\mu \lambda} \epsilon_{\rho \nu}
\right) \left(
\delta_{ij}\delta_{gh} + \delta_{ig}\delta_{jh} + \delta_{ih}\delta_{jg}
\right) \left[
\tfrac{\bar{r}-r}{(2v_F\Delta)^2} 
\right] 
\end{align}
where the we have used the identities Eq.\eqref{eqn:sigma1221}) and the relation
\begin{align}
&\int_{-\pi}^\pi \frac{\diff \theta_\kk}{2\pi} e_i(\theta_\kk) e_j(\theta_\kk) e_g(\theta_\kk) e_h(\theta_\kk) \nonumber \\
&= \tfrac{1}{8} \left( 
\delta_{ij}\delta_{gh} + \delta_{ig}\delta_{jh} + \delta_{ih}\delta_{jg}
\right).
\end{align}
Thus we need non-trivial band curvature $r \neq \bar{r}$ or varying DOS for this coefficient (coupling constant) to be non-zero. This diagram then leads to the following contribution to the effective Lagrangian
\begin{align*}
\delta\mathcal{L}_\text{eff}^{(4)} &= - \frac{1}{4} I_{\mu i ,\nu j, \rho g, \lambda h} \Gamma_{\mu i} \Gamma_{\nu j} \Gamma_{\rho g} \Gamma_{\lambda h} \\
& = \frac{1}{8} \left[\frac{r-\bar{r}}{(2v_F\Delta)^2}\right] \left[
\left(\Gamma_{\mu i}\Gamma_{\mu i} \right)^2  + 2 \left(\Gamma_{\mu i}\Gamma_{\mu j}\Gamma_{\nu j}\Gamma_{\nu i}\right)
\right]
\end{align*}
which is manifestly $O(2)$$_\text{iso}\times$$O(2)$$_\text{rot}$ symmetric. We further simplify this by defining the parameter
\begin{align}
\lambda = \frac{1}{2} \left[\frac{r-\bar{r}}{(2v_F\Delta)^2}\right] \approx - \frac{N''(0)}{24}
\end{align}
which gives after some manipulation 
\begin{align}
\delta\mathcal{L}_\text{eff}^{(4)} = \frac{3\lambda}{4} (\Gamma_{\mu i}\Gamma_{\mu i})^2 - \frac{\lambda}{4} \left( \epsilon_{\mu \nu}\epsilon_{ij} \Gamma_{\mu i}\Gamma_{\nu j}\right)^2
\label{eqn:L4}
\end{align}
where we have used  $$(\Gamma_{\mu 1}\Gamma_{\mu 2})^2 = (\Gamma_{\mu 1}\Gamma_{\mu 1}) (\Gamma_{\nu 2}\Gamma_{\nu 2}) - (\epsilon_{\mu \nu}\Gamma_{\mu 1}\Gamma_{\nu 2})^2$$ in the previous expression for $\delta\mathcal{L}_\text{eff}^{(4)}$. 

As it stands the the first term on the RHS of Eq.\eqref{eqn:L4}) is parametrically larger than the second which would suggest that only one type of symmetry broken-phase, the $\beta$-phase will be preferred. Also for stability reasons of the free-energy, it appears that we will always require that $\lambda>0$ or equivalently $N''(0)<0$. However by working in a \emph{fixed density} scheme as opposed to a fix chemical potential, we can supplement $\delta \mathcal{L}^{(4)}_\text{eff}$ with a counter-term at the quartic order in $\Gamma$ which will lead to a richer phase diagram.   
 
\subsubsection{Fixed Density}

\begin{figure}
	\boxed{\includegraphics[width=0.3\textwidth]{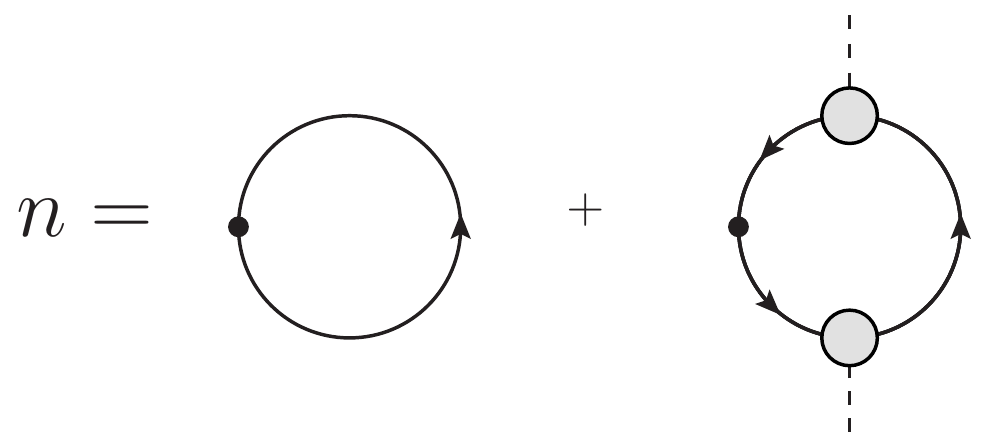}}
	\caption{The density $n$ with its perturbative correction $\delta n$ due to $\Gamma$ at order $O(\Gamma^2)$. }
	\label{fig:density}
\end{figure}

Following Refs.[\onlinecite{wu2007fermi},\onlinecite{sun2008time}], we next determine the fixed density corrections to the quartic interaction in $\Gamma$. The field $\Gamma_{\mu i}(q=0) \neq 0$ will lead to distortions of the Fermi-surfaces, which in turn leads to changes in the total density. Now the density $n$ with $O(\Gamma^2)$ corrections is shown in Fig.\ref{fig:density} which evaluates to
\begin{align}
n &= n_0 + (2v_F\Delta) r + \frac{N'(0)}{2}(\Gamma_{\mu i}\Gamma_{\mu i}) 
\end{align}
where $n_0 = 2 \int_{-\infty}^{0}\diff \xi \; N(\xi) $ is the reference density when there is zero splitting $\Delta=0$. Then we parametrize 
\begin{align}
n_0(\delta \mu)= 2 \int_{-\infty}^{\delta \mu} \diff \xi\; N(\xi) \approx n_0 + 2N(0)\delta \mu
\end{align} 
as a change in the ``bare" density due to a small change in chemical potential $\delta \mu$. Demanding that $n$ remain constant at its $\Gamma=0$ value, we have to leading order the constraint equation
\begin{align}
&n_0(\delta \mu) + (2v_F\Delta) r (\delta \mu) + \frac{N'(+\delta \mu)}{2} (\Gamma_{\mu i}\Gamma_{\mu i}) \nonumber \\
&= n_0 + (2v_F\Delta)r 
\end{align}
with $r(\delta \mu)$ being the new measure of average DOS after a change in chemical potential. It has the form 
\begin{align}
r(\delta \mu ) = \tfrac{1}{2v_F\Delta} \int_{-v_F\Delta}^{v_F\Delta} \diff \xi \; N(\xi-\delta \mu) \approx r - N'(0)\delta \mu.  
\end{align}
To leading order in smallness in $\Gamma_{\mu i}\Gamma_{\mu i}$ we can ignore the variation in $N'(+\delta \mu)$. Then we have to $O(\delta \mu)$ from solving the constraint equation
\begin{align}
\delta \mu = -\frac{N'(0)}{2\left[(2v_F\Delta)N'(0)+2N(0)\right]} \Gamma_{\mu i} \Gamma_{\mu i}
\end{align}
which leads to a correction in $r$ as 
\begin{align}
\delta r = - \frac{1}{2}\left[\frac{N'(0)^2}{2N(0)-(2v_F\Delta)N'(0)}\right](\Gamma_{\mu i}\Gamma_{\mu i})
\end{align}
Hence, the effective action acquires an additional ``counter-term" arising from the quadratic term $-(r/2)\Gamma_{\mu i}\Gamma_{\mu i}$ as
\begin{align}
-\frac{\delta r(\Gamma)}{2}(\Gamma_{\mu i}\Gamma_{\mu i}) 
&= \frac{1}{4} \left[\frac{N'(0)^2}{2N(0)-(2v_F\Delta)N'(0)}\right](\Gamma_{\mu i}\Gamma_{\mu i})^2 \nonumber \\
&\approx \frac{1}{8} \left(\frac{N'(0)^2}{N(0)}\right)(\Gamma_{\mu i}\Gamma_{\mu i})^2
\end{align}
in the approximation that $\left|\frac{N(0)}{N'(0)}\right|\gg 2v_F\Delta$; that is in the large DOS limit. Also note that this correction is regular in the $\Delta =0$ limit. 

Finally, including the constant density correction with the previously calculated direct quartic interaction yields the following total quartic contribution to the effective action 
\begin{align}
\delta \mathcal{L}_\text{eff}^{(4)} =\left( \frac{\alpha'}{4} + \frac{3\lambda}{4}\right) (\Gamma_{\mu i}\Gamma_{\mu i})^2 - \frac{\lambda}{4} (\epsilon_{\mu\nu}\epsilon_{ij} \Gamma_{\mu i}\Gamma_{\nu j})^2
\end{align}
where $\alpha' = \frac{N'(0)^2}{2N(0)}$ is newly defined parameter. This leads then the expressions quoted in Eq.\eqref{eqn:couplings}).

\section{Harmonic expansion of the Fermi-liquid density-density propagator}

Calculations will often involve angular $\int \diff \theta_k$ integrals with the density-density propagator $(v_F \delta q - iq_0)^{-1}$, with $\delta q= \hat{\kk}\cdot \qq = |\qq| \cos \theta_{kq}$. We present here a closed form harmonic expansion for this function, which can then be used to evaluate the desired angular integrals. Defining the following geometrical quantities
\begin{align}
&\| q \| = \sqrt{q_0^2 + v_F^2 |\qq|}, \quad 
 x = \frac{q_0}{\|q\|} , \quad y = \frac{v_F |\qq|}{\|q\|} \nonumber \\ 
&\tan \alpha = \frac{y}{|x|}, \quad \tan(\tfrac{\alpha}{2}) = \frac{y}{1+|x|} = \frac{v_F |\qq|}{\|q\|+ |q_0|}
\end{align}
we have the expansion
\begin{widetext}
\begin{align}
&\frac{1}{v_F \delta q - iq_0} = \frac{i\, \text{sign}(q_0)}{\|q\|}
\left(
1 + 2 \sum_{n=1}^\infty (-1)^n \e^{i\frac{n\pi}{2} \text{sign}(q_0)} \tan^n\left(\tfrac{\alpha}{2}\right) \cos(n\theta_{kq})
\right) 
\label{eqn:harmonic_exp}
\end{align}
\end{widetext}
This expansion can be derived by using the Jacobi-Anger identity\cite{abramowitz1966handbook} to expand the RHS of 
\begin{align}
\frac{1}{v_F\delta q -iq_0 } &= \left(\frac{1}{\| q \|}\right) \frac{1}{y \cos \theta_{kq} -i x} \nonumber \\
&= \frac{i \text{sign}(q_0)}{\|q\|} \int_{0}^{\infty} \diff s \; 
\e^{-|x|s} \e^{-i \text{sign}(x)ys\cos \theta_{kq} } 
\end{align}
in terms of Bessel functions $J_m(z)$. It turns out the integrals can be evaluated by Laplace transforms which gives the stated expression after some simplification using $x^2+y^2=1$.

We note that more general expansions for the function $[v_F(\hat{\kk})\hat{\kk}\cdot \qq -iq_0]^{-1}$ than Eq.\eqref{eqn:harmonic_exp}) can be derived for anisotropic Fermi velocities $v_F(\hat{\kk})$. Although tedious, this can be achieved by through a generalized Jacobi-Anger expansion of $\e^{i f(\theta_{k}) \cos \theta_{kq}}$ using multivariate Bessel functions\cite{dattoli1991note} and their associated Laplace transforms. Extensions to three dimensions are also relatively straightforward.  

\section{Nambu-Goldstone action for a broken \texorpdfstring{$U(N)$}{} symmetry}\label{app:NG_UN}

For the curious reader, we sketch here the derivation of the Nambu-Goldstone action in the case of $U(N)$ symmetry of a non-relativistic $N$-component boson. This may be regarded as the symmetry broken action of Eq.\eqref{eqn:Leffphi}) in the $\lambda=0$ limit for an enlarged symmetry group such that $\phi({\bf x},\tau) \in \mathbb{C}^N$. When $\rho<0$ with $\alpha>0$, the $U(N)$ symmetry is broken and the vacuum is set by the equilibrium density $n_0$ 
\begin{align}
\langle\phi({\bf x},\tau)\rangle &= \phi_0 \in \mathbb{C}^N \\
n_0 &= \phi_0^\dagger \phi_0 =\frac{|\rho|}{\alpha}.
\end{align}

Now the field $\phi$ as a representation of the $U(N)$ group is \emph{not simply transitive}. Hence we need to quotient out the stabilizer of $\phi_0$ which brings the symmetry group down to $SU(N)$. This stabilizer is expressed in terms of the central $U(1)$ subgroup in $U(N)$ (which is given by the determinant map) and the Cartan subgroup generated by the following $su(N)$ generator
\begin{align}
\hat{\Sigma} = \frac{2 \phi_0 \phi_0^\dagger }{(\phi_0^\dagger \phi_0)} - \mathbbm{1}.
\end{align}
The stabilizer of $\phi_0$ is then simply given by 
\begin{align}
H_{\phi_0} \equiv \{ \exp(i \theta[\mathbbm{1}-\hat{\Sigma}]) : \theta \in \mathbb{R} \} \subset {U}(N).
\end{align}
Hence the order parameter target space is the following coset space
\begin{align}
{U}(N)\slash H_{\phi_0} \cong {SU}(N).
\end{align}
Now the group quotient operation removes the generator ($\mathbbm{1}-\hat{\Sigma}$) by enforcing $\mathbbm{1}\equiv\hat{\Sigma}$ at the level of the Lie algebra. Hence as a matter of convenience, we will continue to use the fundamental representation of $SU(N)$ acting on $\phi$, but with the understanding that the Cartan subgroup generated by $\hat{\Sigma}$ is to be identified with the central U(1) of $U(N)$$\cong {U}(1) \times$ $SU(N)$. 

Next we parametrize the first order fluctuations of $\phi$ about $\phi_0$ by the following
\begin{align}
\phi &= \sqrt{\tfrac{n_0 + \delta n}{n_0}} \e^{i \pi_A T^A} \phi_0 
= \phi_0 + \delta \phi \\
\delta \phi &= i \pi_A T^A \phi_0 + \left(\tfrac{\delta n}{ 2 n_0}\right)\phi_0
\end{align}
where the first term in $\delta \phi$ corresponds to the SU(N) fluctuation given as a Lie algebra valued field $\pi_A T^A$ and the second term to the amplitude fluctuation expressed as a density variation $\delta n$. The matrices $\{T^A\}$ are a traceless Hermitian basis of $su(N)$ with the structure constants $f^{AB}{}_C$ given by 
\begin{align}
[T^A,T^B] = i f^{AB}{}_C T^C. 
\end{align} 
To the level of Gaussian fluctuations in the order parameter, we have the following expanded Lagrangian
\begin{align}
\mathcal{L}_\text{NG}^{(0)} & =i(\phi_0^\dagger T^A \phi_0 ) \left(\tfrac{\delta n}{n_0}\right)\partial_\tau \pi_A 
+ \tfrac{i}{2}f^{AB}{}_C (\phi_0^\dagger T^C \phi_0) \pi_A \partial_\tau \pi_B \nonumber \\
&+(\phi_0^\dagger T^{(A}T^{B)}\phi_0) \nabla \pi_A \cdot \nabla \pi_B \nonumber \\
& +\left(\tfrac{1}{4n_0}\right) \nabla(\delta n) \cdot \nabla(\delta n) 
+ \left(\tfrac{\alpha}{2}\right) \delta n^2
\label{eqn:Lgold0}\end{align} 
where boundary terms have been tacitly omitted. We next proceed to integrate out the massive $\delta n$ amplitude modes and note that the functional measures $\mathscr{D}\phi^\dagger \mathscr{D}\phi$ and $\mathscr{D}\pi_A  \mathscr{D}(\delta n)$ agree up to irrelevant constants. We then have the following final lowest order (in gradients) effective free Lagrangian for the Nambu-Goldstone fields
\begin{align}
\mathcal{L}_\text{NG}^{(0)} &= 
G^{AB} \partial_\tau \pi_A \partial_\tau \pi_B  + i \Omega^{AB} \pi_A \partial_\tau \pi_B \nonumber \\
&+ K^{AB} \nabla \pi_A \cdot \nabla \pi_B 
\end{align}
where the isospin tensors are given by the following order parameter expectations
\begin{align}
G^{AB} &= \tfrac{(\phi^\dagger_0 T^A \phi_0)(\phi^\dagger_0 T^B \phi_0)}{2|\rho|} = G^{BA}\\
K^{AB} &= (\phi^\dagger_0 T^{(A} T^{B)} \phi_0) = K^{BA}\\ 
\Omega^{AB}&=\tfrac{1}{2}f^{AB}{}_C(\phi_0^\dagger T^C \phi_0) = - \Omega^{BA}.
\end{align}
In the $N=1$ case which is the single component non-relativistic superfluid, $G$ and $K$ are scalars and $\Omega=0$. However, in the general SU($N$) symmetric non-relativistic boson case, the Goldstone Lagrangian $\mathcal{L}_\text{NG}^{(0)}$ seems to include the additional `Berry' phase term $i \Omega^{AB} \pi_A \partial_\tau \pi_B$. The fact that $\Omega\neq 0$ is the reason for the modified counting rule for the Nambu-Goldstone modes.\cite{nielsen1976count,watanabe2012unified} 

\section{Gradient Expansions}\label{app:gradient}
In this appendix we describe the gradient expansion in $q$ which operates by isolating the singularity at $q\rightarrow 0$. This is necessary because neither a Taylor nor Laurent expansion in $q$ is possible with the response kernels of a gapless metallic model. In many ways the gradient expansion resembles that of an Operator Product Expansion\cite{wilson1969non-lagrangian,kadanoff-1969,Polyakov-1970} of coincident poles in complex frequency space as $q_0 \rightarrow 0$. We first describe the simplest example of the response bubble encountered in the density-density response.

\subsection{Density-density example}

\begin{figure}
	\boxed{\includegraphics[width=6cm]{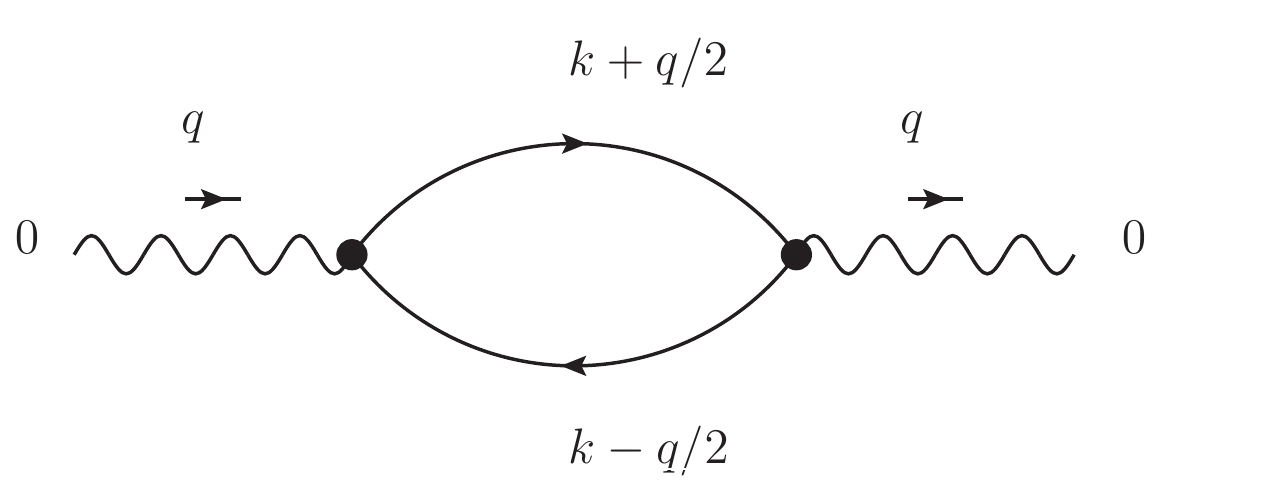}}
	\caption{The conventional density-density response bubble}
	\label{fig:density2}
\end{figure}

In calculating the bare or conventional (gapless) density-density response (shown in Fig.\ref{fig:density2}) we encounter the following expression obtained from linearizing $\xi_{\kk\pm \qq/2} = \xi_\kk \pm v_F \delta q /2$ and  $n_F(\xi_{\kk\pm \qq/2})$,
\begin{align}
&\int \frac{\diff^2 k}{(2\pi)^2} \; \frac{1}{\beta} \sum_{k_0} G(ik_0^+, \kk^+ )G(ik_0^-,\kk^-) \nl
&= \int \frac{\diff^2 k}{(2\pi)^2} \left(
\frac{v_F \delta q}{v_F \delta q - i q_0} 
\right) n'_F(\xi_\kk) \nonumber  \\
&= \int \frac{\diff^2 k}{(2\pi)^2} \left(
\frac{v_F \delta q}{v_F \delta q - i q_0} 
\right) \frac{1}{\beta} \sum_{k_0} G(ik_0,\kk)^2 
\end{align}
where $\delta q = \qq \cdot \hat{\kk}$ and in the last line we have re-substituted the Matsubara sum by using the relation between Matsubara summed powers of the (bare) Green's function and derivatives of $n_F$
\begin{align}
\frac{1}{\beta} \sum_{k_0} G(ik_0,\kk)^m = \frac{1}{(m-1)!}\,  n^{(m-1)}_F(\xi_\kk), \quad m>0.
\end{align} 
Also we let $G (z,\kk) = (z-\xi_\kk)^{-1}$ be the single particle Green's function. This suggest the following identity 
\begin{align}
G(ik_0^+,\kk^+) G(ik_0^-,\kk^-)  \equiv  
\left(
\frac{v_F \delta q}{v_F \delta q - i q_0} 
\right) G(ik_0,\kk)^2
\label{eqn:GpGm}
\end{align}
that is valid only inside the Matsubara frequency sum, and in the limit of $q \rightarrow 0$. Thus in this sense it resembles an OPE in complex frequency space, where in the limit $q\rightarrow 0$, the complex energy poles of the $G$'s coincide to form a double pole. Moreover in performing the Matsubara sum, a contour integral is invoked which furthers the analogy with OPE's encountered in conformal field theories. Amusingly it is the opposite of a short distance expansion, since small $q$ corresponds to long wavelengths and low frequencies. 

In deriving gradient expansion terms of the many response kernels, we will need more replacement identities between different combinations and powers of Green's functions under a Matsubara frequency sum. These can all be derived in much the same way as the above argument suggest. 
As an example, consider a meromorphic function $F(z)$ which is never singular at $z=k_0$ and the following Matsubara frequency summation using the contour integral
\begin{align}
&\frac{1}{\beta} \sum_{k_0} G(ik_0^+,\kk^+) G(ik_0^-,\kk^-) F(ik_0) \nonumber \\
&= \frac{1}{v_F \delta q - iq_0 } \oint_{-\mathcal{C}} \tfrac{\diff z}{i2\pi} \left(
\frac{1}{z+i\tfrac{q_0}{2}-\xi_{\kk^+}} -\tfrac{1}{z-i\tfrac{q_0}{2}-\xi_{\kk^-} } 
\right) \nonumber \\ &\hspace{3.5cm}\times F(z) n_F(z) 
\end{align}
where $\mathcal{C}$ is the contour which surrounds (counterclockwise) the poles of $n_F(z)$ located at the odd Matsubara frequencies $z=i k_0 = (2n+1)\pi/ k_B T$. Let $(-\mathcal{C})'$ be the contour $(-\mathcal{C})$ but which excludes a neighbourhood which contains $\xi_{\kk^\pm} \mp i\tfrac{q_0}{2}$. 
Then from linearizing by $(q_0,\qq)$ 
\begin{widetext}
\begin{align*}
&\frac{1}{\beta} \sum_{k_0} G(ik_0^+,\kk^+) G(ik_0^-,\kk^-) F(ik_0) \\
&=\frac{1}{v_F \delta q - i q_0}  
\left\{
F(\xi_{\kk^+} -i\tfrac{q_0}{2}) n_F(\xi_{\kk^+}) 
-F(\xi_{\kk^-}+i \tfrac{q_0}{2}) n_F(\xi_{\kk^-})
\right\}  \\
&+ \frac{1}{v_F\delta q -i q_0} \oint_{-\mathcal{C}'} \frac{\diff z}{i2\pi } \left[
G(z+i\tfrac{q_0}{2} , \kk^+) - G(z-i\tfrac{q_0}{2},\kk^-) 
\right] F(z) n_F(z) \\
& = F'(\xi_\kk) n_F(\xi_\kk) + \left(\frac{v_F \delta q }{v_F \delta q - i q_0}\right) F(\xi_\kk) n_F'(\xi_\kk) 
+ \left(\frac{1}{v_F \delta q - i q_0 }\right)
\oint_{-\mathcal{C'}} \frac{\diff z}{i2\pi} \left[
G(z+i\tfrac{q_0}{2},\kk^+) - G(z-i\tfrac{q_0}{2},\kk^-) 
\right] F(z) n_F(z) \\ 
& = F'(\xi_\kk) n_F(\xi_\kk) + \left(\frac{v_F \delta q }{v_F \delta q - i q_0}\right) F(\xi_\kk) n_F'(\xi_\kk) 
+ \left(\frac{1}{v_F \delta q - i q_0 }\right) \oint_{-\mathcal{C}'}\frac{\diff z}{i2\pi}
(v_F \delta q - i q_0 ) G(z,\kk)^2 F(z) n_F(z) \\ 
&= F'(\xi_\kk) n_F(\xi_\kk) + F(\xi_\kk) n_F'(\xi_\kk) + \left( \frac{v_F \delta q}{v_F \delta q -iq_0} -1 \right) F(\xi_\kk) n_F'(\xi_\kk) 
+ \oint_{-\mathcal{C'}} \frac{\diff z}{i2\pi} G(z,\kk)^2 F(z) n_F(z)   \\ 
&= \left(\frac{iq_0}{v_F \delta q -i q_0}\right) F(\xi_\kk) n_F'(\xi_\kk) 
+ \oint_{-\mathcal{C}} \frac{\diff z}{i2\pi} G(z,\kk)^2 F(z) n_F(z) \\ \end{align*}
\end{widetext}
\begin{align*}
&= \oint_{-\mathcal{C}} \frac{\diff z}{i2\pi} \left[
\frac{iq_0 F(\xi_\kk)}{v_F \delta q - i q_0} G(z,\kk)^2 +G(z,\kk)^2 F(z) 
\right]n_F(z)
\end{align*}
where in the second equality we used the analyticity of $F$ in the neighborhood $z=\xi_{\kk^{\pm}}$ and Taylor series expanded in $q$. Likewise in the third equality we used the analyticity of the $G$'s under the integral in the same neighborhood of $z=\xi_{\kk^{\pm}}$ to series expand to lowest order in $q$. For the penultimate equality we can consolidate the first two terms of the previous line into the whole contour integral $-\mathcal{C}$. Since this expansion holds within the linearized approximations, we can infer the OPE-like long distance expansion
\begin{align}
&G(ik_0^+,\kk^+) G(ik_0^-,\kk^-) F(ik_0) \nonumber \\
&= \left(\frac{iq_0 }{v_F \delta q - iq_0}\right) F(\xi_\kk) G(ik_0, \kk)^2 
+G(ik_0,\kk)^2 F(ik_0) 
\end{align} 
where the two sides are meant to agree only when, \\
(i) inside the Matsubara frequency summation of $i k_0$, \\
(ii) in the linearized approximation of the dispersion $\xi_\kk^\pm$ and $n_F(\xi_{\kk^\pm})$, \\
(iii) to lowest most singular order in the small expansion of $q$ and (iv) $F(z)$ is analytic in the neighborhood of $z= \xi_{\kk}$. Note also both sides agree in the $q=0$ limit only when the order of limits is taken such that $q_0 \rightarrow 0$ first then $|\qq| \rightarrow 0$. 

Finally, we can derive more intricate gradient expansions involving multiple Green's functions which are valid only inside Matsubara frequency sums using the same methods as above. In a way, this is just a systematic way to perform the contour integral around each individual pole separately. The standard procedure is as follows:
\begin{itemize}
	\item[0)]{ Convert the Matsubara sum into a contour integral in the conventional way.}
	\item[1)]{ Deform and isolate the contour around the desired singularity eg. $z=\xi_{n \kk^{\pm}}$.}
	\item[2)]{Perform the contour and collect the residues just around that pole.}
	\item[3)]{Using the assumed analyticity properties of the remaining factors of the integrand, Taylor expand about $q=0$.}
	\item[4)]{Re-organize terms such that the resulting terms may be re-incorporated into an integral using the original contour.}
	\item[5)]{Identify terms inside to contour integral to derive the desired relation.}
\end{itemize}
These expansions are then used to derive the explicit forms of the response kernels in Appendix \ref{app:all_responses}. Note also that after the expansion, the terms are factored into angular $\theta_\kk$ and isotropic factors, making it easier to carry out the $\int \diff^2 k$ integrations.  


\section{Damping of the Nambu-Goldstone modes}\label{app:NBG_damping}

\begin{figure}
\includegraphics[width=0.49\textwidth]{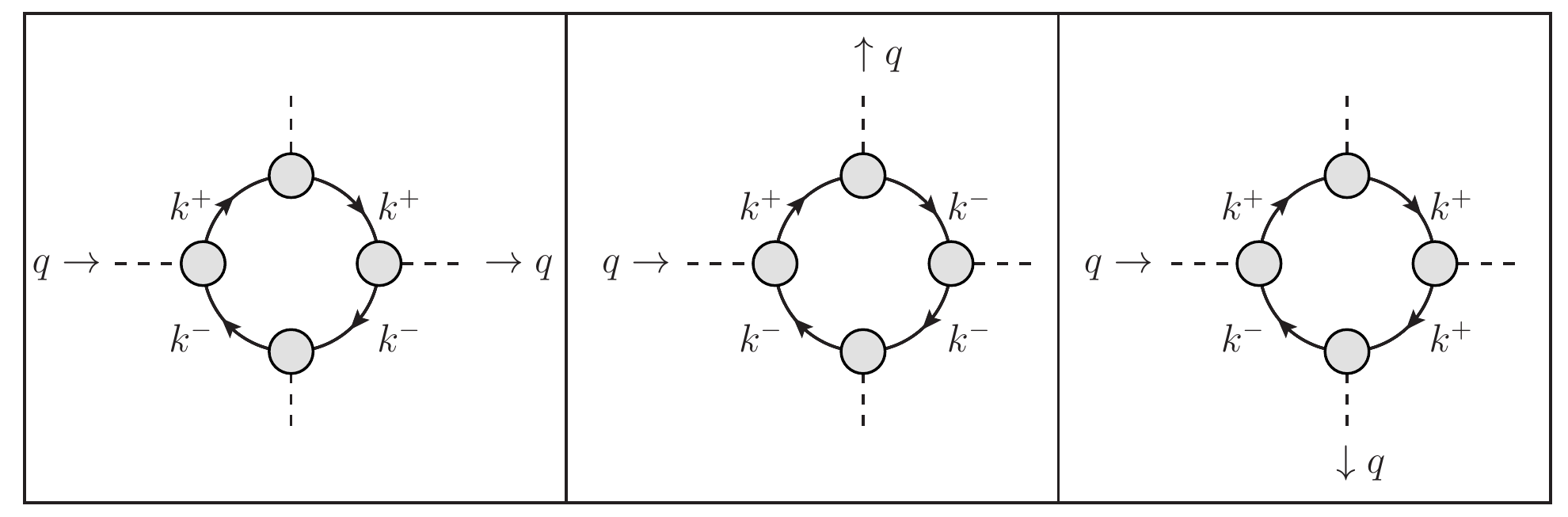}
\caption{The simplest quartic interaction terms at finite momentum $q$ that leads to damping from inter-band scattering in the broken symmetry mean-field phases. Here $k^{\pm} = k\pm q/2$ where $k$ is the internal loop momentum. The diagrammatic notation follows that of Fig.\ref{fig:FeynmanRules} in Appendix \ref{app:derive_Seff}.}
\label{fig:quartic_gradient}
\end{figure}

In this appendix, we will derive an approximate form of this damping which appears in the RPA corrected effective action using the mean-field Green's functions in either broken symmetry phase. The mean-field Green's function can be conveniently parametrized as
\begin{align}
G(ik_0,\kk) 
&= [ik_0 - \xi_\kk - v_F\Delta \sigma^z - k_F^{-2} \langle\Gamma_{\mu i}\rangle (\kk^T\tau^i \kk) \sigma^\mu]^{-1} \nonumber \\
&= [ik_0 - \xi_\kk - \Delta_\kk \, a_\alpha(\kk)\sigma^\alpha]^{-1}
\end{align}
where the indices range over $\mu=x,y$ but $\alpha=x,y,z$, and the uniform mean-field is given as $\langle \Gamma_{\mu i}\rangle\neq 0$. The quantity $\Delta_\kk >0$ is the mean-field corrected gap between the Fermi surfaces and is given as
\begin{align}
\Delta_\kk^2 = (v_F\Delta)^2 + k_F^{-4}\langle \Gamma_{\mu i}\rangle \langle \Gamma_{\mu j}\rangle (\kk^T\tau^i \kk) (\kk^T\tau^j \kk) 
\end{align}
such that $\vec{a}(\kk)\cdot \vec{a}(\kk) = 1$. The mean-field energy bands are in turn given as $E_{(1,2)\kk} = \xi_\kk \pm \Delta_\kk$. We further express $G(ik_0,\kk)$ in terms of its spectral projectors as
\begin{align}
G(ik_0,\kk) 
&=  G_1(ik_0,\kk)\,P_1(\kk)  + G_2(ik_0,\kk)\,P_2(\kk)
\end{align}
where $G_{n}(ik_0,\kk) = ik_0 -E_{n}(\kk)$ and 
\begin{align}
P_1(\kk) &= \tfrac{1}{2} \left( \mathbbm{1} + a_\alpha(\kk)\sigma^\alpha \right) \nonumber \\
P_2(\kk) &= \tfrac{1}{2} \left( \mathbbm{1} - a_\alpha(\kk)\sigma^\alpha \right).
\end{align}
Next, allowing for fluctuations in $\Gamma_{\mu i}$ leads to coupling to the mean-field Hamiltonian through the following interaction term
\begin{align}
\int\left[k_F^{-2} (\kk^T\tau^i \kk) \psi^\dagger_{\kk^{-}}\sigma^\mu\psi_{\kk^{+}} \right] \delta\Gamma_{\mu i}(-q)
\end{align}
where $\delta \Gamma$ is the fluctuation and the integral is understood to mean to be over $\kk$ and $(q_0,\qq)$. Here $\kk$ is understood to be the average momentum between $\psi$ and $\psi^\dagger$. Hence, the computation of the susceptibility bubble in the mean-field phase proceeds in much the same way as the computation of the Feynman diagram of Fig.\ref{fig:Gamma2} in Appendix \ref{app:derive_Seff}. The key difference now being that the fermion-boson vertex coupling is through the fluctuation $\delta \Gamma$ and the fermion propagators have now acquired uniform mean-field self-energy corrections in $G(ik_0,\kk)$ as given above. We quote below the final result for the intra-band contribution to the susceptibility kernel 
$J^{(2)}(q)_{\mu i \nu j} =  \langle \delta\Gamma_{\nu j}(-iq_0,-\qq)\; \delta\Gamma_{\mu i}(iq_0,\qq)\rangle_\text{MF}$, 
\begin{widetext}
\begin{align}
(J_\text{intra}^{(2)})(q)_{\mu i \nu j} 
&=\frac{1}{k_F^4}
\idk{} \sk [(\kk^+)^T\tau^i (\kk^+)] [(\kk^-)^T\tau^j (\kk^-)] 
\sum_{m=1}^2 \left(\frac{\qq \cdot \nabla E_{m\kk}}{\qq \cdot \nabla E_{m\kk} - iq_0 }\right) 
\text{Tr}\left[ \sigma^\mu P_m(\kk^+) \sigma^\nu P_m(\kk^-) \right]
\delta(E_{m\kk}).
\end{align}\\
\end{widetext}
To proceed we make the following simplifying approximations
\begin{align*}
&(\kk^\pm)^T\tau^i (\kk^\pm) \approx \kk^T\tau^i \kk\\ 
&\qq \cdot \nabla E_{m\kk} \approx v_F \qq \cdot \hat{\kk} = v_F \delta q \\
&\text{Tr}\left[ \sigma^\mu P_m(\kk^+) \sigma^\nu P_m(\kk^-) \right] \approx 
\text{Tr}\left[ \sigma^\mu P_m(\kk) \sigma^\nu P_m(\kk) \right]
\end{align*}
which amounts to moving all the complexities of non-zero $q$ to the particle-hole excitation kernel in the large curly parentheses. The effects of the broken symmetry manifests as a non-zero trace which is normally zero in the normal phase. This expression then simplifies to 
\begin{align}
&(J_\text{intra}^{(2)})(q)_{\mu i \nu j} 
\approx  
\frac{1}{k_F^4}\idk{} \sk [\kk^T\tau^i \kk] [\kk^T\tau^j \kk] \nonumber \\ 
&\hspace{1.6cm}\times\left(1+ \frac{iq_0 }{v_F \delta q - iq_0 }\right)
a_\mu(\kk) a_\nu(\kk) \sum_{m=1}^2 \delta(E_{m\kk}) 
\end{align} 
where for convenience we have separated the static and dynamic parts out, and used the expression relating $P_m(\kk)$ and $\vec{a}(\kk)$. Finally, we approximate $\kk \approx k_F \hat{\kk}$ and  the delta functions localized on the Fermi surfaces by $\delta(E_{(1,2)\kk})\approx \delta(\xi_\kk \pm  v_F\Delta)$. Carrying out the radial integration with the density of states function $N(\xi)$ and substitute $\Delta_\kk \approx v_F\Delta$ in the denominator gives the leading order term
\begin{align}
&(J_\text{intra}^{(2)})(q)_{\mu i \nu j} - (J_\text{intra}^{(2)})(0)_{\mu i \nu j} 
\approx \frac{2 \bar{r}\; \langle\Gamma_{\mu l}\rangle \langle \Gamma_{\nu p}\rangle}{(v_F \Delta)^2} \nonumber  \\
&\times\int \frac{\diff \theta_\kk}{2\pi}
e_i(\theta_\kk)e_j(\theta_\kk)e_l(\theta_\kk)e_p(\theta_\kk)  
\left(\frac{iq_0 }{v_F \delta q - iq_0 }\right)
\end{align}
with $\bar{r} = \tfrac{1}{2}(N(+v_F\Delta)+N(-v_F\Delta))$ and the harmonic functions $e_1(\theta_\kk) = \cos(2\theta_\kk)$ and $e_2(\theta_\kk) = \sin(2\theta_\kk)$. Now the static part is uninteresting and is a mass renormalization that according to Goldstone's theorem must vanish for the relevant broken symmetry. Thus we focus only on the dynamic contribution. The final integral can be performed exactly using the expansion Eq.\eqref{eqn:harmonic_exp}) which leads to $n=0,4,8$ angular harmonic contributions. The higher angular harmonics are needed to distinguish between longitudinal and transverse modes, but are sub-leading compared to the isotropic $n=0$ mode. Thus we make a final approximation by computing the leading order $n=0$ angular integral which gives
\begin{align}
&(J_\text{intra}^{(2)})(q)_{\mu i \nu j} - (J_\text{intra}^{(2)})(0)_{\mu i \nu j} \nonumber \\
&\approx-\frac{2 \bar{r}\;C_{ijlp}\langle\Gamma_{\mu l}\rangle \langle \Gamma_{\nu p}\rangle}{(v_F \Delta)^2}\left(\frac{|q_0|}{\sqrt{v_F^2|\qq|^2+q_0^2}}\right) 
\end{align}
where we have defined the tensor
\begin{align}
C_{ijlp} = \frac{1}{8} \left( \delta_{ij}\delta_{lp} + \delta_{il}\delta_{jp} + \delta_{ip}\delta_{jl}\right). 
\end{align}
Finally analytically continuing to real frequencies $iq_0 \rightarrow \omega + i0^+$ and in the branch where $s= \frac{|\omega|}{v_F |\qq|}<1$, yields to $O(s)$ 
\begin{align}
&(J_\text{intra}^{(2)})(\omega,\qq)_{\mu i \nu j} - (J_\text{intra}^{(2)})(0)_{\mu i \nu j} \nonumber \\
&\approx \left(\frac{-i\omega}{v_F |\qq|}\right) \left[ \frac{2\bar{r}}{(v_F \Delta)^2}\right]C_{ijlp}\langle\Gamma_{\mu l}\rangle\langle \Gamma_{\nu p}\rangle
\end{align}
which is an imaginary self-energy correction to the correlator $\langle \delta\Gamma_{\nu j}(-\omega,-\qq)\; \delta\Gamma_{\mu i}(\omega,\qq)\rangle$ and is the form quoted in the main text. Note that the inter-band fluctuations also contribute to this correlator and are already accounted for in the effective actions Eq.\eqref{eqn:L_alpha}) and Eq.\eqref{eqn:L_beta}) to lowest order. 

We remark that in principle this damping channel could have be deduced from the trace-log expansion of Eq.\eqref{eqn:Seffall}). The $O(\langle\Gamma\rangle^4)$ dependence of this term indicates that it has its origins in the gradient expansion of the quartic interaction terms. A straightforward estimation of the Feynman diagrams shown in Fig.\ref{fig:quartic_gradient} shows that this is the case. However these terms, as are irrelevant in the renormalization group sense in the normal phase (cf. Section \ref{sec:RG}). Due to the $z=2$ dynamical scaling, the factor $iq_0/|\qq|$ has scaling dimensions of a gradient and thus quartic and higher order couplings appearing together with factors of $iq_0/|\qq|$ are necessarily irrelevant.

\begin{widetext}

\section{Explicit Forms of the Linear EM Response Kernels}\label{app:all_responses}

For convenience we list here some definitions for parameters and notation used in expression the response kernels. 
\begin{align*}
&\xi_1(\kk) = \xi_\kk + v_F\Delta,& 
&\xi_2(\kk) = \xi_\kk - v_F\Delta,& 
&v_F = \left.\frac{\partial \xi_\kk}{\partial |\kk|}\right|_{\xi_\kk =0}&
\\%
& r = \frac{1}{2v_F \Delta} \int_{-v_F\Delta}^{+v_F\Delta} \; N(\xi) \diff\xi,& 
&\bar{r} = \frac{N(+v_F\Delta)+N(-v_F\Delta)}{2},& 
& s = \frac{N(+v_F\Delta)-N(-v_F\Delta)}{2 v_F\Delta}& 
\\%
& \lql = \sqrt{v_F^2|\qq|^2 + q_0^2}&
& G_{1,2} = \frac{1}{iq_0 - \xi_{1,2}(\kk)}& 
& v_F \delta q = v_F \qq \cdot \hat{\kk} \equiv v_F |\qq| \cos(\theta_\kk -\theta_\qq) &\end{align*}
and the quadrupole tensors $\tau^1 \equiv\sigma^z,\tau^2 \equiv \sigma^x$ such that $\hat{\kk}^T \tau^i\hat{\kk} = (\cos 2\theta_k, \sin 2\theta_k)_i$.

\subsection{Conventional Polarization Diagrams}\label{K0}

These diagrams do not couple $\Gamma_{\mu i}$ and represent the ``bare" electromagnetic response.
\Hline
\begin{align*}
K^{00}_0(q)=\feny{width=2.5cm}{diagrams/feynman/qq.pdf}
&=-e^2 \idk \sk \left(\kervq\right) (G_1^2 +G_2^2) \nonumber \\
&=2e^2 \bar{r} \left(1-\frac{|q_0|}{\lql}\right)
\end{align*}
\Hline

\begin{align*}
K^{0b}_0(q)=\feny{width=2.5cm}{diagrams/feynman/qj.pdf}
&= e^2 \idk \sk \partial_b \xi_\kk \left(\kervq\right) (G_1^2 +G_2^2) \\
&= -2i e^2 \bar{r} v_F \frac{|q_0|}{\lql} \left(\kertan\right)
\begin{pmatrix} \cos \theta_q \\ \sin \theta_q \end{pmatrix}_{b}
\end{align*}
\Hline

\begin{align*}
K^{ab}_0(q)=\left(
\feny{width=2.5cm}{diagrams/feynman/jj.pdf}-
\feny{width=1.8cm}{diagrams/feynman/tadpole_m.pdf}
\right)
&= -e^2 \idk \sk \partial_a \xi_\kk \partial_b \xi_\kk 
\left(\keriq\right) (G_1^2 +G_2^2) \\
&= e^2 \bar{r} v_F^2  \frac{|q_0|}{\lql} \left[
-\delta_{ab}  + \left(\kertan\right)^2 
\begin{pmatrix}
\cos 2\theta_q & \sin 2\theta_q \\
\sin 2\theta_q & -\cos 2\theta_q
\end{pmatrix}_{ab}
\right]
\end{align*}
\Hline 
\\ \\
where $(\cos \theta_q, \sin \theta_q)^T = \hat{\qq}$ and 
$\begin{pmatrix}
\cos 2\theta_q & \sin 2\theta_q \\
\sin 2\theta_q & -\cos 2\theta_q
\end{pmatrix}_{ab} = 2 \hat{q}_a \hat{q}_b - \delta_{ab}
$. Also $K^{b0}_0(q)= K^{0b}_0(q)$. 
\\ \\ 
The Ward identity $q_a K^{ab}_0(q) + iq_0 K^{0b}_0(q)=0$ can be easily checked in the integrated and un-integrated forms. Coupling to the $\Gamma$ nematic field will introduce more complicated harmonics in the EM response kernels as well as a Hall conductivity (Chern-Simons) contribution. Also when analytically continuing to the retarded branch, we use the following 
\begin{align*}
|q_0| \rightarrow -i \omega + 0^+, \qquad 
\lql = \left\{
\begin{matrix}
\sqrt{v_F^2 |\qq|^2 -\omega^2}, \quad \left(\frac{\omega}{v_F |\qq|}\right)^2 < 1 \\
-i \sqrt{\omega^2 - v_F^2 |\qq|^2},\quad \left(\frac{\omega}{v_F |\qq|}\right)^2 > 1
\end{matrix}
\right.
\end{align*}

\subsection{Response of \texorpdfstring{$K_2^{\mu b}$}{} and its Ward identity}

These are diagrams which have an outgoing (to the right) current vertex that originates from the quadrupole vertex ( shaded blobs). They produce the first non-trivial Hall response. 

\Hline
\begin{align*}
K_2^{ab}(iq_0,\qq) &= \left(
\feny{width=2.5cm}{diagrams/feynman/gg.pdf}
-\feny{width=1.8cm}{diagrams/feynman/tadpole_gg.pdf} +
\feny{width=2.5cm}{diagrams/feynman/gj_gj_up.pdf}+
\feny{width=2.5cm}{diagrams/feynman/gj_gj_down.pdf}
\right) \\
&= -\left(\frac{4e^2}{k_F^4}\right) \idk \sk \left[
+\,2iq_0 (\tau^i\kk)_a (\tau^j\kk)_b \left(\frac{G_1 - G_2 }{(2v_F\Delta)^2}\right) i\epsilon_{\mu \nu} \right. \\
&\hspace{4.5cm}\left.
+\left(\keriq\right) \partial_a \xi_\kk\;(\kk^T \tau^i \kk)(\tau^j \kk)_b \left(\frac{G_1^2-G_2^2}{2v_F\Delta} \right)\delta_{\mu \nu} \right. \\
&\hspace{4.5cm}\left.
-\left(\keriq\right)\partial_a\xi_\kk\;(\qq^T \tau^i \kk)(\tau^j\kk)_b \left(\frac{G_1^2 +G_2^2}{2v_F\Delta}\right) i\epsilon_{\mu \nu}
\right] \Gamma_{\mu i}\Gamma_{\nu j}\\
&= -\left(\frac{4e^2}{k_F^2}\right) \left(\frac{r}{2v_F\Delta}\right)
(\epsilon_{ij}\epsilon_{\mu\nu}\Gamma_{\mu i}\Gamma_{\nu j})\; q_0 \epsilon_{ab} \\
&\hspace{0.5cm}
+\left(\frac{2e^2}{k_F^2}\right) \left(\frac{\bar{r}}{2v_F\Delta}\right) (\epsilon_{ij}\epsilon_{\mu\nu}\Gamma_{\mu i}\Gamma_{\nu j}) q_0 \left(1-\frac{|q_0|}{\lql}\right) \left\{
\epsilon_{ab} + 
\begin{pmatrix}
-\sin 2\theta_q & \cos 2\theta_q \\
\cos 2\theta_q & \sin 2\theta_q 
\end{pmatrix}_{ab}
\right\} \\
&\hspace{0.5cm}
+\left(\frac{e^2 s}{k_F}\right)\Gamma_{\mu i}\Gamma_{\mu j} \left(\frac{v_F |\qq|}{\lql}\right) \\ 
&\hspace{1cm}{\times \Bigg\{  }+\delta_{ij}\delta_{ab}\\
&\left. \hspace{1.5cm}
-\left(\kertan\right)^2 \left[
\delta_{ij} \begin{pmatrix}
\cos 2\theta_q & \sin 2\theta_q \\ \sin 2\theta_q & -\cos 2\theta_q 
\end{pmatrix}_{ab}	
+\sigma_{ij}^z \begin{pmatrix}
\cos 2\theta_q & -\sin 2\theta_q \\ -\sin 2\theta_q & -\cos 2 \theta_q
\end{pmatrix}_{ab}
+\sigma_{ij}^x \begin{pmatrix}
\sin 2\theta_q & \cos 2\theta_q \\ \cos 2\theta_q & -\sin 2\theta_q 
\end{pmatrix}_{ab}
\right] \right. \\
&\left. \hspace{1.5cm} 
-\left(\kertan\right)^4 
\left[
\sigma^z_{ij} \begin{pmatrix}
\cos 4\theta_q & -\sin 4\theta_q \\ \sin 4\theta_q & \cos 4\theta_q 
\end{pmatrix}_{ab}
+\sigma^x_{ij} \begin{pmatrix}
\sin 4\theta_q & \cos 4\theta_q \\ -\cos 4\theta_q & \sin 4\theta_q 
\end{pmatrix}_{ab}
\right]
\right. \Bigg\}
\end{align*}
\\
\Hline 
\begin{align*}
K_2^{0b}(iq_0,\qq) &= \left(
\feny{width=2.5cm}{diagrams/feynman/qg_gj_up.pdf} +
\feny{width=2.5cm}{diagrams/feynman/qg_gj_down.pdf}
\right) \\
&= \left(\frac{4e^2}{k_F^4}\right) \idk \sk \left[ \left(\kervq\right)(\kk^T\tau^i\kk)(\tau^j\kk)_b\left(\frac{G_1^2-G_2^2}{2v_F\Delta}\right)\delta_{\mu\nu} \right.\\
&\hspace{4.5cm}\left.
+2(\qq^T \tau^i \kk)(\tau^j\kk)_b \left(\frac{G_1-G_2}{(2v_F\Delta)^2}\right) i\epsilon_{\mu\nu} \right. \\
&\hspace{4.5cm}\left. -\left(\kervq\right) (\qq^T\tau^i\kk)(\tau^j\kk)_b \left(\frac{G_1^2+G_2^2}{2v_F\Delta}\right) i\epsilon_{\mu\nu}
\right]\Gamma_{\mu i}\Gamma_{\nu j}\\
&= -i \left(\frac{4e^2}{k_F^2} \right) \left(\frac{r}{2v_F\Delta}\right) (\epsilon_{ij}\epsilon_{\mu\nu}\Gamma_{\mu i} \Gamma_{\nu j} )\; q_a \epsilon_{ab} \\
&\hspace{0.25cm}+i\left(\frac{4e^2}{k_F^2}\right)\left(\frac{\bar{r}}{2v_F\Delta}\right)(\epsilon_{ij}\epsilon_{\mu\nu}\Gamma_{\mu i} \Gamma_{\nu j} ) \left(1 -\frac{|q_0|}{\lql}\right) q_a \epsilon_{ab} \\
&\hspace{0.25cm}+i\left(\frac{2e^2s}{k_F}\right)(\Gamma_{\mu i }\Gamma_{\mu j}) \frac{q_0}{\lql} \left\{ 
\left(\kertan\right)\delta_{ij} \begin{pmatrix} \cos \theta_q \\ \sin \theta_q\end{pmatrix}_b
\right. \\
& \hspace{4cm}\left.+\left(\kertan\right)^3 \left[
\sigma^z_{ij} \begin{pmatrix} -\cos 3\theta_q \\ \sin 3\theta_q \end{pmatrix}_b 
- \sigma^x_{ij} \begin{pmatrix}
\sin 3\theta_q \\ \cos 3\theta_q 
\end{pmatrix}_b
\right] \right\}
\end{align*}
\Hline 
\\ \\  
where a tedious calculation will show a line by line cancellation in the Ward identity $q_a K^{ab}_2 + iq_0 K^{0b}_2 = 0$. Moreover the first lines of either $K_2$ contribute to a Hall (Chern-Simons) response. 

\subsection{Response of \texorpdfstring{$K_1^{\mu \nu}$}{} }

The remaining set of response kernels which are order $O(\Gamma^2)$ are as follows

\Hline
\begin{align*}
K_1^{ab}(q) &=  \left(
\feny{width=2.5cm}{diagrams/feynman/jgjg.pdf}
- \feny{width=1.8cm}{diagrams/feynman/tadpole_m_gg.pdf}
\right) +
\left(
\feny{width=2.5cm}{diagrams/feynman/jg_jg_up.pdf} 
+\feny{width=2.5cm}{diagrams/feynman/jg_jg_down.pdf}
\right) \nl &+
\left(
\feny{width=2.5cm}{diagrams/feynman/jggj_up.pdf} 
+\feny{width=2.5cm}{diagrams/feynman/jggj_down.pdf} 
\right)
\end{align*}
where 
\begin{flalign*}
&\left(
\feny{width=2.5cm}{diagrams/feynman/jgjg.pdf}
- \feny{width=1.8cm}{diagrams/feynman/tadpole_m_gg.pdf}
\right)  &\\ &=
\frac{2e^2}{k_F^4} \idk \sk 
\left\{
(\kk^T \tau^i\kk) (\kk^T \tau^j \kk) \; \partial_a \xi_\kk \partial_b \xi_\kk \, 
	\left[G_1 G_2 (G_1^2 + G_2^2) - \left( \frac{iq_0}{v_F\delta q -iq_0 }\right)\frac{G_1^2+G_2^2}{(2v_F\Delta)^2}\right] \right. &\\
&\left.\hspace{3.5cm}+2 (\tau^i\kk)_a (\kk^T \tau^j \kk)\; \partial_b \xi_\kk \left[\frac{G_1^2-G_2^2}{2v_F\Delta}\right] 
\right\}\Gamma_{\mu i}\Gamma_{\mu j} &\\
\end{flalign*}
\begin{flalign*}
&\left(
\feny{width=2.5cm}{diagrams/feynman/jg_jg_up.pdf} 
+\feny{width=2.5cm}{diagrams/feynman/jg_jg_down.pdf} 
\right) &\\ & = 
-\frac{4e^2}{k_F^4} \idk \sk \partial_b \xi_\kk \; 
\left\{
	(\tau^i\kk)_a (\kk^T\tau^j\kk)
		\left[ 
			\frac{G_1^2-G_2^2}{2v_F\Delta}+\left(\frac{iq_0}{v_F\delta q-iq_0}\right) \frac{G_1^2-G_2^2}{2v_F\Delta}
		\right]	\Gamma_{\mu i}\Gamma_{\mu j} \right. &\\
&\left.\hspace{4.5cm}-(\tau^i\kk)_a (\qq^T\tau^j\kk) \left(\frac{iq_0}{v_F\delta q-iq_0}\right) \left(\frac{G_1^2+G_2^2}{2v_F\Delta}\right) i\epsilon_{\mu\nu} \Gamma_{\mu i}\Gamma_{\nu j}
\right\} &\\
\end{flalign*}
\begin{align*}
&\left(
\feny{width=2.5cm}{diagrams/feynman/jggj_up.pdf} 
+\feny{width=2.5cm}{diagrams/feynman/jggj_down.pdf} 
\right) \\ &= 
-\frac{2e^2}{k_F^4} \idk \sk (\kk^T \tau^i \kk)(\kk^T \tau^j \kk)\; \partial_a \xi_\kk \partial_b \xi_\kk \; 
\left\{
	G_1 G_2(G_1^2+G_2^2) + \left(\frac{iq_0}{v_F\delta q- iq_0 }\right) 
		\left[ G_1 G_2 (G_1^2+G_2^2) - \frac{2G_1 G_2}{(2v_F\Delta)^2}\right]
\right\}\Gamma_{\mu i} \Gamma_{\mu j} \\
&+\frac{4e^2}{k_F^2} \idk \sk (\qq^T \tau^i \kk)(\kk^T \tau^j \kk) \; \partial_a \xi_\kk \partial_b \xi_\kk 
\left(\frac{iq_0}{(v_F\delta q -iq_0)^2}\right) \left[\frac{G_1^2-G_2^2}{2v_F\Delta}\right] \Gamma_{\mu i }\Gamma_{\mu j}
\end{align*}

\Hline

\begin{align*}
K_1^{0b}(q) &=
\left(\feny{width=2.5cm}{diagrams/feynman/qgjg.pdf} 
+\feny{width=2.5cm}{diagrams/feynman/qggj_up.pdf}
+\feny{width=2.5cm}{diagrams/feynman/qggj_down.pdf}\right)
\end{align*}
where
\begin{flalign*}
&\left(\feny{width=2.5cm}{diagrams/feynman/qgjg.pdf}\right) &\\ &=
\frac{2e^2}{k_F^4} \idk \sk (\kk^T\tau^i\kk )(\kk^T\tau^j\kk )\; \partial_b \xi_\kk 
	\left[
		G_1^2 G_2^2 + \left(\frac{iq_0}{v_F\delta q- iq_0 }\right) \frac{G_1^2+G_2^2}{(2v_F\Delta)^2} 
	\right]\Gamma_{\mu i} \Gamma_{\mu j}
\end{flalign*}
\begin{flalign*}
&\left(\feny{width=2.5cm}{diagrams/feynman/qggj_up.pdf}
+\feny{width=2.5cm}{diagrams/feynman/qggj_down.pdf}\right) &\\ &=
\frac{2e^2}{k_F^4}\idk \sk (\kk^T\tau^i\kk )(\kk^T\tau^j\kk )\;\partial_b \xi_\kk \left(\frac{iq_0}{v_F\delta q -iq_0}\right)
	\left[
		G_1 G_2(G_1^2+G_2^2) - \frac{2G_1 G_2}{(2v_F\Delta)^2}
	\right]\Gamma_{\mu i}\Gamma_{\mu j} 
&\\
&-\frac{4e^2}{k_F^4} \idk \sk (\qq^T\tau^i\kk)(\kk^T\tau^j \kk) \; \partial_b \xi_\kk 
	\left(\frac{iq_0}{(v_F\delta q -iq_0)^2}\right)\left[\frac{G_1^2-G_2^2}{2v_F\Delta}\right]\Gamma_{\mu i }\Gamma_{\mu j}
\end{flalign*}

\Hline 

\begin{align*}
K_1^{00}(q) &=
\left(\feny{width=2.5cm}{diagrams/feynman/qgqg.pdf} 
+\feny{width=2.5cm}{diagrams/feynman/qggq_up.pdf}
+\feny{width=2.5cm}{diagrams/feynman/qggq_down.pdf}\right)
\end{align*}

where 
\begin{flalign*}
&\left(\feny{width=2.5cm}{diagrams/feynman/qgqg.pdf}\right) &\\ &=
-\frac{2e^2}{k_F^4} \idk \sk (\kk^T\tau^i\kk )(\kk^T\tau^j\kk )
	\left[
		G_1^2 G_2^2 + \left(\frac{iq_0}{v_F\delta q- iq_0 }\right) \frac{G_1^2+G_2^2}{(2v_F\Delta)^2} 
	\right]\Gamma_{\mu i} \Gamma_{\mu j}
\end{flalign*}
and 
\begin{flalign*}
&\left(\feny{width=2.5cm}{diagrams/feynman/qggq_up.pdf}
+\feny{width=2.5cm}{diagrams/feynman/qggq_down.pdf}\right) &\\ &=
-\frac{2e^2}{k_F^4} \idk \sk (\kk^T \tau^i \kk)(\kk^T \tau^j \kk)\;  
\left\{
	G_1 G_2(G_1^2+G_2^2) + \left(\frac{iq_0}{v_F\delta q- iq_0 }\right) 
		\left[ G_1 G_2 (G_1^2+G_2^2) - \frac{2G_1 G_2}{(2v_F\Delta)^2}\right]
\right\}\Gamma_{\mu i} \Gamma_{\mu j} \\
&+\frac{4e^2}{k_F^2} \idk \sk (\qq^T \tau^i \kk)(\kk^T \tau^j \kk) 
\left(\frac{iq_0}{(v_F\delta q -iq_0)^2}\right) \left[\frac{G_1^2-G_2^2}{2v_F\Delta}\right] \Gamma_{\mu i }\Gamma_{\mu j}
\end{flalign*}

\Hline

These response kernels can be evaluated using the same methods of those used to evaluate $K_2^{\mu \nu}(q)$, but they do not add much to the discussion. However, from a tedious calculation, in their un-integrated form they satisfy the necessary Ward identities $q_a K_1^{a\mu}(q)+iq_0 K_1^{0\mu}(q)=0$.  

\end{widetext}


\bibliographystyle{apsrev}
\bibliography{references.bib}

\begin{thebibliography}{78}
\bibitem[{sof()}]{soft}
\bibinfo{note}{{Here we take the term soft mode to mean a light boson near the
  instability but not exactly massless.}}
\bibitem[{tec({\natexlab{a}})}]{technically}
\bibinfo{note}{{Technically speaking when the $\psi$ fermion is spinful and
  isospin is physical spin, the $O(2)$$_\text{iso}\times$$O(2)_\text{rot}$
  symmetry group needs to be broken down to a single $O(2)$ with the constraint
  that $\theta \equiv \varphi/2$. This is because a physical rotation from the
  isometry group of $\mathbb{E}^2$ has an action on $\psi$ which is now a
  spin-1/2 irreducible representation of the same isometry group.}}
\bibitem[{cav()}]{caveat}
\bibinfo{note}{{There is a caveat when attempting to integrate out a gapless
  fermionic theory at $T=0$. This is due to the presence of zero modes in the
  fermionic determinant. In the present situation, we shall adopt the
  convention of a finite $T>0$ prior to performing the integration. Thereafter
  we take the $T\rightarrow 0^+$ limit.}}
\bibitem[{con()}]{convenience}
\bibinfo{note}{{As a matter of convenience, we will continue to use the
  fundamental representation of $SU(2)$ acting on $\phi$, but with the
  understanding that the Cartan subgroup of isospin rotations in the $\vec{s}$
  direction is identified with the central $U(1)$ of $U(2)$$\cong {U}(1)
  \times$ $SU(2)$.}}
\bibitem[{der()}]{derivation}
\bibinfo{note}{{In Appendix \ref{app:NG_UN}, we present the derivation of the
  broken symmetry action of the $U(N)$ symmetric case for general $N$.}}
\bibitem[{wor()}]{work}
\bibinfo{note}{{We work in the zero temperature limit and cutoffs in Matsubara
  frequency are not required.}}
\bibitem[{div()}]{divided}
\bibinfo{note}{{We have divided by the space-time volume factors.}}
\bibitem[{tec({\natexlab{b}})}]{technically2}
\bibinfo{note}{{Technically $j_1^a$ is only gauge invariant up to linear order
  in $(k+eA)$ which is typical of linear response calculations.}}
\end{thebibliography}

\end{document}